\documentclass{SciPost}

\hypersetup{
    colorlinks,
    linkcolor={red!50!black},
    citecolor={blue!50!black},
    urlcolor={blue!80!black}
}

\usepackage[bitstream-charter]{mathdesign}
\newcommand*{\matminus}{\mathbin{\text{--}}}
\DeclareSymbolFont{usualmathcal}{OMS}{cmsy}{m}{n}
\DeclareSymbolFontAlphabet{\mathcal}{usualmathcal}

\usepackage{cite}
\usepackage{graphicx, psfrag, dsfont, amssymb}
\usepackage{float}
\usepackage{caption, comment}
\usepackage[mathscr]{euscript}
\usepackage{dynkin-diagrams}
\usepackage{appendix}

\usepackage{coolstr}
\usepackage{hyperref}
\usepackage[normalem]{ulem}
\hypersetup{pdfauthor={Name}}
\usepackage{color}
\usepackage{chngcntr}
\counterwithin{table}{subsection}
\usepackage{booktabs,caption}
\usepackage{lscape}
\usepackage{slashed}
\definecolor{Mygrey}{gray}{0.8}
\definecolor{Mywhite}{gray}{1.0}

\newcommand{\be}{\begin{equation}}
\newcommand{\ee}{\end{equation}}
\newcommand{\bea}{\begin{eqnarray}}
\newcommand{\eea}{\end{eqnarray}}
\usepackage{multirow,longtable,enumerate,bm}
\usepackage{mismath}
\usepackage[flushleft]{threeparttable}

\usepackage{tikz}
\usepackage{tikz-cd}
\usetikzlibrary{calc,topaths,decorations,decorations.pathmorphing,arrows,decorations.markings,cd}
\tikzset{
->-/.style args={#1rotate#2}{decoration={markings, mark=at position #1 with {\arrow[scale=1.5,rotate = #2 ]{stealth}}}, postaction={decorate}}
}

\tikzset{line/.style={line width=0.25mm},
curve/.style={line,smooth,tension=1},
->-/.style={decoration={
  markings,
  mark=at position #1 with {\arrow[>=stealth]{>}}},postaction={decorate}},
-<-/.style={decoration={
  markings,
  mark=at position #1 with {\arrow[>=stealth]{<}}},postaction={decorate}},
}

\usetikzlibrary{arrows.meta,calc,decorations.markings,bending,positioning}

\newcommand\bi{\begin{itemize}}
\newcommand\ei{\end{itemize}}

\newcommand\tb{{\tilde b}}

\newcommand\ZZ{\hbox{Z\kern-.4emZ}}
\newcommand\sZZ{\hbox{\sevenfont Z\kern-.4emZ}}

\newcommand{\Comment}[1]{{}}

\makeatletter
\newcommand*{\owedge}{%
  \mathbin{%
    \mathpalette\@owedge{}%
  }%
}
\newcommand*{\@owedge}[2]{%
  \sbox0{$#1\oplus\m@th$}%
  \dimen2=.5\dimexpr\wd0-\ht0-\dp0\relax 
  \dimen@=\dimexpr\ht0+\dp0\relax
  \def\lw{.04}
  \def\radius{.5-\lw/2}%
  \kern\dimen2 
  \tikz[
    line width=\lw\dimen@,
    line join=round,
    x=\dimen@,
    y=\dimen@,
    baseline=\dimexpr-.5\dimen@+\dp0\relax,
  ]
  \draw
    (0,0) circle[radius=\radius]
    (225:\radius) -- (0,.5-\lw) -- (-45:\radius)
  ;%
  \kern\dimen2 
}
\makeatother

\definecolor{Mygrey}{gray}{0.8}
\definecolor{Mywhite}{gray}{1.0}

\colorlet{dred}{red!70!black!100!}

\def\IB{\relax{\rm I\kern-.18em B}}

\def\ID{\relax{\rm I\kern-.18em D}}
\def\IE{\relax{\rm I\kern-.18em E}}
\def\IF{\relax{\rm I\kern-.18em F}}
\def\II{\relax{\rm I\kern-.18em I}}

\def\Id{\relax{1\kern-.32em 1}}
\def\IG{\relax\hbox{$\inbar\kern-.3em{\rm G}$}}
\def\IR{\relax{\rm I\kern-.18em R}}

\numberwithin{equation}{section}

\newcommand{\AD}[1]{\textcolor{blue}{\textsf{[AD: #1]}}}

\newcommand{\AK}[1]{\textcolor{brown}{\textsf{[AK: #1]}}}

\begin{document}


\begin{center}{\Large \textbf{\color{scipostdeepblue}{
OPE and correlation functions in a generally covariant form\\
}}}\end{center}

\begin{center}\textbf{
Arpit Das,\textsuperscript{a,b,c,e$\spadesuit$} 
Anatoly Konechny\textsuperscript{d,e$\clubsuit$} and Naveen Balaji Umasankar \textsuperscript{f$\blacklozenge$}
}\end{center}

\begin{center}
{\bf a} School of Mathematics, University of Edinburgh, Edinburgh, EH9 3FD, U.K.
\\
{\bf b} Higgs Centre for Theoretical Physics, University of Edinburgh, Edinburgh EH8 9YL, U.K.
\\
{\bf c} Department of Astrophysics and High Energy Physics, S.N. Bose National Centre for Basic Sciences, Salt Lake, Kolkata 700106, India
\\
{\bf d} Department of Mathematics, Heriot-Watt University
Edinburgh EH14 4AS, United Kingdom
\\
{\bf e} Maxwell Institute for Mathematical Sciences
Edinburgh, United Kingdom
\\
{\bf f} Department of Physics, Yale University, 217 Prospect St, New Haven, CT 06511
\\[\baselineskip]
$\spadesuit$ \href{mailto:email2}{\small adas95@bose.res.in}\,,\quad
$\clubsuit$ \href{mailto:email1}{\small A.Konechny@hw.ac.uk}\,,\quad
$\blacklozenge$ \href{mailto:email1}{\small naveen.umasankar@yale.edu}
\end{center}

\begin{abstract}
In this paper we consider some general aspects of Euclidean conformal field theories on curved spaces of dimension $D\ge 3$. We first look at the OPE of scalar primary fields on conformally flat spaces. Extending the results of \cite{Konechny:2026bqg}, we give a general construction of the descendants' contributions to the OPE in the scalar channel. We then discuss CFTs on non-conformally flat spaces in the ambient space formalism. We investigate the short-distance behaviour of three-point functions on such spaces using the general ansatz proposed by Parisini, Skenderis, and Withers \cite{Parisini:2022wkb, Parisini:2023nbd}. We find that some additional corrections need to be added to the ansatz to ensure the existence of a local covariant OPE.   
\end{abstract}

\tableofcontents

\section{Introduction}\label{sec:intro}

Quantum field theory (QFT) draws much of its predictive organising power from a single structural principle: locality. Observables are constructed from fields defined at points on spacetime, and the dynamics couple these fields only through their behaviour in arbitrarily small neighbourhoods. The physical content of locality, however, runs deeper than the statement that interactions are pointlike. Its sharpest expression is that the entire short-distance structure of a theory, the manner in which operators interfere, mix, and generate one another as their insertion points are brought together, is controlled by a \textit{local algebraic law}, one that is insensitive to the global state in which the theory happens to be prepared. The operator product expansion (OPE) is the precise formulation of that law, and it is arguably the most economical statement of what it means for a QFT to be local (see e.g. \cite{Hollands:2002rz,Hollands:2006ag,Pinamonti:2009zqj} for some earlier works pertaining to OPEs on curved spacetimes and \cite{Hollands:2023txn}, for a general discussion).

 In a CFT on flat space, the conformal group is large enough to fix the whole short-distance structure, which may obscure the fact that the OPE is a consequence of locality rather than of symmetry. Once the same theory is placed on a generic curved manifold, there are no conformal Killing vectors, and what survives is diffeomorphism covariance and Weyl covariance. The working hypothesis of this paper is that these two covariances suffice to fix the kinematic structure of the short-distance expansion: a covariant OPE exists, organised in powers of the geodesic distance between the insertion points, with the kinematic tensors built out of the unit tangent to the connecting geodesic together with the metric and curvature at the expansion point, so that every ingredient is a covariant object living at a single point. Broadly speaking, our goal is a clean separation of kinematics from dynamics.
On conformally flat backgrounds, we show that this separation is complete. The model-dependent OPE coefficients are the flat-space ones, and, apart from one-point functions which carry the dependence on the state, the entire effect of the background sits in the universal, computable curvature dressings---a precise version of the statement that curved space does not give us a new theory, just the same theory read in a different chart. On general curved backgrounds, the situation is subtler. New structures proportional to the Weyl and Cotton tensors appear, and whether their coefficients are universal kinematic dressings of the flat-space data or genuinely new model-dependent input is not settled a priori. Formulating and sharpening this question is one of the aims of this work.

In CFTs the OPE is particularly powerful because it is a convergent expansion with coefficients being power functions (see \cite{Pappadopulo:2012jk}  for a review). On a general curved
background, however, we shall use the OPE only as a formal local
short-distance expansion and defer the corresponding convergence analysis to a future work.

On a flat Euclidean space ${\mathbb R}^{D}$ there is a conserved charge corresponding to dilations and 
the OPE of two scaling fields ${\cal O}_{1}$, ${\cal O}_{2}$ is organised by the scaling dimension:
\be
{\cal O}_{i}(x_1){\cal O}_{j}(x_2) = \sum_{k} |x_{1}-x_2|^{-\Delta_{1}-\Delta_{2}+\Delta_{k}} C_{ij}^{(k)}P^{\mu_{1}\dots \mu_{n_{k}}}(\hat x_{12}) 
{\cal O}_{\mu_{1}\dots \mu_{n_{k}}}^{(k)}(x_1)
\ee
Here we separated the model dependent OPE coefficients $ C_{ij}^{(k)}$ and the tensors $P^{\mu_{1}\dots \mu_{n_{k}}}(\hat x_{12})$ that contract with spin $n_{k}$ scaling fields. These tensors are homogeneous functions of the coordinates and are constructed from the flat metric tensor $\delta_{\mu\nu}$ and the unit-length displacement vector 
\be
\hat x_{12}^{\mu} = \frac{x_{1}^{\mu}-x_{2}^{\mu}}{|x_1-x_2|} \, . 
\ee
One can arrange the expansion so that these tensors are purely kinematic (model independent). To that end we utilise the global conformal symmetry further by organising the OPE into conformal families. The coefficients of the descendant fields are then fully fixed by the conformal symmetry, see e.g. formula \eqref{C_flat} in the main body of the paper that encodes the contributions of all of the descendant fields for the scalar channel. 

In this paper we are interested in CFTs on a Riemannian manifold $({\cal M}, g_{\mu\nu})$ of dimension $D\ge 3$. Such theories are of interest in cosmology and holography approaches to quantum gravity. Even when the main interest is a CFT or a perturbed CFT on a flat space the radial quantisation is performed on a cylinder $S^{D-1}\times{\mathbb R}$ which is a curved space for $D\ge 3$.

The basic observables for a CFT on $({\cal M}, g_{\mu\nu})$ are correlation 
functions 
\be
\langle {\cal O}_{1}(x_1) \dots {\cal O}_{n}(x_{n}) \rangle_{g_{\mu \nu}} 
\ee
They should satisfy the diffeomorphism and Weyl covariance equations.
Given a diffeomorphism 
$$
x \mapsto \varphi(x)
$$
the correlation functions should satisfy 
\be
 \langle {\cal O}_{1}(\varphi(x_1)) \dots {\cal O}_{n}(\varphi(x_n)) \rangle_{g_{\mu \nu}} = \langle {\cal O}_1(x_1) \dots  {\cal O}_n(x_n)\rangle_{\varphi^{*}g_{\mu \nu}} 
 \ee
where for simplicity we assume the fields ${\cal O}_i$ are scalars. If these fields are primary, then for a given Weyl transformation
$$
g_{\mu \nu}(x)\mapsto \Lambda^2(x)g_{\mu\nu}
$$
the correlation functions transform as 
\be \label{Weyl_covariance}
 \langle {\cal O}_{1}(x_1) \dots {\cal O}_{n}(x_{n})\rangle_{\Lambda^2 g_{\mu \nu}}  = \prod_{i=1}^{n} \Lambda^{-\Delta_{i}}(x_{i}) 
       \langle {\cal O}_{1}(x_1) \dots {\cal O}_{n}(x_{n})\rangle_{g_{\mu \nu}}
       \ee
       where $\Delta_{i}$ are the scaling dimensions. 
If $({\cal M}, g_{\mu\nu})$ admits some conformal Killing vectors, both transformations can be combined to give a symmetry equation with fixed metric. Generically, there are no conformal Killing vectors, and the Weyl covariance equation \eqref{Weyl_covariance} becomes the main constraint on CFT correlators.

\begin{figure}[h!]
    \centering
   \begin{tikzpicture}[scale=1.3,>=latex]
    \begin{scope}
     \draw[cyan, very thick] (0,0)  to[bend right] (1.7,0.4);
   \draw[cyan, very thick] (1.7,0.4)  to[bend left] (2.5,0.6);
    \draw[cyan, very thick] (2.5,0.6)  to[bend right] (2.9,0.7);
     \draw[cyan, very thick] (2.9,0.7)  to[bend right] (2.9,1);
         \draw[cyan, very thick] (2.9,1)  to[bend left] (3,2.2);
    \draw[black,thick,->] (0,0) --(0.5,-0.14);
   \end{scope}
   \fill[black] (0,0)circle (1.5pt);
    \fill[black] (3,2.2)circle (1.5pt);
    \draw (-0.3,-0.05) node{$x_1$};
     \draw (3.35,2.2) node{$x_2$};
        \draw (2,0.3) node{$ d$};
     \draw (0.5,-0.35) node{$ {\bf t}$};
    \end{tikzpicture}
   \caption{The geodesic connecting $x_1$ and $x_2$ and the tangent vector $t^{\mu}$.}
   \label{geodesic_pic1}
   \end{figure}
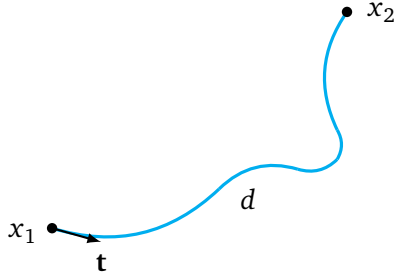

With OPE being a purely local concept, it must exist when the two insertion points are sufficiently close to each other. The diffeomorphism and Weyl covariance then impose constraints on the OPE. To implement the diffeomorphism invariance, it was suggested in \cite{Konechny:2026bqg} to organise the OPE as an expansion in powers of the geodesic distance $d=d(x_1,x_2)$ between the insertion points $x_1$ and $x_2$. Furthermore,  the tensors $P^{\mu_{1}\dots \mu_{n_{k}}}(\hat x_{12})$ present in the flat space OPE should be covariantised as well. If we choose $x_1$ as the expansion point, the curved space  tensors $P^{\mu_{1}\dots \mu_{n_{k}}}(t^{\mu},g_{\alpha \beta})$ should be constructed out of  $t^{\mu}$ -- the normalised outward tangent vector at $x_1$ to the geodesic connecting $x_1$ to $x_2$ and on the metric and curvature tensors at $x_1$, see Figure \eqref{geodesic_pic1} for illustration. 

Since the tensors $P^{\mu_{1}\dots \mu_{n_{k}}}(t^{\mu},g_{\alpha \beta})$ should be nonsingular, the curvature and its derivatives will bring in a positive degree of homogeneity under locally constant Weyl rescalings, i.e., those with $\Lambda$ being constant in some neighbourhood of $x_1$. It makes sense to fix the degree of homogeneity as $N\ge 0$ and to denote the corresponding tensors as  $P^{\mu_{1}\dots \mu_{n_{k}}}_{N}(t^{\mu},g_{\alpha \beta})$. The resulting covariant OPE then has the form 
\be \label{curved_OPE}
{\cal O}_{i}(x_{1}) {\cal O}_{j}(x_2) = \sum_{k}\sum_{N=0}^{\infty}  C_{ij}^{(k,N)} \frac{1}{ d^{\Delta_{i} + \Delta_{j} - \Delta_{k}-N}} 
 P^{\mu_{1}\dots \mu_{n_{k}}}_{N}(t^{\mu},g_{\alpha \beta}) {\cal O}_{\mu_{1}\dots \mu_{n_{k}}}^{(k)}(x_1) 
\ee
 where we assume that the expansion is organised so that the tensors $P^{\mu_{1}\dots \mu_{n_{k}}}_{N}$ are purely kinematic while the OPE coefficients $C_{ij}^{(k,N)}$ carry all model-dependent information.
While the covariant quantities $d$ and $t^{\mu}$, to the best of our knowledge, have not been used to analyse  general CFTs before, they are basic objects in the heat kernel and Hadamard parametrix techniques used to analyse short distance behaviour of free theories' propagators on curved spaces, see e.g. \cite{Vassilevich:2003xt}, \cite{Decanini:2005gt}, \cite{Decanini:2005eg}
for a review. In that literature one uses the Synge's function $\sigma(x_1,x_2)=d^2(x_1,x_2)/2$ and its derivative at $x_1$
$$
\nabla^{\mu}\sigma(x_1,x_2)=-d(x_1,x_2)t^{\mu} \, .
$$

While we are not concerned at this stage with the convergence issues, we would like to be able to treat the curvature terms perturbatively. This suggests that to have a well-controlled OPE of the form \eqref{curved_OPE}, we need the distance $d(x_1,x_2)$ to be much smaller than the geodesic distances to other insertions when using \eqref{curved_OPE} in a particular correlation function, and we also need it to be much smaller than the local curvature radius at the centre of expansion
\be
d(x_1,x_2)\ll R_{\rm curv}(x_1) \, .
\ee
 
Two conceptual questions arise in regard to expansion \eqref{curved_OPE}: how does one separate the kinematic tensors from the dynamic OPE coefficients, and what is the relationship of the curved theory data (e.g. the OPE coefficients) to that of the flat one? 
A natural place to start investigating these questions is the conformally flat manifolds. A primary field $\hat {\cal O}_i$ on $({\cal M}, g_{\mu\nu})$ 
with $g_{\mu\nu}=\Lambda^{2}(x)\delta_{\mu\nu}$ is related to the flat space field ${\cal O}_i$ by
\be
\hat {\cal O}_i(x) = \Lambda^{-\Delta_i}(x){\cal O}_i(x) \, .
\ee
For a globally conformally flat manifold, we can carry over the two- and three-point functions from flat space. This was used in \cite{Konechny:2026bqg} to calculate some universal curvature terms in the OPE. For the OPE of two scalar primaries $\hat {\cal O}_{1}$, 
$\hat {\cal O}_{2}$ the covariant OPE in the scalar channel 
 $\hat {\cal O}_3$  was found to start as    
 \bea \label{cov_exp2}
\hat {\cal O}_{1}(x_1)\hat {\cal O}_{2}(x_2) &&= \frac{C_{123}}{ d^{\Delta_{1} + \Delta_{2}-\Delta_{3}}}\Bigl(  
\hat {\cal O}_{3}(x_1) + \alpha_{123}
d  t^{\mu}\nabla_{\mu}\hat {\cal O}_{3}(x_1)  \nonumber \\
&& + \beta_{123}
d^2  t^{\alpha} t^{\beta}\nabla_{\alpha}\nabla_{\beta}\hat {\cal O}_{3}(x_1)    + 
\gamma_{123}
d^2   g^{\alpha \beta} \nabla_{\alpha}\nabla_{\beta}\hat {\cal O}_{3}(x_1) \nonumber \\
&& + A_{123} d^2  R \hat {\cal O}_{3}(x_1) + B_{123} d^2  R_{\alpha \beta} t^{\alpha} t^{\beta}  \hat {\cal O}_{3}(x_1) +  \dots \Bigr)
\eea 
where $R$, $R_{\alpha \beta}$ are the scalar and Ricci curvatures for the metric $g_{\mu\nu}$ evaluated at $x_1$. The coefficients $\alpha_{123}$, $\beta_{123}$, $\gamma_{123}$ standing at the covariantised derivative terms are the same as the flat space descendants coefficients while the coefficients $A_{123}$ and $B_{123}$  at the curvature terms can be expressed as linear combinations of $\alpha_{123}$, $\beta_{123}$, $\gamma_{123}$. In section \ref{confflat_sec} we will show how the expansion in \eqref{cov_exp2} can be systematically rearranged so that only the flat space descendant coefficients appear. 

In the identity channel $\hat {\cal O}_3={\mathbb 1}$ there are no descendants and only the curvature terms appear. Formula \eqref{cov_exp2} then reduces to 
\be \label{2pt_gen_correction_1}
 \hat {\cal O}_{1}(x_1) \hat {\cal O}_{1}(x_2) =   
 \frac{1}{ d^{2\Delta_{1}}}\left( 1 + 
\frac{\Delta_{1}}{6} d^2  P_{\mu \nu}(x_1) t^{\mu} t^{\nu} + \dots \right)
\ee
where 
\be \label{Schouten_1}
  P_{\mu \nu} = \frac{1}{D-2}\left(  R_{\mu \nu} - \frac{1}{2(D-1)} g_{\mu \nu} R \right) 
\ee
is the normalised Schouten tensor.

In the present paper we extend the results of \cite{Konechny:2026bqg} in two directions. Firstly, for conformally flat manifolds, we find a more efficient way to generate the expansion \eqref{cov_exp2} to arbitrary order using the explicit formula for the flat space descendants operator. 
All terms come proportional to the flat space OPE coefficients.
The conformally flat descendant operator is obtained through three separate maps. The Weyl rescaling factor $\Lambda(x_2)$ at the other end of the geodesic 
and the affine displacement vector $x_{2}^{\mu}-x_{1}^{\mu}$ each need to be expressed in terms of the geodesic data and curvature derivatives at the expansion point $x_1$. The third map involves expressing the flat space descendants in terms of the covariant derivatives of the curved space operator $\hat O_{3}$. This last map can be described in terms of conformal tractor bundles that also partly clarify the Weyl covariance properties of the descendant towers on a curved space.  We explain this in detail in section \ref{confflat_sec}. 

In section \ref{ambient_sec} we relax the assumption of conformal flatness. On a general background, there is no flat-space parent theory to Weyl-transform, and one needs an independent construction of the correlators whose short-distance limit the OPE must reproduce. For this we use the recent ambient-space proposal of \cite{Parisini:2022wkb,Parisini:2023nbd} for constructing CFT correlation functions on general manifolds using the ambient space construction of \cite{AST_1985__S131__95_0}, \cite{Fefferman:2007rka}.
 The main ingredient in \cite{Parisini:2022wkb, Parisini:2023nbd} for constructing correlation functions is the Weyl covariant quantity $\widetilde X_{ij}$ that is assigned to any pair of points on the light-like subspace of the ambient space. After reviewing the basics of the ambient space approach, we study the short-distance expansion of $\widetilde X_{ij}$, which we then use to analyse the short-distance behaviour of the ansatz proposed in \cite{Parisini:2022wkb,Parisini:2023nbd} for the 2- and three-point functions. In section \ref{sec:ansatz_OPE} we show that this short-distance behaviour admits an OPE description only provided certain additional corrections are included into the ansatz. 

We conclude the paper with section \ref{discussion_sec} where we offer a further discussion of our results and future directions.

\section{OPE on conformally flat spaces} \label{confflat_sec}
\subsection{Conformally flat space descendants operator}
We start by reminding some basic facts about descendants' contributions to OPE on flat space.

Let ${\cal O}_{i}$, $i=1,2,3$ be scalar primary fields with scaling dimensions $\Delta_{i}$ and normalised so that on a flat space ${\mathbb R}^{D}$
\begin{equation} \label{2pt_flat}
    \langle {\cal O}_{i}(x_1){\cal O}_{j}(x_2)\rangle_{\delta_{\mu\nu}}=\frac{\delta_{ij}}{|x_1-x_2|^{2\Delta_{i}}}\, .
\end{equation}
Here and elsewhere we use $\langle \dots \rangle_{g_{\mu\nu}}$ to denote a correlation function on a space with metric $g_{\mu\nu}$.
The three-point function of these fields has the form 
\begin{equation}\label{3pt_flat}
    \langle {\cal O}_{1}(x_1){\cal O}_{2}(x_2){\cal O}_{3}(x_3)\rangle_{\delta_{\mu\nu}}=\frac{C_{123}}{|x_1 - x_3|^{2\alpha_{2}}|x_1-x_2|^{2\alpha_{3}}|x_2-x_3|^{2\alpha_1}}
\end{equation}
where $\vert x_{i}-x_{j}\vert$ are the Euclidean distances, $C_{123}$ is the OPE structure constant and 
\begin{equation}
    \alpha_{1}=\frac{1}{2}(\Delta_{2}+\Delta_{3}-\Delta_{1}) \, , \quad  \alpha_{2}=\frac{1}{2}(\Delta_{1}+\Delta_{3}-\Delta_{2}) \, , \quad  \alpha_{3}=\frac{1}{2}(\Delta_{1}+\Delta_{2}-\Delta_{3}) \, .
    \end{equation}
The form of the three-point function fixes the descendants' contributions to the OPE, which can be written as  
\begin{equation}
{\cal O}_2(x_2){\cal O}_1(x_1)
=
\frac{C_{123}}{|x_1-x_2|^{2\alpha_3}}
C_{\mathrm{flat}}(\Delta x^\mu){\cal O}_3(x_1)
+\cdots .
\label{OPE_flat}
\end{equation}
where $ C_{\rm flat}(\Delta x^{\mu})$ is an infinite order formal differential operator. This operator can be written explicitly as \cite{PETKOU1996180} 
\begin{eqnarray}\label{C_flat}
  C_{\rm flat}(\Delta x^{\mu}) &=&\frac{1}{B(\alpha_1,\alpha_2)}\int\limits_{0}^{1} \!\! ds\, s^{\alpha_{1}-1}(1-s)^{\alpha_{2}-1}
\sum_{m=0}^{\infty} \frac{\left[-\frac{1}{4}\Delta x^{2}s(1-s)\right]^{m}}{m! (\Delta_{3}+1-D/2)_{m}} e^{s\Delta x\cdot \partial}(\partial^{2})^m \nonumber \\
&=& 1 + \frac{\alpha_1}{\Delta_{3}}\Delta x\cdot \partial + \frac{\alpha_1 \alpha_{2}}{2\Delta_{3}(\Delta_{3}+1)(D-2\Delta_{3}-2)} \Delta x^2 \partial^2  \nonumber \\
&& \ \ \ + \frac{\alpha_{1}(\alpha_{1}+1)}{2\Delta_{3}(\Delta_{3}+1)}  \Delta x^{\mu}\Delta x^{\nu}\partial_{\mu}\partial_{\nu} 
+ \dots 
\end{eqnarray}
where
\begin{equation}
\Delta x^{\mu} = x_{2}^{\mu}-x_{1}^{\mu}\, , 
    \end{equation}
the derivative $\partial^{\mu}$ is with respect to $x_{1}$ but is assumed to commute with the $\Delta x^{\mu}$
 vector in the exponential, also
\be
\left(\Delta_{3}+1-\frac{D}{2}\right)_{m} = \frac{\Gamma\left(\Delta_{3}+1-\frac{D}{2}+m\right)}{\Gamma\left(\Delta_{3}+1-\frac{D}{2}\right)} \, .
\ee
Here and elsewhere we assume that the dimension $\Delta_{3}$ is generic so that the denominators in the above expressions do not vanish. If it is not the case, one needs to remove the null fields from the expansion. The defining property of $C_{\mathrm{flat}}(\Delta x^\mu)$ is that
inserting \eqref{OPE_flat} into $\langle{\cal O}_2(x_2){\cal O}_1(x_1){\cal O}_3(x_3)\rangle$ and using \eqref{2pt_flat} reproduces \eqref{3pt_flat}.

We consider next a CFT on a $D$-dimensional Riemannian manifold $({\cal M},  g_{\mu\nu})$ with $D>2$. We will assume that ${\cal M}$ is locally conformally flat and will work in a single coordinate patch with coordinates $x^{\mu}$ in which the metric is 
\begin{equation}\label{metric_transform}
     g_{\mu\nu}=e^{2\sigma(x)}\delta_{\mu\nu} \, .
\end{equation}
Assuming our CFT is invariant under Weyl transformations, the 
primary fields ${\cal O}_{i}$ transform as 
\begin{equation} \label{field_Weyl}
    \hat {\cal O}_{i}(x) = e^{-\Delta_{i}\sigma(x)}{\cal O}_{i}(x)
\end{equation}
where the hatted operators are those on the curved space with metric
\eqref{metric_transform}.

On a globally conformally flat manifold, the 2- and three-point functions of the fields $\hat {\cal O}_{i}$ can be obtained by Weyl transforming \eqref{2pt_flat} and \eqref{3pt_flat} which gives
\begin{equation} \label{2pt_curved}
     \langle \hat {\cal O}_{i}(x_1)\hat {\cal O}_{j}(x_2)\rangle_{g_{\mu\nu}}=\frac{\delta_{ij}e^{-\Delta_{i}(\sigma(x_1)+\sigma(x_2))}}{|x_1-x_2|^{2\Delta_{i}}} \, , 
\end{equation}
\begin{equation}\label{3pt_curved}
    \langle \hat {\cal O}_{1}(x_1)\hat {\cal O}_{2}(x_2)\hat {\cal O}_{3}(x_3)\rangle_{g_{\mu\nu}}=\frac{C_{123}e^{-\Delta_{1}\sigma(x_1)-\Delta_{2}\sigma(x_2)-\Delta_{3}\sigma(x_3)}}{|x_1 - x_3|^{2\alpha_{2}}|x_1-x_2|^{2\alpha_{3}}|x_2-x_3|^{2\alpha_1}} \, .
\end{equation}

We denote the complete descendant operator on a conformally flat
background by $C_{\mathrm{cf}}(g_{\mu\nu},t^\mu,d)$ and define it
through the OPE
\begin{equation}
\hat{\mathcal O}_2(x_2)\hat{\mathcal O}_1(x_1)
=
\frac{C_{123}}{\widetilde X_{12}^{\alpha_3}}
C_{\mathrm{cf}}(g_{\mu\nu},t^\mu,d)
\hat{\mathcal O}_3(x_1)+\dots ,
\label{OPE_curved}
\end{equation}
where  
\be\label{tildeX12}
\widetilde X_{12} = |x_1-x_2|^2e^{\sigma(x_1)+\sigma(x_2)} \, .
\ee
The subscript $\mathrm{cf}$ records that this operator is fixed by the
locally conformally flat construction of the present section. It should
not be confused with the flat-space operator
$C_{\mathrm{flat}}(\Delta x^\mu)$ appearing in
\eqref{OPE_flat}.

Later in this section we will explain why we chose to include the factor $\widetilde X_{12}^{-\alpha_3}$ into the definition. For now we note that with this factor, in the $x_2\to x_1$ limit we get the correctly normalised leading singularity of the three-point function \eqref{3pt_curved} so that the expansion of $C_{\mathrm{cf}}(g_{\mu\nu},t^\mu,d)$ starts with~1. Moreover, we will argue that this operator depends on the geodesic distance $ d$ between $x_{1}$ and $x_{2}$, the unit tangent vector $ t^{\mu}$, and the curvature tensor calculated with respect to the metric $ g_{\mu\nu}$.

Using the relation between the three-point functions \eqref{3pt_curved} and the definitions \eqref{OPE_flat}, \eqref{OPE_curved} we obtain the relation between the descendant operators
\begin{equation} \label{CC}
    C_{\mathrm{cf}}(g_{\mu\nu},t^\mu,d) \hat {\cal O}_{3} = e^{-\alpha_{2}\sigma(x_1)-\alpha_{1}\sigma(x_2)}C_{\rm flat}(\Delta x^{\mu}){\cal O}_{3} \, .
    \end{equation}
Although the expressions \eqref{2pt_curved}, \eqref{3pt_curved} are only valid on a globally conformally flat manifold, the relation \eqref{CC} only needs local conformal flatness. It can be derived from the definitions \eqref{OPE_flat}, \eqref{OPE_curved} and the transformation law \eqref{field_Weyl}. 
    
\begin{figure}[h!]
    \centering
   \begin{tikzpicture}[scale=1.3,>=latex]
    \begin{scope}
     \draw[cyan, very thick] (0,0)  to[bend right] (1.7,0.4);
   \draw[cyan, very thick] (1.7,0.4)  to[bend left] (2.5,0.6);
    \draw[cyan, very thick] (2.5,0.6)  to[bend right] (2.9,0.7);
     \draw[cyan, very thick] (2.9,0.7)  to[bend right] (2.9,1);
         \draw[cyan, very thick] (2.9,1)  to[bend left] (3,2.2);
    \draw[black,thick,->] (0,0) --(0.5,-0.14);
   \end{scope}
   \fill[black] (0,0)circle (1.5pt);
    \fill[black] (3,2.2)circle (1.5pt);
    \draw (-0.3,-0.05) node{$x_1$};
     \draw (3.35,2.2) node{$x_2$};
       \draw (1.2,1.2) node{$\Delta x^{\mu}$};
        \draw (2,0.3) node{$ d$};
     \draw[black,thick,->] (0,0)--(3,2.2);
     \draw (0.5,-0.35) node{$ {\bf t}$};
    \end{tikzpicture}
   \caption{Two geodesics connecting $x_1$ and $x_2$:  with metric $\delta_{\mu\nu}$ (black) and  with metric 
   $ g_{\mu\nu}=e^{2\sigma(x)}\delta_{\mu\nu}$ (cyan).}
   \label{geodesic_pic}
   \end{figure}
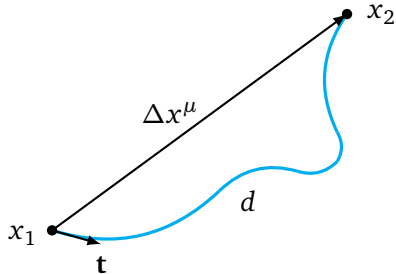

We are going to use \eqref{CC} to obtain an explicit formula for 
 $C_{\mathrm{cf}}(g_{\mu\nu},t^\mu,d)$ which is written entirely in terms of covariant quantities. To this end, we note that the flat space displacement vector $\Delta x^{\mu}$ can be expressed  as a series in the distance 
 $ d$ with the coefficients being constructed out of $t^{\mu}$ and derivatives of $\sigma(x)$, see Figure \ref{geodesic_pic} for illustration.
Generating this expansion is a purely geometric problem which can be  obtained, for example by solving the geodesic equation for the metric $g_{\mu\nu}$ perturbatively in the order of derivatives of $\sigma(x)$ 
at $x=x_1$ (see the appendix in \cite{Konechny:2026bqg}). One can first obtain a series for $|\Delta x|$. It  starts as 
\begin{equation} \label{d_exp}
  |\Delta x| =  d e^{-\sigma(x_1)}\Bigl( 1 - \frac{ d}{2}\partial_{\mu}\sigma t^{\mu} - \frac{ d^2}{24}\Bigl[ 4\partial_{\mu}\partial_{\nu}\sigma t^{\mu} t^{\nu} - 13 ( \partial_{\mu}\sigma t^{\mu})^2 + 5e^{-2\sigma(x_1)}(\partial\sigma)^2\Bigr]+\bigo(d^3) \Bigr)
\end{equation}
where the derivatives of $\sigma$ are all evaluated at $x_1$.
 From this expansion we can obtain an expansion for $\Delta x^{\mu}$
 \begin{eqnarray}\label{dx_exp}
   &&  \Delta x^{\mu} = - |\Delta x|\frac{\partial |\Delta x|}{\partial x^{\mu}_1}=  dt^{\mu}\Bigl[
     1-d(t\cdot \partial \sigma) +\frac{4}{3}d^2(t\cdot \partial \sigma)^2 - \frac{d^2}{3}e^{-2\sigma}(\partial\sigma)^2 -\frac{d^2}{3}\partial_{\mu}\partial_{\nu}\sigma t^{\mu} t^{\nu}\Bigr] \nonumber \\ 
     && +e^{-2\sigma(x_1)}\left(-\frac{2}{3}d^3 (t\cdot \partial \sigma)\partial^{\mu}\sigma+\frac{d^2}{2} \partial^{\mu}\sigma + 
     \frac{d^3}{6}\partial_{\nu}\partial^{\mu}\sigma t^{\nu}\right) + {\cal O}(d^4)
 \end{eqnarray}
where all index contractions are done using the metric $\delta_{\mu\nu}$.
 
 The expansion \eqref{dx_exp} can be substituted into 
 $C_{\rm flat}(\Delta x^{\mu})$. Furthermore, in the factor $e^{-\alpha_{2}\sigma(x_1)-\alpha_{1}\sigma(x_2)}$ we can use the covariant Taylor expansion for $\sigma(x_2)$ along the geodesic in the  metric $ g_{\mu\nu}$ to write it as 
 \begin{equation}
     \sigma(x_2) = e^{ d  t^{\mu} \nabla_{\mu}} \sigma \Bigr|_{x_1} = \sigma(x_1) +  d t^{\mu}  \nabla_{\mu}\sigma(x_1) + \frac{ d^2}{2} t^{\mu} t^{\nu}  \nabla_{\mu} \nabla_{\nu}\sigma(x_1) +\bigo(d^3) 
 \end{equation}
 where $\nabla_{\alpha}$ stands for the covariant derivative in metric $g_{\mu\nu}$.
 Finally, we need to express the derivatives 
 $\partial_{\mu_{1}}\dots \partial_{\mu_{n}}{\cal O}_{3}$ in terms of the covariant derivatives  $ \nabla_{\mu_{1}}\dots  \nabla_{\mu_{n}}\hat {\cal O}_{3}$ and the derivatives of $\sigma$. This last task is an entirely local problem at the point $x_1$ without any reference to the geodesic. This task can be accomplished simply by writing
\begin{equation}
\partial_{\mu_1}\cdots\partial_{\mu_n}{\cal O}_3(x)
=
\partial_{\mu_1}\cdots\partial_{\mu_n}
\left(
e^{\Delta_3\sigma(x)}
\hat{\cal O}_3(x)
\right).
\end{equation}
differentiating and then covariantising the derivatives using 
\begin{equation}
    \partial_{\mu}V_{\nu} = \nabla_{\mu}V_{\nu} +  \Gamma_{\mu \nu}^{\alpha} V_{\alpha},\qquad  \Gamma_{\mu \nu}^{\alpha}=\delta_{\mu}^{\alpha}\partial_{\nu}\sigma + \delta^{\alpha}_{\nu}\partial_{\mu}\sigma-\delta_{\mu\nu}\partial^{\alpha}\sigma \, . 
\end{equation}
We use the convention in which the indices of partial derivatives of $\sigma$ are raised with the flat metric $\delta_{\mu\nu}$.
This way we get 
\begin{equation} \label{OO1}
    \partial_{\mu}{\cal O}_3 = e^{\Delta_{3}\sigma}(\Delta_3 \partial_{\mu}\sigma \hat {\cal O}_3 +  \nabla_{\mu}\hat{\cal O}_3)\, , 
\end{equation}
\begin{equation}\label{OO2}
 \partial_{\mu}\partial_{\nu}{\cal O}_3 = e^{\Delta_{3}\sigma}[\nabla_{\mu}\nabla_{\nu} + (\Delta_3+1)(\partial_{\mu}\sigma  \nabla_{\nu}+\partial_{\nu}\sigma\nabla_{\mu})- \delta_{\mu\nu}\partial^{\alpha}\sigma\nabla_{\alpha}+\Delta_3\partial_{\mu}\partial_{\nu}\sigma +\Delta^2_{3} \partial_{\mu}\sigma\partial_{\nu}\sigma]\hat {\cal O}_3
\end{equation}
We note that while $d$, $ t^{\mu}$ and $ \nabla_{\mu}$ are fully covariant quantities on the space $({\cal M}, g_{\mu\nu})$ the derivatives of $\sigma$  in general are not. We find however, that substituting all of the expansions into \eqref{CC} and collecting terms at the same order of derivatives, at the first three orders all apparently non-covariant terms cancel out and we are left with a covariant expansion where all derivatives of $\sigma$ are put into the Schouten tensor and its covariant derivatives. In section \ref{ambient_sec}, we will give a general argument for the general covariance of $C_{\mathrm{cf}}(g_{\mu\nu},t^\mu,d)$ for $D\ge 3$.

To facilitate a more efficient method of constructing $C_{\mathrm{cf}}(g_{\mu\nu},t^\mu,d)$ we use special conformal coordinates 
at the point $x_{1}$ for which 
\begin{equation} \label{special}
    \sigma(x_1)=0\, , \qquad \partial_{\mu}\sigma(x_1)=0 
    \end{equation}
    and the metric is locally of the form \eqref{metric_transform}. Such a system of coordinates always locally exists. 
    This can be shown as follows. 
    Suppose we have conformally flat coordinates $x^{\mu}$ in which the metric tensor has the form \eqref{metric_transform}. We can choose a conformal diffeomorphism that keeps the metric in the same form and arranges for \eqref{special} to hold. Namely, start with a translation that brings $x_1$ to the origin, followed by a dilation that ensures the first condition in \eqref{special}. This gives the transformation 
    \be \label{conf1}
x^{\mu} \mapsto y^{\mu} = e^{\sigma(x_1)}(x^{\mu}-x_{1}^{\mu}) \, .
    \ee
    Finally, perform a special conformal transformation 
    \be \label{conf2}
y^{\mu} \mapsto z^{\mu} = \frac{y^{\mu}-b^{\mu} y^2}{1-2b\cdot y + b^2 y^2}
    \ee
    with
    \begin{equation}
b_\mu
=
\frac12
\left.
\frac{\partial}{\partial y^\mu}
\left[
\sigma\!\left(x_1+e^{-\sigma(x_1)}y\right)-\sigma(x_1)
\right]
\right|_{y=0}
=
\frac12e^{-\sigma(x_1)}
\partial_\mu\sigma(x_1).
\end{equation}
   to bring the metric to the form 
   \begin{equation}
g'_{\mu\nu}(z)=e^{2\sigma'(z)}\delta_{\mu\nu}.
\end{equation}
    where 
    \begin{equation}
\sigma'(z)
=
\sigma\!\left(
x_1+e^{-\sigma(x_1)}y(z)
\right)
-\sigma(x_1)
+
\ln\!\left(
1-2b\cdot y(z)+b^2y(z)^2
\right),
\label{sigmaprime}
\end{equation}
with $y(z)$ being the inverse of the special conformal transformation given in \eqref{conf2}, and
\begin{equation}
z_1=0,
\qquad
\sigma'(z_1)=0,
\qquad
\partial_{z^\mu}\sigma'(z_1)=0.
\end{equation}
We will assume that $x_{2}$ is sufficiently close to $x_1$ to be in the special conformal patch. In these coordinates, any derivative of $\sigma$ at $x_{1}$ can be expressed via the Schouten tensor $ P_{\mu\nu}$ and its derivatives. This can be obtained using 
    \begin{equation} \label{hatP}
         P_{\mu\nu}(x)=  - \partial_{\mu}\partial_{\nu}\sigma(x) + \partial_{\mu}\sigma(x)\partial_{\nu}\sigma(x) - \frac{1}{2}\partial_{\alpha}\sigma(x)\partial^{\alpha}\sigma(x)  \delta_{\mu\nu} 
    \end{equation}
and taking its covariant derivatives at $x=x_1$. For example, we obtain from evaluating \eqref{hatP} and its first derivative at $x=x_1$
\begin{equation}
    \partial_{\mu}\partial_{\nu}\sigma(x_1)=- P_{\mu\nu}(x_1) \, , \qquad 
     \partial_{\mu}\partial_{\nu}\partial_{\lambda}\sigma(x_1) = -\nabla_{\mu} P_{\nu \lambda}(x_1) \, .
\end{equation}
Note that since the Cotton tensor vanishes on a conformally flat manifold, 
\begin{equation}
\nabla_{\mu} P_{\nu \lambda} = \nabla_{\nu} P_{\mu \lambda}
  \end{equation}
  is in fact a totally symmetric tensor. Furthermore, taking two covariant derivatives of \eqref{hatP} and using
  \begin{equation}
      \partial_{\alpha} \Gamma^{\rho}_{\beta \mu}(x_1) =
      -\delta_{\beta}^{\rho} P_{\alpha \mu} -\delta_{\mu}^{\rho}P_{\alpha\beta}+ g_{\beta\mu} P_{\alpha}^{\rho}
  \end{equation}
  we get 
\begin{eqnarray}    \partial_{\alpha}\partial_{\beta}\partial_{\mu}\partial_{\nu}\sigma(x_1)&=&- \nabla_{\alpha}\nabla_{\beta} P_{\mu\nu}(x_1)+ 2[ P_{\alpha\mu} P_{\beta\nu}(x_1)+ P_{\beta\mu} P_{\alpha\nu}(x_1)+ P_{\alpha\beta} P_{\mu\nu}(x_1)]\nonumber \\ && - g_{\mu\nu} P_{\alpha\lambda} P^{\lambda}_{\beta}(x_1)
     - g_{\beta\mu} P_{\nu \lambda} P^{\lambda}_{\alpha}(x_1)- g_{\beta\nu} P_{\mu\lambda} P^{\lambda}_{\alpha}(x_1) \, .
\end{eqnarray}  
In general we have 
\begin{equation}
     \nabla_{\alpha_1}\dots \nabla_{\alpha_{n}} P_{\mu\nu}(x_1) = 
    -\partial_{\alpha_1}\dots \partial_{\alpha_n}\partial_{\mu}\partial_{\nu}\sigma(x_1) + \dots
\end{equation}
where the omitted terms are expressions that depend on the lower-order derivatives of $\sigma$ at $x_1$. Hence, all derivatives of $\sigma$ at $x_1$ can be expressed recursively in terms of the Schouten tensor and its covariant derivatives evaluated at $x_1$. 
  
Using the last statement, we can now argue that the quantity $\widetilde X_{12}$ 
defined in \eqref{tildeX12} is diffeomorphism invariant. To show that, we first note that 
$\widetilde X_{12}$  is invariant under the conformal transformations $x^{\mu}\mapsto y^{\mu}\mapsto z^{\mu}$ defined in  
  \eqref{conf1}, \eqref{conf2}:
  \be
|x_1-x_2|^2e^{\sigma(x_1)+\sigma(x_2)}
=
|z_1-z_2|^2e^{\sigma'(z_1)+\sigma'(z_2)}.
\ee
  This is easy to check using \eqref{sigmaprime} and 
  \be
|z_{2}|^2 = \frac{|y_{2}|^2}{1-2b\cdot y_{2} + b^2 y_{2}^2} \, .
  \ee
  Hence we can evaluate $\widetilde X_{12}$ in the special conformal coordinates \eqref{special}. As we explained above, we can use the geodesic between $x_1$ and $x_2$ to express $\widetilde X_{12}$ as a power series in $d$ with coefficients being constructed out of the derivatives of $\sigma(x)$ evaluated at $x_1$ and contracted with $t^{\mu}$. Since in the special coordinates all derivatives of $\sigma$ can be expressed via $P_{\mu\nu}$ and its covariant derivatives, we get a fully covariant expression. In section \ref{ambient_sec} 
  we will explain, following \cite{Parisini:2022wkb,Parisini:2023nbd}, 
  how to describe $\widetilde X_{12}$ in the language of ambient space and will derive the covariant expansion for this quantity in a different way. For now however we want to argue that the existence of such an expansion is sufficient to argue that $C_{\mathrm{cf}}(g_{\mu\nu},t^\mu,d)$ is diffeomorphism invariant and depends only on the quantities shown as its arguments. This follows from the fact that the two-  and three-point functions \eqref{2pt_curved}, \eqref{3pt_curved} can be rewritten as 
  \begin{eqnarray} \label{23pt_curved2}
    \langle \hat {\cal O}_{i}(x_1)\hat {\cal O}_{j}(x_2)\rangle_{g_{\mu\nu}}&=&\frac{\delta_{ij}}{\widetilde X_{12}^{\Delta_{i}}} \, ,
   \nonumber \\
    \langle \hat {\cal O}_{1}(x_1)\hat {\cal O}_{2}(x_2)\hat {\cal O}_{3}(x_3)\rangle_{g_{\mu\nu}}&=&\frac{C_{123}}{\widetilde X_{13}^{\alpha_{2}}
    \widetilde X_{12}^{\alpha_{3}}\widetilde X_{23}^{\alpha_1}} \, .
\end{eqnarray}
with each $\widetilde X_{ij}$ depending on the corresponding covariant data.
Substituting (\ref{OPE_curved}) into the three-point function  (\ref{23pt_curved2}) we obtain 
\be \label{C_defining}
C_{\mathrm{cf}}(g_{\mu\nu},t^\mu,d) \widetilde X_{13}^{-\Delta_{3}}= \frac{1}{\widetilde X_{13}^{\alpha_2}\widetilde X_{23}^{\alpha_1}}
\ee
that can be taken as the defining equation for $C_{\mathrm{cf}}(g_{\mu\nu},t^\mu,d)$.
Since $\widetilde X_{13}, \widetilde X_{23}$ each depend on the covariant data so must  
the operator $C_{\mathrm{cf}}(g_{\mu\nu},t^\mu,d)$. 

The upshot of the above discussion is that $C_{\mathrm{cf}}(g_{\mu\nu},t^\mu,d)$ indeed depends on the covariant data and hence can be computed in the special coordinates \eqref{special}. In these coordinates we obtain 
  \begin{equation} \label{new_dx}
      \Delta x^{\mu} = dt^{\mu} + \frac{d^3}{3}P_{tt}t^{\mu}-\frac{d^3}{6}P_{t}^{\mu} + \bigo(d^4) \, ,
  \end{equation}
  \begin{equation}\label{new_prefactor}
     e^{-\alpha_{2}\sigma(x_1)-\alpha_{1}\sigma(x_2)}=1 + \frac{\alpha_{1}}{2}d^2 P_{tt} + \frac{\alpha_1}{6}d^3\nabla_{t}P_{tt} + \bigo(d^4) 
  \end{equation}
  where for brevity, here and elsewhere,  we write 
  \be 
  \nabla_{t}P_{tt}\equiv t^{\mu}t^{\nu}t^{\alpha}\nabla_{\mu}P_{\nu\alpha} \, .
  \ee
   With the routine we outlined above, both expansions can be continued, each generating terms that depend only on the covariant quantities on $({\cal M},g_{\mu\nu})$. The last ingredient needed to write out a covariant expansion for $C_{\mathrm{cf}}(g_{\mu\nu},t^\mu,d)$ is the expressions of the derivatives of ${\cal O}_{3}$ of the kind given in \eqref{OO1}, \eqref{OO2}. In the special coordinates, those expressions reduce to 
  \begin{equation} \label{OO1_new}
    \partial_{\mu}{\cal O}_3(x_1) =  \nabla_{\mu}\hat{\cal O}_3(x_1)\, , 
\end{equation}
\begin{equation}\label{OO2_new}
 \partial_{\mu}\partial_{\nu}{\cal O}_3 = [\nabla_{\mu}\nabla_{\nu} -\Delta_3 P_{\mu\nu}]\hat {\cal O}_3(x_1) 
\end{equation}
  that now look fully covariant. Clearly the straightforward approach outlined above can be used to generate the expressions for all such derivatives. Let ${\mathbb T}$ 
  be the resulting map that gives the covariant expression for the derivatives of ${\cal O}_{3}$ in terms of the covariant derivatives of $\hat {\cal O}_3$. 
  To summarise our findings in this subsection, we write the action of $C_{\mathrm{cf}}$ as
\be \label{C_new}
 C_{\mathrm{cf}}(g_{\mu\nu},t^\mu,d) \hat {\cal O}_{3} = F(g_{\mu\nu},t^{\mu},d){\mathbb T}C_{\rm flat}(\Delta x^{\mu}( g_{\mu\nu},  t^{\mu},  d)){\cal O}_{3}  \\
 \ee
where we denoted 
\begin{equation}   F(g_{\mu\nu},t^{\mu},d)=  e^{-\alpha_{2}\sigma(x_1)-\alpha_{1}\sigma(x_2)}\, , 
\end{equation}
\begin{equation}
\Delta x^{\mu}( g_{\mu\nu},  t^{\mu},  d)=\Delta x^{\mu}
\end{equation}
calculated in the special coordinates \eqref{special}. The explicit expansion up to the order $d^3$ is  
\begin{align}\label{C_new_exp}
C_{\mathrm{cf}}(g_{\mu\nu},t^\mu,d)\hat {\cal  O}_3
&=
\Bigg\{
1
+\alpha\Bigg[
d\nabla_t
+\frac{\Delta_3}{2}d^2P_{tt}
-\frac{d^3}{6}P_t{}^\mu\nabla_\mu
-\frac{d^3}{6}P_{tt}\nabla_t
+\frac{\Delta_3}{6}d^3\nabla_tP_{tt}
\Bigg]
\nonumber\\
&
+\beta d^2t^\mu t^\nu
\Big[
\nabla_\mu\nabla_\nu-\Delta_3P_{\mu\nu}
+d(\Delta_3+1)P_{\mu\nu}\nabla_t
\Big]
\nonumber\\
& +\gamma d^2(\Box-\Delta_3P)
\nonumber\\
&
+\delta d^3t^\mu t^\nu t^\rho
\Big[
\nabla_{(\mu}\nabla_\nu\nabla_{\rho)}
-(3\Delta_3+2)P_{(\mu\nu}\nabla_{\rho)}
+g_{(\mu\nu}P_{\rho)}{}^\lambda\nabla_\lambda
-\Delta_3\nabla_{(\mu}P_{\nu\rho)}
\Big]
\nonumber\\
&
+\epsilon d^3t^\mu
\Big[
\nabla_\mu(\Box-\Delta_3P)
+(D-2\Delta_3-2)P_\mu{}^\nu\nabla_\nu
\Big]+ \bigo(d^4)
\Bigg\}\hat {\cal O}_{3}
\end{align}
where the round brackets at the indices stand for the normalised symmetrisation,
\be
\Box = g^{\mu\nu}\nabla_{\mu}\nabla_{\nu} 
\ee
and 
\be
\alpha = \frac{\alpha_1}{\Delta_{3}}\, , \qquad \beta = \frac{\alpha_{1}(\alpha_1 + 1)}{2\Delta_{3}(\Delta_{3}+1)}\, ,  \qquad \gamma = \frac{\alpha_1\alpha_2}{2\Delta_3(\Delta_3+1)(D-2\Delta_3-2)}\, , 
\ee
\be
\delta = \frac{\alpha_{1}(\alpha_{1} + 1)(\alpha_{1} + 2)}{6\Delta_{3}(\Delta_{3} + 1)(\Delta_{3} + 2)}\, , \qquad \epsilon = \frac{\alpha_{1}(\alpha_{1} + 1)\alpha_{2}}{2\Delta_{3}(\Delta_{3} + 1)(\Delta_{3} + 2)(D - 2\Delta_{3} - 2)}
\ee
are the flat space descendant coefficients. Clearly, in \eqref{C_new} one can strip off the $\hat {\cal O}_{3}$ on both sides and write this as a differential operator. The first and second order terms in \eqref{C_new_exp} match formula \eqref{cov_exp2} obtained in \cite{Konechny:2026bqg}, after a rearrangement.

We note that the conformally flat descendant operator
$C_{\mathrm{cf}}$ in \eqref{C_new} and \eqref{C_new_exp}
comes out as an expansion with terms proportional to the flat-space
OPE coefficients of the descendant operators. Thus the net result can be interpreted as a rule for replacing the flat space descendants by the curved space ones, which are differential operators of the same order with covariant coefficients. The replacement rule, however, has three separate ingredients: the prefactor $ F(g_{\mu\nu},t^{\mu},d)$ that is a function, the expression replacing $\Delta x^{\mu}$ 
 and the map ${\mathbb T}$. While the first two ingredients depend on both points $x_1$ and $x_2$, the mapping ${\mathbb T}$ is local, and hence it makes more sense to call its images as the curved space descendants.
 In the next subsection, we explain 
  a more elegant  and conceptually rich approach to constructing the map ${\mathbb T}$ based on conformal tractor bundles and the Thomas operator. 
  
\subsection{Tractor bundles associated with primary operators and the map ${\mathbb T}$}\label{sec:tractor_bundles}

Given a $D$-dimensional Riemannian manifold ${\cal M}$ with a conformally flat metric 
  $g_{\mu\nu}$ it can be locally embedded into the light cone of the pseudo-Euclidean space ${\mathbb R}^{D+1,1}$, see e.g. the detailed discussion in \cite{Parisini:2023nbd}. Vector fields in ${\mathbb R}^{D+1,1}$ with certain homogeneity properties when restricted to the embedded neighbourhood  ${\cal U}\subset {\cal M}$ give rise to conformal tractor bundles on ${\cal U}$. Sections of such bundles transform 
  in a specific way under Weyl transformations. In this section we are not going to attempt a general exposition of conformal tractors (we refer the interested reader to lecture notes  \cite{curry2015introductionconformalgeometrytractor}) but rather give a concrete sequence of constructions needed to describe the mapping ${\mathbb T}$ from the previous subsection. 

  Given a primary ${\cal O}$ of dimension $\Delta$ we will use it as a seed for constructing sections of tractor bundles of degrees $\Delta+n$, $n\in {\mathbb N}$ which in physics terms can be thought of as multiplets whose components are constructed from derivatives of $\cal O$ which form a representation under the Weyl transformations. We define a tractor of weight  $\Delta+1$ associated with ${\cal O}$ as a $D+2$-dimensional vector $V_{A}$, $A=0,\dots, D+1$ with the components $A=1, \dots, D$ associated with the tangent bundle ${\rm T}{\cal M}$, that can be  written as a column 
\be \label{V}
 V=\left( \begin{array}{c}
 -\Delta \kappa_{0} {\cal O}\\
  \kappa_{0} \nabla_{\mu}{\cal O}\\
 -(\Box -\Delta P){\cal O}
 \end{array}\right)  
\ee
  where $P=g^{\mu\nu}P_{\mu\nu}$ is the trace of the Schouten tensor, and we use the following notation 
 \be
 \kappa_n = (D-2(\Delta+n) - 2) \, , 
 \ee
 \be
 \Box= g^{\alpha\beta}\nabla_{\alpha}\nabla_{\beta} \, .
 \ee
 
 As a rule, we will use Greek letters $\mu,\nu,\rho,\ldots$ for spacetime indices on $(\mathcal{M}, g_{\mu\nu})$. Capital Roman letters $A,B,C,\ldots$ denote tractor indices, while $M,N,\ldots$ denote ambient-space indices. In a choice of metric splitting, a tractor index decomposes schematically as $A=(+,\mu,-)$; thus, the middle tractor slot carries an ordinary spacetime index. Lowercase Roman letters $i,j,k,\ldots$ are reserved for operator, conformal-family, and insertion-point labels.
  
  Under a Weyl transformation 
 \be \label{gen_Weyl_metric}
 \hat g_{\mu \nu} = e^{2\sigma} g_{\mu \nu}
 \ee
 we have 
 \be
 \hat \Gamma_{\mu \nu}^{\alpha} = \Gamma_{\mu\nu}^{\alpha} + \delta_{\mu}^{\alpha}\Upsilon_{\nu} + \delta_{\nu}^{\alpha}\Upsilon_{\mu}
 -g_{\mu \nu} g^{\alpha \beta}\Upsilon_{\beta} \, , 
 \ee
 \be
 \hat P_{\mu\nu}=P_{\mu\nu} - \nabla_{\mu}\Upsilon_{\nu} + \Upsilon_{\mu}\Upsilon_{\nu} - \frac{1}{2}\Upsilon_{\alpha}\Upsilon_{\beta}g^{\alpha\beta} g_{\mu\nu} 
 \ee
 where 
 \be
 \Upsilon_{\mu} = \partial_{\mu}\sigma \, .
 \ee
 Using these, we find the transformation rule for $V$  
 \be \label{tractor_transform1}
 \left( \begin{array}{c}
 -\Delta   \kappa_{0} \hat{\cal O}\\
   \kappa_{0} \nabla_{\nu}\hat{\cal O}\\
 -(\hat\Box -\Delta \hat P)\hat {\cal O}
 \end{array}\right) = e^{-(\Delta+1) \sigma} \left( \begin{array}{ccc}
 e^{\sigma}&0&0 \\
 e^{\sigma}\Upsilon_{\nu} & e^{\sigma} \delta_{\nu}^{\mu} & 0\\
 -\frac{1}{2}e^{-\sigma}\Upsilon^2 & -\Upsilon^{\mu}e^{-\sigma} & e^{-\sigma} 
 \end{array}\right)  \left( \begin{array}{c}
 -\Delta  \kappa_{0} {\cal O}\\
   \kappa_{0} \nabla_{\mu}{\cal O}\\
 -(\Box -\Delta P){\cal O}
 \end{array}\right) 
 \ee
We will denote 
\be \label{U}
U=(U^{A}_{B})=\left( \begin{array}{ccc}
 e^{\sigma}&0&0 \\
 e^{\sigma}\Upsilon_{\nu} & e^{\sigma} \delta_{\nu}^{\mu} & 0\\
 -\frac{1}{2}e^{-\sigma}\Upsilon^2 & -\Upsilon^{\mu}e^{-\sigma} & e^{-\sigma} 
 \end{array}\right) 
\ee
The above construction defines the action of the Thomas operator, which we denote $D_{A}^{[\Delta]}$, on our generating primary field of weight $\Delta$:
\be
D_{A}^{[\Delta]}{\cal O}\equiv V_{A} \, .
\ee
Thomas operator can be iterated to construct tensors $W=V_{A_1\dots A_n}$ which are assigned weight $n+\Delta$, $n\in{\mathbb N}$. Given such a tensor, we define the action of the Thomas operator on it as 
\be \label{gen_DA}
D_{A}^{[\Delta+n]}W = \left( \begin{array}{c}
 -(\Delta+n) \kappa_{n} W\\
  \kappa_{n} \nabla_{\mu}^{\cal T}W\\
 -(\Box^{\cal T} -(\Delta +n)P)W
 \end{array}\right) 
\ee
where $\nabla_{\mu}^{\cal T}$ is the coupled Levi-Civita--tractor connection and 
\be
\Box^{\cal T} = g^{\mu\nu}\nabla_{\mu}^{\cal T}\nabla_{\nu}^{\cal T}\, .
\ee
The connection $\nabla_{\alpha}^{\cal T}$ is defined by
\be
\nabla_{\alpha}^{\cal T} V_{B_1\dots B_n} = \partial_{\alpha}V_{B_1\dots B_n} + (M_{\alpha})^{A_1}_{B_1}V_{A_1\dots B_n} + \dots + 
(M_{\alpha})^{A_n}_{B_n}V_{B_1\dots A_n}
\ee
where 
\be
(M_{\alpha})_{B}^{A} = \left( \begin{array}{ccc}
 0&-\delta_{\alpha}^{\mu} & 0\\
P_{\alpha \nu} &  - \Gamma^{\mu}_{\alpha \nu} & g_{\alpha \nu} \\
0 & -P_{\alpha}^{\mu} &  0
\end{array}
\right) \, . 
\ee
The crucial property of this connection is the relation 
\be \label{Ucommutator}
(\partial_{\alpha} + \hat M_{\alpha})U = U(\partial_{\alpha} + M_{\alpha}) 
\ee
where $\hat M_{\alpha}$ is constructed using the Weyl transformed quantities: $\hat P_{\mu\nu}$, $\hat \Gamma_{\alpha \nu}^{\mu}$, $\hat g_{\mu\nu}$, 
and $U$ is given by \eqref{U}. Assuming that the tractor tensor $V_{B_1\dots B_n}$ transforms according to 
\be \label{gen_tractor_transform}
 \hat V_{B_1\dots B_n} = e^{-(\Delta + n)\sigma}U_{B_1}^{A_1}\dots U_{B_n}^{A_n}V_{A_1\dots A_n}
\ee
we get from \eqref{gen_DA} and \eqref{Ucommutator}, performing a similar computation to the one leading to formula  \eqref{tractor_transform1}, 
\be \label{DW1}
\hat D_{B}^{[\Delta + n]}\hat V_{B_1\dots B_n} = e^{-(\Delta + n + 1)\sigma}  
U_{B}^{A}U_{B_1}^{A_1}\dots U_{B_n}^{A_n}D_{A}^{[\Delta+n]}V_{A_1\dots A_n} \, .
\ee
Equation \eqref{DW1} together with formula \eqref{tractor_transform1} proves by induction that all 
tractor tensors generated from the primary ${\cal O}$ by successive applications of the Thomas operator indeed transform as in \eqref{gen_tractor_transform}. As the order of derivatives increases with each application of the Thomas operator, the resulting tensors encode the entire jet of derivatives of ${\cal O}$. 

We now specialise to the conformally flat case \eqref{metric_transform} and resort to the notation we used in the previous subsection with $g_{\mu \nu}$ denoting the curved metric with the corresponding fields $\hat {\cal O}_{i}$ and ${\cal O}_{i}$ denoting the primaries on the flat space. 
In this case, in the special coordinates \eqref{special} the transformation rule \eqref{gen_tractor_transform} implies  a sequence of matching equations 
\be \label{matching1}
\hat D_{A_{n}}^{[\Delta+n-1]}\dots \hat D_{A_{2}}^{[\Delta+1]} \hat D_{A_{1}}^{[\Delta]}\hat {\cal O}(x_1) =  D_{A_{n}}^{[\Delta+n-1]}\dots  D_{A_{2}}^{[\Delta+1]}  D_{A_{1}}^{[\Delta]} {\cal O}(x_1) 
\ee
from which we can obtain the action of the map ${\mathbb T}$:
\be
{\mathbb T} D_{A_{n}}^{[\Delta+n-1]}\dots  D_{A_{2}}^{[\Delta+1]}  D_{A_{1}}^{[\Delta]} {\cal O} = \hat D_{A_{n}}^{[\Delta+n-1]}\dots \hat D_{A_{2}}^{[\Delta+1]} \hat D_{A_{1}}^{[\Delta]}\hat {\cal O} \, .
\ee

Let us illustrate the matching equations \eqref{matching1} in the first three iterations. From \eqref{V} we get 
\be \label{first_matching}
V=\left( \begin{array}{c}
 -\Delta  \kappa_{0} {\cal O}\\
   \kappa_{0} \partial_{\alpha}{\cal O}\\
 -\partial^{\mu}\partial_{\mu}{\cal O}
 \end{array}\right) = \left( \begin{array}{c}
 -\Delta  \kappa_{0} \hat {\cal O}\\
   \kappa_{0} \nabla_{\alpha}\hat {\cal O}\\
 -(\Box -\Delta P)\hat {\cal O}
 \end{array}\right) = \hat V
\ee
that gives 
\be \label{Tvalues1}
{\mathbb T} {\cal O} = \hat {\cal O} \, , \quad 
{\mathbb T} \partial_{\mu}{\cal O} = \nabla_{\mu} \hat {\cal O}\, , 
\quad {\mathbb T} \partial^{\mu}\partial_{\mu} {\cal O} = (\Box -\Delta P) \hat {\cal O} \, . 
\ee
Applying one more Thomas operator, we get 
\be\label{TT2}
\hat D_{B}^{[\Delta+1]}\hat V=\left( \begin{array}{c}
 -(\Delta+1)  \kappa_{1} \hat V\\
  \kappa_{1}  \nabla^{{\cal T},g}_{\alpha}\hat V\\
 -   (\Box^{\cal T} -(\Delta+1)  P)\hat V
 \end{array}\right)
 =\left( \begin{array}{c}
 -(\Delta+1)  \kappa_{1} V\\
  \kappa_{1} \nabla^{\cal T}_{\alpha}V\\
 -   \delta^{\mu\nu}\nabla^{\cal T}_{\mu}\nabla_{\nu}^{\cal T} V
 \end{array}\right) =  D_{B}^{[\Delta+1]} V
 \ee
where we used the notation $\nabla^{{\cal T},g}_{\alpha}$ for the tractor covariant derivative with metric $g_{\mu\nu}$ and 
$\nabla^{{\cal T}}_{\alpha}$ for the flat one. The top equation in 
\eqref{TT2} is equivalent to \eqref{first_matching}. The middle equation gives, after using \eqref{Tvalues1}, two new identities 
\bea \label{Tsecond_der}
&& {\mathbb T}\partial_{\mu}\partial_{\nu}{\cal O} = [\nabla_{\mu}\nabla_{\nu} -\Delta P_{\mu\nu}]\hat {\cal O}\, , 
\nonumber \\
&& {\mathbb T} \partial_{\nu}\partial^2 {\cal O} =  \nabla_{\nu} ( \Box - \Delta  P)\hat {\cal O} + \kappa_{0}  P_{\nu}^{\rho} \nabla_{\rho} \hat {\cal O} \, . 
\eea
Finally, we can consider the bottom component of \eqref{TT2}. Looking at the components of that equation, we only find one new value of ${\mathbb T}$ on a certain 4th-order derivative combination, which we are not going to present here. 

To find 
${\mathbb T}\partial_{\alpha}\partial_{\beta}\partial_{\gamma}{\cal O}$ we need to apply one more Thomas operator and go to the middle component equation. We have 
\be \label{Thomas3}
  \hat D_{C}^{[\Delta+2]} \hat D_{B}^{[\Delta+1]} \hat D_{A}^{[\Delta]}  {\cal \hat O} = \left( \begin{array}{c}
 -(\Delta+2)  \kappa_{2} \hat D_{B}^{[\Delta + 1]}\hat V\\
  \kappa_{2} \nabla^{{\cal T},g}_{\rho}\hat D_{B}^{[\Delta + 1]}\hat V\\
 -   (\Box^{\cal T} -(\Delta+2) P)\hat D_{B}^{[\Delta + 1]}\hat V
 \end{array}\right)
  \ee
The middle component in the last equation, after dropping the $\kappa_{2}$ factor, can be expanded as 
$$
\nabla^{{\cal T},g}_{\rho} 
\left( \begin{array}{c}
 -(\Delta+1)  \kappa_{1} \hat V \\
  \kappa_{1} \nabla^{{\cal T},g}_{\nu}\hat V \\
 -   (\Box^{\cal T} -(\Delta+1) P)\hat V
 \end{array}\right)  = \left( 
 \begin{array}{c}
 -(\Delta+2)  \kappa_{1} \nabla^{{\cal T},g}_{\rho}\hat V \\
\lbrack \kappa_{1} \nabla^{{\cal T},g}_{\rho}  \nabla^{{\cal T},g}_{\nu} - g_{\rho\nu} (\Box^{\cal T} -(\Delta+1) P) -(\Delta+1)\kappa_{1} P_{\rho\nu}\rbrack \hat V\\
 \lbrack -   \nabla^{{\cal T},g}_{\rho}(\Box^{\cal T} -(\Delta+1) P)-P_{\rho}^{\lambda}\kappa_{1}\nabla^{{\cal T},g}_{\lambda}\rbrack \hat V
 \end{array}
  \right)
$$
We can now use the matching condition for the middle component of the last expression. The resulting equation can be simplified using the
 bottom component of equation \eqref{TT2} to yield
 \be
  \nabla^{{\cal T}}_{\rho}  \nabla^{{\cal T}}_{\nu}V = [\nabla^{{\cal T},g}_{\rho}  \nabla^{{\cal T},g}_{\nu}  -(\Delta+1) P_{\rho\nu}]\hat V \, .
 \ee
Going for the middle component one more time, we finally obtain 
 \be \label{T3derivatives}
 \begin{split}
 {\mathbb T}\partial_{\mu}\partial_{\nu}\partial_{\rho}{\cal O} & =  \nabla_{\mu} \nabla_{\nu} \nabla_{\rho}\hat {\cal O} 
  -(\Delta+1) P_{\mu\nu} \nabla_{\rho}\hat {\cal O}  -(\Delta+1) P_{ \mu\rho} \nabla_{\nu}\hat {\cal O} \,
  \\
&   +  g_{\nu\rho} P_{\mu}^{\lambda} \nabla_{\lambda}\hat {\cal O} - \Delta \nabla_{\mu}( P_{\nu\rho}\hat {\cal O})
    \end{split}
  \ee
(It can be checked, using the vanishing of the Cotton tensor,  that the last expression is totally symmetric under permutations of $\mu,\nu,\rho$). Note that the trace of equation \eqref{T3derivatives} gives the second formula in \eqref{Tsecond_der}.

From the above worked examples we can discern that in general we can obtain 
\be
{\mathbb T} \partial_{\mu_1}\dots \partial_{\mu_n}{\cal O}
\ee
from the matching equation
\be \label{gen_matching2}
\hat D_{\mu_1}^{[\Delta + n-1]}\hat D_{\mu_2}^{[\Delta + n-2]} \dots \hat D_{\mu_n}^{[\Delta ]}\hat {\cal O} =  D_{\mu_1}^{[\Delta + n-1]}D_{\mu_2}^{[\Delta + n-2]} \dots  D_{\mu_n}^{[\Delta ]} {\cal O}
\ee
that is, by taking the middle component for each Thomas operator in the sequence.
There are two possible strategies to extract ${\mathbb T} \partial_{\mu_1}\dots \partial_{\mu_n}{\cal O}$ from the matching equation \eqref{gen_matching2}. 
The right-hand side of \eqref{gen_matching2} in addition to  $\partial_{\mu_1}\dots \partial_{\mu_n}{\cal O}$ contains a combination of partial traces of this quantity. Such partially traced quantities appear in the bottom components of the shorter sequences of Thomas operators. Hence, if one keeps track of all matching equations from the start of the process one can use the expressions for partially traced quantities to simplify the middle component of \eqref{gen_matching2} in such a way that on the right-hand side it only contains $\partial_{\mu_1}\dots \partial_{\mu_n}{\cal O}$. Thus, in the above sample calculations we used the last equation in \eqref{Tvalues1} to obtain  \eqref{Tsecond_der}, and subsequently we used the second equation in \eqref{Tsecond_der} to obtain \eqref{T3derivatives}.

Alternatively, one can write out the matching equation \eqref{gen_matching2} and then extract $\partial_{\mu_1}\dots \partial_{\mu_n}{\cal O}$ by taking partial traces of this equation. Note that as $\partial_{\mu_1}\dots \partial_{\mu_n}{\cal O}$ is fully symmetric in its indices, we can start with the symmetrised version of equation \eqref{gen_matching2}. To illustrate the second method, let us start with 
\bea \label{Nabc}
&& N_{\mu\nu\rho}\equiv\hat D_{(\mu}^{[\Delta+2]}\hat D_{\nu}^{[\Delta+1]}\hat D_{\rho)}^{[\Delta]}\hat {\cal O} = \kappa_1\kappa_{2}\Bigl[ \kappa_{0}
\nabla_{(\mu}\nabla_{\nu}\nabla_{\rho)} - \kappa_{0}\Delta(\nabla_{(\mu}P_{\nu\rho)}) \nonumber \\
&& -\kappa_{0}(3\Delta+2)P_{(\mu\nu}\nabla_{\rho)}-3g_{(\mu\nu}\nabla_{\rho)}(\Box - \Delta P) 
-2\kappa_{0}g_{(\mu\nu}P_{\rho)}^{\lambda}\nabla_{\lambda}\Bigr] \hat {\cal O}
\eea
where the round brackets at the indices stand for the normalised symmetrisations. 
On the flat space this expression reduces to 
\be
N_{\mu\nu\rho}=\kappa_1\kappa_{2}\lbrack \kappa_{0} \partial_{\mu}\partial_{\nu}\partial_{\rho} 
-3g_{(\mu\nu}\partial_{\rho)}\partial^{2} \rbrack {\cal O} \, .
\ee
Taking the trace of the last expression and substituting it back, we find 
\be
 \partial_{\mu}\partial_{\nu}\partial_{\rho}{\cal O} = \frac{1}{\kappa_{0}\kappa_{1}\kappa_{2}}\left[ N_{\mu\nu\rho} - \frac{3g_{(\mu\nu}N^{\lambda}_{\rho)\lambda}}{2(\Delta +2)}\right]
\ee
that, using \eqref{Nabc}, gives 
\bea \label{T3derivatives2}
 && {\mathbb T}\partial_{\mu}\partial_{\nu}\partial_{\rho}{\cal O} =  \nabla_{(\mu} \nabla_{\nu} \nabla_{\rho)}\hat {\cal O} 
  -(3\Delta+2) P_{(\mu\nu} \nabla_{\rho)}\hat {\cal O}    +  g_{(\mu\nu} P_{\rho)}^{\lambda} \nabla_{\lambda}\hat {\cal O} - \Delta (\nabla_{(\mu} P_{\nu\rho)})\hat {\cal O}
  \nonumber \\
&& 
  \eea
  that is a symmetrised version of \eqref{T3derivatives}.

  We finish this section by making a remark that while the tractor method of constructing ${\mathbb T}$ is conceptually more interesting and may lead to more connections with tractor calculus, the direct method described in the previous subsection is just as efficient. Either method can be easily implemented algorithmically. The construction above relied on conformal flatness at every step; we now ask what replaces it on a general background. 
  
\section{The ambient space approach to general spaces} \label{ambient_sec}

\subsection{Review}\label{review_subsec}
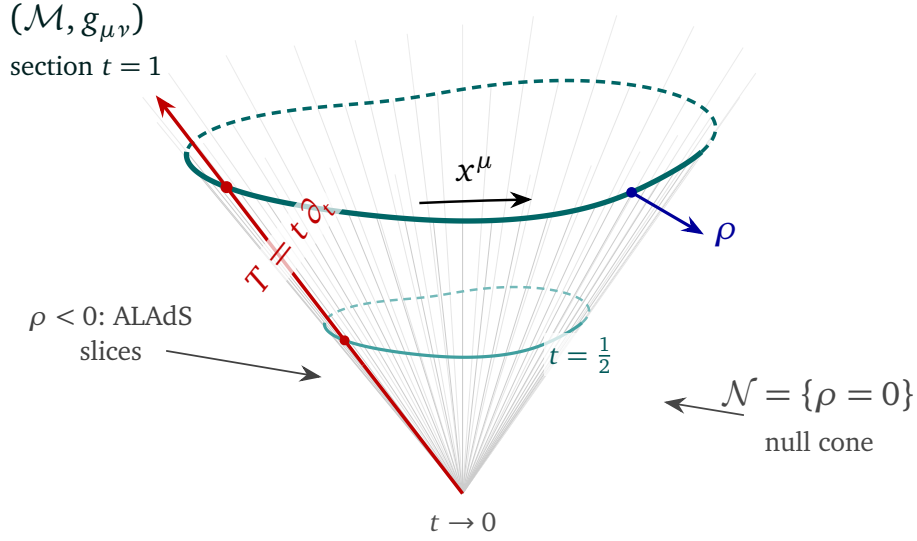
\begin{figure}[t]
\centering
\resizebox{0.82\textwidth}{!}{%
\begin{tikzpicture}[>={Stealth[length=2.2mm]},
  declare function={
    rr(\t)  = 1 + 0.14*sin(2*\t+40) + 0.09*sin(3*\t-60);
    px(\t)  = 2.35*rr(\t)*cos(\t);
    py(\t)  = 0.72*rr(\t)*sin(\t);
    H       = 3.4;
  }]
 
  \foreach \t in {0,7.5,...,352.5}{
    \pgfmathsetmacro{\shade}{ (sin(\t)>0) ? 20 : 40 }
    \draw[gray!\shade, line width=0.25pt]
      (0,0) -- ({px(\t)},{H+py(\t)});
  }
 
  \foreach \t in {0,15,...,345}{
    \draw[gray!16, line width=0.25pt]
      ({px(\t)},{H+py(\t)}) -- ({1.16*px(\t)},{1.16*(H+py(\t))});
  }
 
  \draw[teal!55, line width=0.7pt, dash pattern=on 2.2pt off 1.6pt,
        domain=0:180, samples=90, smooth]
    plot ({0.5*px(\x)}, {0.5*(H+py(\x))});
  \draw[teal!75, line width=0.9pt, domain=180:360, samples=90, smooth]
    plot ({0.5*px(\x)}, {0.5*(H+py(\x))});
  \node[teal!70!black, font=\scriptsize, fill=white, fill opacity=0.75,
        text opacity=1, inner sep=1pt] at (1.18,1.38) {$t=\tfrac12$};
 
  \draw[teal!80!black, line width=1.0pt, dash pattern=on 2.6pt off 1.8pt,
        domain=0:180, samples=120, smooth]
    plot ({px(\x)}, {H+py(\x)});
  \draw[teal!80!black, line width=1.5pt, domain=180:360, samples=120, smooth]
    plot ({px(\x)}, {H+py(\x)});
 
  \pgfmathsetmacro{\ta}{206}
  \coordinate (P) at ({px(\ta)},{H+py(\ta)});
  \draw[red!75!black, line width=1.0pt] (0,0) -- ($0.42*(P)$);
  \draw[red!75!black, line width=1.0pt, ->] ($0.42*(P)$) -- ($1.30*(P)$);
  \fill[red!75!black] ($0.5*(P)$) circle (1.4pt);
  \fill[red!75!black] (P) circle (1.7pt);
  \node[red!75!black, font=\small, fill=white, fill opacity=0.7, text opacity=1,
        inner sep=1pt, rotate=52] at ($0.86*(P)+(0.30,-0.14)$) {$T=t\,\partial_t$};
 
  \pgfmathsetmacro{\tb}{322}
  \coordinate (Q) at ({px(\tb)},{H+py(\tb)});
  \draw[blue!60!black, line width=0.9pt, ->]
    (Q) -- ($(Q)+(0.72,-0.42)$) node[right=-1pt, font=\small] {$\rho$};
  \fill[blue!60!black] (Q) circle (1.5pt);
 
  \draw[black, line width=0.7pt, ->, shorten <=1pt]
    ({px(258)},{H+py(258)+0.16})
    .. controls ({px(272)},{H+py(272)+0.20}) .. ({px(288)},{H+py(288)+0.18});
  \node[font=\small] at ({px(273)},{H+py(273)+0.52}) {$x^{\mu}$};
 
  \node[teal!25!black, font=\small, align=left]
    at ({px(168)-1.15},{H+py(168)+0.95}) {$(\mathcal{M},g_{\mu\nu})$\\[-2pt]
      \scriptsize section $t=1$};
  \node[gray!55!black, font=\small, align=center] at (3.55,0.78)
    {$\mathcal{N}=\{\rho=0\}$\\[-2pt]\scriptsize null cone};
  \draw[gray!55!black, line width=0.5pt, shorten >=1.5pt, ->]
    (2.80,0.78) -- (1.95,0.92);
  \node[gray!60!black, font=\scriptsize, below=1.5pt] at (0,0) {$t\to 0$};
 
  \node[gray!50!black, font=\scriptsize, align=center] at (-3.5,1.6)
    {$\rho<0$:\ ALAdS\\[-2pt]slices};
  \draw[gray!50!black, line width=0.5pt, shorten >=1.5pt, ->]
    (-2.95,1.42) -- (-1.35,1.15);
 
\end{tikzpicture}}
\caption{The ambient space of a conformal class $[g]$. The null cone $\mathcal{N}=\{\rho=0\}$ is ruled by the dilation orbits of the homothety $T=t\,\partial_t$ (one orbit shown in red), and its sections are the conformal representatives, with the physical space $(\mathcal{M},g_{\mu\nu})$ recovered at $t=1$ and its homothetic images at $t<1$. The coordinate $\rho$ moves off the cone, and the region $\rho<0$ is foliated by the asymptotically locally AdS slices.}
\label{fig:ambient_geometry}
\end{figure}

The embedding space realises the conformal group of $\mathbb{R}^{D}$ as the Lorentz group of $\mathbb{R}^{1, D+1}$ and represents conformally flat backgrounds as sections of the null cone $X^{2} = 0$. On a general background, this linearisation of conformal symmetry is lost, but Weyl covariance survives as the universal kinematic constraint, and Fefferman and Graham \cite{Fefferman:2007rka} showed that it too admits a geometric realisation. To the conformal class $[g]$ of a $D$-dimensional metric, we can canonically associate a $(D+2)$-dimensional Lorentzian \textit{ambient space} carrying a metric $\widetilde{g}$ that is Ricci flat to the appropriate order and admits a homothety, that is a vector field $T$ satisfying $\mathcal{L}_{T}\widetilde{g} = 2\widetilde{g}$ together with $\widetilde{\nabla}_{A}T_{B} = \widetilde{g}_{AB}$. In Gaussian null coordinates, $\widetilde{X}^{M} = (t, x^{\mu}, \rho)$ adapted to the cone, the ambient metric takes the form
\begin{align}\label{general_metric}
    \widetilde{g} = 2\rho\, dt^{2} + 2t\,dt\,d\rho + t^{2}g_{\mu\nu}(x, \rho)\, dx^{\mu}\,dx^{\nu},\qquad T = t\partial_{t},
\end{align}
where the null cone sits at $\rho = 0$, with the degenerate induced metric $t^{2}g_{\mu\nu}(x)dx^{\mu}dx^{\nu}$; it is the curved analogue of the Minkowski light cone, swept out by rescalings of the physical metric. The physical space is recovered in the section $t = 1$, $\rho = 0$, with $g_{\mu\nu}(x,0) = g_{\mu\nu}(x)$ a chosen representative of the conformal class. The coordinate $t$ parametrises the dilation orbits generated by $T$, and $\rho$ measures the departure from the cone. The gauge is chosen so that the curves of constant $(t, x)$ and of constant $(x, \rho)$ are ambient geodesics, which is what fixes the $t$-dependence in \eqref{general_metric} completely. The geometry described by \eqref{general_metric} is summarised in Figure \ref{fig:ambient_geometry}. 
The only free function in \eqref{general_metric} is $g_{\mu\nu}(x, \rho)$, and Ricci flatness determines it order by order in $\rho$. The equations $\widetilde{R}_{\mu\nu} = \widetilde{R}_{\mu\rho} = \widetilde{R}_{\rho\rho} = 0$ form a non-linear radial evolution system, and its formal solution is the Fefferman--Graham expansion
\begin{align}\label{eq:FG}
    g_{\mu\nu}(x,\rho) = g_{\mu\nu}(x) + 2\rho\, P_{\mu\nu}(x) + \ldots + \rho^{\frac{D}{2}}g_{(D)\mu\nu}(x) + \ldots,
\end{align}
where the displayed terms are the only ones we shall need, and the linear coefficient is universal and is given by the Schouten tensor \eqref{Schouten_1}. All coefficients below $\rho^{D/2}$ are local functions of $g_{\mu\nu}(x)$, computable recursively. At order $\rho^{D/2}$ the expansion ceases to be determined; the trace and divergence of $g_{(D)\mu\nu}(x)$ are fixed, but its transverse traceless part is free integration data, and in even $D$ a logarithmic term $h_{(D)}$ appears, while in odd $D$ half-integer powers of $\rho$ enter. This undetermined coefficient is not a defect of the construction. It is the second piece of physical data, and its interpretation is the subject of the next section.

Two special cases are worth mentioning. If $g_{\mu\nu}$ is an Einstein metric, the expansion truncates at order $\rho^{2}$ in closed form. And the ambient space is Riemann flat, hence locally the embedding space, if and only if the boundary metric is conformally flat with vanishing $g_{(D)}$ data, except in $D = 2$, where every four-dimensional ambient space is automatically flat. The ambient Riemann tensor therefore measures precisely the failure of the physical situation to be conformally trivial, which is what makes it the natural expansion parameter for correlators.

\paragraph{The hyperbolic slicing and the state.} The similarity between \eqref{general_metric} and the near boundary expansion of holography is not a coincidence. Under the coordinate change
\begin{align}\label{eq:hybol}
    \rho = -\frac{r^{2}}{2},\qquad t = \frac{s}{r},
\end{align}
covering the region $\rho<0$ inside the cone. The ambient metric \eqref{general_metric} becomes an exact cone over the $(D+1)$-dimensional space,
\begin{align}\label{cover_metric}
    \widetilde{g} = -ds^{2} + s^{2}\left(\frac{dr^{2} + g_{\mu\nu}(x,r)\,dx^{\mu}\,dx^{\nu}}{r^{2}}\right),\qquad T = s\partial_{s},
\end{align}
where the metric in parentheses is called the asymptotically locally AdS (ALAdS) metric in the Fefferman--Graham gauge, $ds^{2}_{\rm ALAdS}$, and Ricci flatness is equivalent to the statement that $ds^{2}_{\rm ALAdS}$ solves the vacuum Einstein equations with a negative cosmological constant. The region $\rho>0$ is similarly foliated by asymptotically locally de Sitter slices. 

This dictionary settles the interpretation of free data $g_{(D)\mu\nu}$. In holographic renormalisation, the transverse traceless part of $g_{(D)}$ encodes the expectation value of the stress tensor of the dual CFT, with $g_{\mu\nu}(x)$ its source. The proposal of \cite{Parisini:2022wkb, Parisini:2023nbd} is to attach to a CFT on a background $g$ in a state with given $\langle T_{\mu\nu}\rangle$ the ambient space whose ALAdS slices carry exactly that data. 

Although the identification passes through the AdS/CFT dictionary, the
resulting formalism solves the purely kinematic Weyl-covariance constraint
already stated in \eqref{Weyl_covariance}. In the notation of that equation \footnote{For local Weyl transformations we retain the notation of
subsection~\ref{sec:tractor_bundles}. From this section onward, \({\cal O}_i\) denotes the operator on the chosen curved representative \(g_{\mu\nu}\). The hats in section~2 were used only when it was necessary to distinguish the curved-space operator from its flat-space representative.},
\begin{equation}
\big\langle {\cal O}_1(x_1)\cdots {\cal O}_n(x_n)\big\rangle_{\Lambda^2 g}
=
\prod_{i=1}^{n}\Lambda(x_i)^{-\Delta_i}
\big\langle {\cal O}_1(x_1)\cdots {\cal O}_n(x_n)\big\rangle_g .
\label{corr_conformal}
\end{equation}
The output of \eqref{corr_conformal} applies to any CFT, holographic or not, in the same way that the embedding space serves free theories as well as holographic ones. Checks at finite temperature against the thermal OPE bear this out \cite{Parisini:2023nbd}. The slicing also supplies exact ambient metrics on demand. Any exact ALAdS solution in Fefferman--Graham gauge, fibred over $s$ as in \eqref{cover_metric}, is an exact Ricci-flat ambient space. The planar AdS black hole gives the ambient space of the thermal state, and pure hyperbolic space gives the ambient space of the vacuum on conformally flat backgrounds, which is once again flat Minkowski space in disguise.

\paragraph{Weyl transformations as ambient diffeomorphisms.}
The defining property of the construction is that it depends only on the conformal class $[g]$. Concretely, if two representatives are related by $\hat{g} = e^{2\sigma}g$, then the two ambient metrics built from them are related by an ambient diffeomorphism that preserves the form \eqref{general_metric}. Working perturbatively in $\rho$, the diffeomorphism reads \footnote{Here \(t\), without a spacetime index, is the ambient dilation coordinate appearing in \eqref{general_metric}; it should not be confused with the physical unit tangent \(t^\mu\) used from subsection~\ref{sec:amb_geo} onward.}
\begin{equation}
\hat t
=
e^{-\sigma(x)}t
\left[
1-\frac12\Upsilon_\mu\Upsilon^\mu\rho+\bigo(\rho^2)
\right],
\qquad
\hat x^\mu=x^\mu+\Upsilon^\mu\rho+\bigo(\rho^2),
\qquad
\hat\rho=e^{2\sigma(x)}\rho+\bigo(\rho^2).
\label{diffeos}
\end{equation}
with \(\Upsilon_\mu=\partial_\mu\sigma\) and indices moved with $g$. On the cone itself, the transformation is nothing but a local rescaling of $t$, the curved counterpart of changing the section of the Minkowski light cone. It follows that the $t$-homogeneity of any ambient quantity restricted to the cone measures its Weyl weight; a scalar homogeneous of degree $-\Delta$ in $t$ restricts to an object transforming with weight $\Delta$ in the sense of \eqref{corr_conformal}. Ambient diffeomorphism invariants with definite homogeneity are therefore automatically Weyl covariant.

Restricting ambient tensors to the cone produces the tractor calculus
on the physical space. To avoid introducing a second choice of tractor
slots, we use throughout the covariant slot ordering fixed in
subsection~\ref{sec:tractor_bundles}. Thus, a weight-\((\Delta+1)\) tractor \(V_B\) transforms as
\begin{equation}
\hat V_B
=
e^{-(\Delta+1)\sigma}U_B{}^A V_A,
\qquad
U_B{}^A
=
\begin{pmatrix}
e^\sigma & 0 & 0 \\[2mm]
e^\sigma\Upsilon_\nu & e^\sigma\delta_\nu{}^\mu & 0 \\[2mm]
\matminus\dfrac12e^{-\sigma}\Upsilon^2
&
\matminus e^{-\sigma}\Upsilon^\mu
&
e^{-\sigma}
\end{pmatrix}.
\label{linear_matrix}
\end{equation}
 In the same slot convention, the tangential ambient covariant
derivative restricts to the tractor connection
\begin{equation}
\nabla_\mu^T V_B
=
\partial_\mu V_B+(M_\mu)_B{}^A V_A,
\qquad
(M_\mu)_B{}^A
=
\begin{pmatrix}
0 & \matminus\delta_\mu{}^\rho & 0 \\[1mm]
P_{\mu\nu} & \matminus\Gamma_{\mu\nu}{}^\rho & g_{\mu\nu} \\[1mm]
0 & \matminus P_\mu{}^\rho & 0
\end{pmatrix}.
\end{equation}
Its compatibility with the Weyl transformation is
\begin{equation}
(\partial_\mu+\hat M_\mu)U
=
U(\partial_\mu+M_\mu),
\end{equation}
which is the identity already established in \eqref{Ucommutator}. The appearance of the Schouten tensor in this connection is the infinitesimal counterpart of the Fefferman--Graham term \(2\rho P_{\mu\nu}\) in \eqref{general_metric}. For the purposes of the correlators, the message is that Weyl covariance, which acts nonlinearly and inhomogeneously on ordinary curvature tensors, is linearised on ambient space, and finding Weyl-covariant structures reduces to representation theory of the matrix \(U_B{}^A\).

\paragraph{Ambient correlators.} To build correlators, we need diffeomorphism-invariant scalars attached to several points of the ambient space. Coordinates of insertion points cannot be contracted across different tangent spaces, and the flat-space trick of using the position vector $X^{M}$ is unavailable. The ambient space, however, carries a canonical vector field, a homothety $T$, which reduces to $X^{M}\partial_{M}$ in the flat space limit. The prescription of Parisini, Skenderis, and Withers in \cite{Parisini:2022wkb, Parisini:2023nbd} is to use $T$ evaluated at the insertions in place of the positions, and to bring everything to a common tangent space by parallel transport along the ambient geodesic connecting the insertions.

Let $\hat{T}_{1}$ denote $T(\widetilde{X}_{1})$ parallel transported to $\widetilde{X}_{2}$. The fundamental bilocal invariant is
\begin{align}
    \widetilde{X}_{12}\equiv -2\hat{T}_{1}\cdot T_{2},
\end{align}
and a short calculation using $\mathcal{L}_{T}\widetilde{g} = 2\widetilde{g}$ and $\widetilde{\nabla}_{A}T_{B} = \widetilde{g}_{AB}$, shows the remarkable identity
\begin{align}
    \widetilde{X}_{12} = \ell\left(\widetilde{X}_{1}, \widetilde{X}_{2}\right)^{2},
\end{align}
where $\ell\left(\widetilde{X}_{1}, \widetilde{X}_{2}\right)$ is the geodesic distance between the two points. Indeed, along an affine geodesic with tangent $\dot{\gamma}$ one finds from $\widetilde{\nabla}_{A}T_{B} = \widetilde{g}_{AB}$ that $\frac{d}{d\lambda}(T\cdot \dot{\gamma}) = \dot{\gamma}^{2}$, so $T$ projected on the geodesic grows linearly, and evaluating the transported vector at the far end produces the squared length. On the flat ambient space with flat boundary, we recover $\widetilde{X}_{12} = X_{12} = (x_{1} - x_{2})^{2}$ on the section $t = 1$, $\rho = 0$, and for general sections the invariant is homogeneous of degree one in each of $t_{1}$ and $t_{2}$, hence carries Weyl weight one at each insertion. It is symmetric in its arguments, and it exists and is unique whenever two points are joined by a unique ambient geodesic; when several geodesics exist, as in thermal states where they wind the thermal circle, each contributes its own invariant and physical correlators involve a sum over them.

The relation to holography is again immediate in the slicing \eqref{cover_metric}. An ambient geodesic between two points on the cone projects to a geodesic of the ALAdS slice with endpoints on its conformal boundary, and the invariant is the renormalised exponential of the ALAdS geodesic length, 
\begin{align}\label{renor_geo_length}
    \frac{1}{\left(\widetilde{X}_{12}\right)^{\Delta}} = \left.\frac{r^{-2\Delta}}{(t_{1}t_{2})^{\Delta}}e^{-\Delta\, L_{\rm AdS}}\right\vert_{r\to 0}.
\end{align}
The right-hand side is the geodesic approximation to a holographic two-point function of a heavy operator. In the ambient formalism, it is promoted from a large-$\Delta$ saddle point to an exact kinematic building block, valid at any $\Delta$, with the dynamical corrections organised separately. Scalar correlators are then written as the flat-space power structure built from the $\widetilde{X}_{ij}$, dressed by Weyl-invariant curvature corrections, 
\begin{eqnarray}\label{corr_ansatz}
    \langle \mathcal{O}(x_{1})\mathcal{O}(x_{2})\rangle_{g} &= &\frac{C_{\Delta}}{\left(\widetilde{X}_{12}\right)^{\Delta}}\left[1 + \mathcal{I}^{(1)}_{2} + \mathcal{I}^{(2)}_{2} + \ldots\right], \nonumber \\
    \langle \mathcal{O}_{1}(x_1)\mathcal{O}_{2}(x_2)\mathcal{O}_{3}(x_3)\rangle_{g} &=& \frac{C_{123}\left[1 + \mathcal{I}^{(1)}_{3} + \mathcal{I}^{(2)}_{3} + \ldots\right]}{\left(\widetilde{X}_{12}\right)^{\alpha_{3}}\left(\widetilde{X}_{13}\right)^{\alpha_{2}}\left(\widetilde{X}_{23}\right)^{\alpha_{1}}},
\end{eqnarray}
with $2\alpha_{k} = \Delta_{i} + \Delta_{j} - \Delta_{k}$ for $\{i,j,k\}$ a permutation of $\{1, 2, 3\}$. The corrections are organised by powers of the ambient curvature, and $\mathcal{I}^{(k)}_{i}$  are the general linear combinations of weight-adjusted invariants with $k$ curvatures. These invariants are constructed by parallel transporting $T$ along the ambient geodesics connecting two insertion points, followed by contracting a number of copies of the transported $T$ with the curvature tensors at the second end. As argued in  \cite{Parisini:2022wkb, Parisini:2023nbd} the ambient identities force $\mathcal{I}^{(1)}_{i} = 0$, so the first corrections of this type are quadratic in curvature. We are going to show in section \ref{sec:ansatz_OPE} that for the three-point functions to be compatible with the existence of local covariant OPE, in addition to    $\mathcal{I}^{(k)}_{i}$, $k\ge 2$ one needs to include some terms $J^{(1)}$ of a different, trilocal nature that are linear in the  Cotton and Weyl tensors. The ingredients feeding that analysis are the short-distance expansions of the invariants $\widetilde{X}_{ij}$ and of their derivatives, which we derive next.

Throughout this section, we expand to linear order in the curvature and through third order in the separations. Thus, we retain the two-derivative structure $P_{\mu\nu}$ and the three-derivative structures $\nabla_\lambda P_{\mu\nu}$, $C_{\mu\nu\lambda}$ and $\nabla_\eta W_{\mu\nu\alpha\lambda}$, while discarding $\bigo(\partial^4)$ curvature-derivative terms and terms quadratic in the curvature. At this accuracy, the only universally determined term in the Fefferman--Graham expansion \eqref{eq:FG} that can contribute is $2\rho P_{\mu\nu}$.

The term $\rho^{D/2}g_{(D)\mu\nu}$ requires a separate power-counting argument. In the hyperbolic slicing \eqref{eq:hybol}, $\rho=-r^2/2$, while a short geodesic whose endpoints are separated by a distance $d$ reaches radial depths $r=\bigo(d)$, as will be explicit in the parametrisation \eqref{eq:semi_circ}. Therefore, we have $\rho^{D/2}=\bigo(d^D)$. The tensor $g_{(D)\mu\nu}$ is held fixed in this short-distance limit, so multiplying by it does not alter the power of $d$. Consequently, its contribution to $\widetilde X_{12}$ begins at relative order $d^D$. Since the leading term in $\widetilde X_{12}$ is of order $d^2$, this corresponds to an absolute correction of order $d^{D+2}$. For $D\geq4$, the correction is therefore of relative order $d^4$ or higher and lies beyond the cubic accuracy retained below. It may consequently be omitted in these dimensions without setting $g_{(D)\mu\nu}$ to zero or otherwise restricting the state. Any logarithmic term present in even dimensions begins at the same relative order, up to the logarithm, and is likewise outside the present truncation.

For $D=3$, by contrast, the $g_{(3)\mu\nu}$ contribution begins at relative order $d^3$, precisely the order at which the $\nabla P$ and Cotton contributions enter. Power counting alone therefore no longer separates the state-dependent Fefferman--Graham datum from the terms determined locally by the background geometry. For the remainder of this section, we set $g_{(3)\mu\nu}=0$ so as to isolate the latter contributions. This is only a temporary restriction: we return to the general $D=3$ case in section~\ref{sec:ansatz_OPE} and explain there why reinstating $g_{(3)\mu\nu}$ cannot provide a universal cancellation of the Cotton-dependent terms.

\subsection{Expansion of the ambient geodesic}\label{sec:amb_geo}
We first expand $\widetilde{X}_{12}$ for two nearby points $x_{1}, x_{2}$ on the section, connected on $M$ by a geodesic of length $d$ with unit tangent $t^{\mu}$ at $x_{1}$. The computation is cleanest in the hyperbolic slicing, using \eqref{renor_geo_length}. We introduce Fermi normal coordinates $(u, y^{\alpha})$ along the boundary geodesic, so that on the geodesic itself $g_{uu} = 1$, $\partial_{\mu}g_{\nu\rho} = 0$, and the two insertions sit at $u = 0$ and $u = d$, $y = 0$. By $ds^{2}_{\rm ALAdS}$ in \eqref{cover_metric}, the slice metric restricted to the $(u,r)$ plane at $y = 0$ reads
\begin{align}
    \left.ds^{2}_{\rm ALAdS}\right\vert_{y=0} = \frac{dr^{2} + \left[1 - r^{2}\,P_{uu}(u) + \bigo(r^{4})\right]du^{2}}{r^{2}},\qquad P_{uu}(u) = P_{tt} + u\,\nabla_{t}P_{tt} + \bigo(u^{2}),
\end{align}
where all curvatures are evaluated at $x_{1}$. At zeroth order in the curvature, the slice is hyperbolic space and the geodesic between the endpoints, regularised at $r = \epsilon$, is the semicircle
\begin{align}\label{eq:semi_circ}
    u(\theta) = \frac{d}{2}(1 - \cos\theta),\qquad r(\theta) = \frac{d}{2}\sin\theta,\qquad ds = \frac{d\theta}{\sin\theta},
\end{align}
with regularised length $L_{0} = 2\,\log(d/\epsilon) + \bigo(\epsilon^{2})$. Since the endpoints are fixed, the first-order shift of the length is obtained by evaluating the metric perturbation on the unperturbed path,
\begin{align}
    \delta L = \frac{1}{2}\int h_{\mu\nu}\,\dot{x}^{\mu}\dot{x}^{\nu}\,ds = -\frac{1}{2}\int P_{uu}(u(\theta))\left(\frac{du}{ds}\right)^{2}ds = -\frac{d^{2}}{8}\int_{0}^{\pi}P_{uu}(u(\theta))\sin^{3}\theta\,d\theta,
\end{align}
where we used $du/ds = \frac{d}{2}\sin^{2}\theta$. The two elementary integrals
\begin{align}
    \int_{0}^{\pi}\sin^{3}\theta\,d\theta = \frac{4}{3},\qquad \int_{0}^{\pi}(1 - \cos\theta)\sin^{3}\theta\,d\theta = \frac{4}{3},
\end{align}
then give
\begin{align}
    L_{\rm AdS} = 2\log\frac{d}{\epsilon} - \frac{d^{2}}{6}P_{tt} - \frac{d^{3}}{12}\nabla_{t}P_{tt} + \bigo(d^{4}).
\end{align}
Feeding this into \eqref{renor_geo_length}, we have
\begin{align}
    \frac{1}{\left(\widetilde{X}_{12}\right)^{\Delta}} = \left.\frac{r^{-2\Delta}}{(t_{1}t_{2})^{\Delta}}\left(\frac{\epsilon}{d}\right)^{2\Delta}\exp\left[\Delta\frac{d^{2}}{6}P_{tt} + \Delta\frac{d^{3}}{12}\nabla_{t}P_{tt} + \bigo(d^{4})\right]\right\vert_{r\to 0}.
\end{align}
Using the usual AdS cutoff relation $\epsilon = r$, the divergent factors cancel, $r^{-2\Delta}\epsilon^{2\Delta} = 1$. Expanding the exponential with $t_{1,2}\to 1$ yields 
\begin{align}
    \frac{1}{\left(\widetilde{X}_{12}\right)^{\Delta}} = \frac{1}{d^{2\Delta}}\left[1 + \frac{\Delta\,d^{2}}{6}P_{tt} + \frac{\Delta\,d^{3}}{12}\nabla_{t}P_{tt} + \bigo(d^{4})\right].
\end{align}
Now, to read off $\widetilde{X}_{12}$, we invert the $-\Delta$ power perturbatively to obtain the central result
\begin{align}\label{X}
    \widetilde{X}_{12} = d^{2}\left[1 - \frac{d^{2}}{6}P_{tt} - \frac{d^{3}}{12}\nabla_{t}P_{tt} + \bigo(d^{4})\right].
\end{align}
Both correction structures are built from the fully symmetric jets $P_{(\mu\nu)}$ and $\nabla_{(\mu}P_{\nu\rho)}$, since $t^{\mu}t^{\nu}$ and $t^{\mu}t^{\nu}t^{\rho}$ project onto total symmetrisations; in particular, the Cotton tensor cannot appear in $\widetilde{X}_{12}$ itself at this order, a fact that will matter below. 

The same expansion follows from an entirely independent route that
never leaves the section. On a conformally flat patch
\(g_{\mu\nu}=e^{2\sigma}\delta_{\mu\nu}\), the Weyl weight of
\(\widetilde X\) and its flat-space limit fix
\begin{equation}
\widetilde X(x,y)
=
e^{\sigma(x)+\sigma(y)}|x-y|^2 .
\label{X_conformally_flat}
\end{equation}
Expanding the right-hand side around \(x\) in terms of the geodesic
distance and tangent of \(g_{\mu\nu}\), and using $P_{\mu\nu}=-\partial_\mu\partial_\nu\sigma+\bigo\big((\partial\sigma)^2\big)$ at linear order, reproduces \eqref{X}, including the sign and
coefficient of the cubic term. We stress that the holographic derivation of \eqref{X} assumed nothing about conformal flatness. Only the universal term $2\rho\,P_{\mu\nu}$ of the Fefferman--Graham expansion entered, and it is present for every conformal class. The conformally flat computation is therefore an independent check of coefficients, not a restriction of the result. One can also see directly that \eqref{X} is complete at this order on a general background. The only tensors available at $x$ are $t^{\mu}$ and the curvature jets, and at two- and three-derivative order the only scalar contractions are $P_{tt}$ and $\nabla_{t}P_{tt}$ since $W_{tttt}$ vanishes by antisymmetry, $C_{ttt} = 0$, and $t^{\mu}t^{\nu}t^{\rho}\nabla_{\mu}P_{\nu\rho}$ projects onto the totally symmetric part of $\nabla P$, from which the Cotton tensor drops out. The Weyl and Cotton tensors thus enter $\widetilde{X}$ only once a second direction is available, i.e., through derivatives with respect to endpoints. We discuss this in detail in  section \ref{sec:ansatz_OPE}. 

Written covariantly for a general pair of points, let
\(d=d(x,y)\) and let \(t^\mu\) be the unit tangent at \(x\) pointing
towards \(y\). Then the expansion reads
\begin{equation}\label{X(x,y)}
\widetilde X(x,y)
=
d^2
\left[
1-\frac{d^2}{6}P_{tt}(x)
-\frac{d^3}{12}\nabla_tP_{tt}(x)
+\bigo(\partial^4,\mathrm{curv}^2)
\right],
\end{equation}
and one checks that this expansion is symmetric under the exchange of endpoints. Transporting the base point to $y$ flips $t\to -t$ and shifts the curvature by its own gradient
\begin{equation}
P_{tt}(y)
=
P_{tt}(x)+d\,\nabla_tP_{tt}(x)+\bigo(d^2),
\end{equation}
so even the quadratic term generates a cubic piece $-\frac{d^{3}}{6}\nabla_{t}P_{tt}$ that combines with the sign flip of the odd term, $+\frac{d^{3}}{12}\nabla_{t}P_{tt}$, to reproduce the original coefficient $-\frac{d^{3}}{12}$. The relative factor of two between the quadratic and the cubic denominators is precisely what endpoint symmetry requires.

For later use, we also record the expansion of the inverse power that appears in the correlator ansatz \eqref{corr_ansatz},
\begin{align}\label{X_inverse_alpha}
    \widetilde{X}_{12}^{-\alpha} = d^{-2\alpha}\left[1 + \alpha\frac{d^{2}}{6}P_{tt} + \alpha\frac{d^{3}}{12}\nabla_{t}P_{tt} + \bigo(d^{4})\right].
\end{align}

\subsection{Free field check from the off-diagonal Hadamard expansion}\label{sec:free_field_Hadamard}
A useful independent check of the covariant OPE comes from the short-distance expansion of the free propagator on a curved background. We stress at the outset that the framework of this subsection applies on an \textit{arbitrary} curved background. No conformal flatness is assumed anywhere in the Hadamard construction itself. Conformally flat backgrounds will appear below only as a special class of examples in which the state-dependent remainder can be computed exactly.

The relevant technology is the off-diagonal Hadamard/DeWitt expansion, because the anisotropic dependence on the separation vector is encoded in the covariant Taylor coefficients of the geometrical biscalars, rather than in the coincident heat-kernel coefficients alone \cite{Decanini:2005eg, Vassilevich:2003xt}. In particular, the coincident coefficient $a_{4}(x)$ in four dimensions shows that the Weyl curvature enters the local short-distance asymptotics, but it does not by itself determine the directional tensor $W_{\mu \rho\nu \lambda}t^{\rho}t^{\lambda}$ that appears naturally in the covariant OPE. 

For the scalar identity channel, this propagator technology is directly relevant. In a free theory, the coefficient of the identity is the propagator itself. We first set up the Hadamard representation for the general operator
\begin{align}
    L = -\nabla^{2} + m^{2} + \xi\,R,
\end{align}
keeping the mass and curvature coupling arbitrary, and only at the end specialising to the conformal values. Carrying $\xi$ and $m$ through the computation makes transparent which pieces of the final Schouten coefficient are geometrical and which are tied to conformal coupling. 

\subsubsection{Hadamard representation on a general background}
Here, \(\sigma(x,x')\) denotes Synge's two-point world function and is
distinct from the $1$-point Weyl exponent \(\sigma(x)\) used above.
We work in a geodesically convex neighbourhood and set
\begin{equation}
2\sigma(x,x')=d(x,x')^2 .
\end{equation}
Let \(t^a\) at \(x\) point along the geodesic from \(x\) towards
\(x'\). With the orientation fixed in section \ref{sec:intro},
\begin{equation}
\sigma_{;\mu}(x,x')=-d\,t_\mu .
\label{synge_def}
\end{equation}
Thus the powers of \(\sigma^{;\mu}\) appearing in the off-diagonal
Hadamard coefficients provide the propagator counterparts of the
powers of \(d\,t^\mu\) in the covariant OPE.

For \textit{odd} $D>2$, the Euclidean Hadamard representation may be written
as
\begin{equation}
G_H(x,x')
=
\frac{\Gamma\!\left(\frac{D}{2}-1\right)}
     {2(2\pi)^{\frac{D}{2}}}
\left[
\frac{U(x,x')}
     {\sigma(x,x')^{\frac{D}{2}-1}}
+W(x,x')
\right],
\qquad
U(x,x')=\sum_{n=0}^{\infty}U_n(x,x')\sigma^n .
\label{Hadamard_odd}
\end{equation}
For even \(D>2\), the corresponding expression is
\begin{equation}
G_H(x,x')
=
\frac{\Gamma\!\left(\frac{D}{2}-1\right)}
     {2(2\pi)^{\frac{D}{2}}}
\left[
\frac{U(x,x')}
     {\sigma(x,x')^{\frac{D}{2}-1}}
+V(x,x')\log\!\left(\frac{2\sigma(x,x')}{\ell^2}\right)
+W(x,x')
\right],
\label{Hadamard_even}
\end{equation}
where
\begin{equation}
V(x,x')
=
\sum_{n=0}^{\infty}V_n(x,x')\sigma^n .
\end{equation}
The normalisation is fixed by the flat-space limit,
\begin{equation}
\frac{\Gamma\!\left(\frac{D}{2}-1\right)}
     {2(2\pi)^{\frac{D}{2}}}
\frac1{\sigma^{\frac{D}{2}-1}}
=
\frac{\Gamma\!\left(\frac D2-1\right)}
     {4\pi^{\frac{D}{2}}d^{D-2}}.
\end{equation}

The leading geometrical coefficient is
\begin{equation}
U_0(x,x')=\Delta(x,x')^{1/2},
\end{equation}
where the Euclidean van Vleck--Morette determinant is
\begin{equation}
\Delta(x,x')
=
\frac{
\det\!\left[-\nabla_\mu\nabla_{\nu'}\sigma(x,x')\right]
}{
\sqrt{g(x)}\sqrt{g(x')}
}.
\end{equation}
This tells us how a thin spray of geodesics out of $x'$ towards $x$ spreads.  In flat
Euclidean space, it spreads exactly as straight lines with $-\nabla_\mu\nabla_{\nu'}\sigma=\delta_{\mu\nu}$, and hence $\Delta=1$. Curvature either focuses or defocuses the spray. There is no additional overall minus sign in this definition.

The coefficients \(U_n\), and \(V_n\) in even dimension, are fixed
locally by the Hadamard transport equations. By contrast, the biscalar
$W(x,x')$ is regular and state-dependent, $W_{0}$ is not fixed by geometrical recursion, and once $W_{0}$ is specified, the higher coefficients of $W$ are determined recursively \cite{Decanini:2005eg}. In this subsection only,
$W(x,x')$ denotes the smooth Hadamard biscalar and is unrelated to
the Weyl tensor \(W_{\mu\nu\rho\lambda}\) used elsewhere in the paper, and admits a Taylor expansion
\begin{align}
    W(x,x') = w(x) + w_{\mu}(x)\sigma^{;\mu} + w_{\mu\nu}(x)\sigma^{;\mu}\sigma^{;\nu} + w_{\mu\nu\rho}\sigma^{;\mu}\sigma^{;\nu}\sigma^{;\rho} + \bigo(d^{4}),
\end{align}
which by \eqref{synge_def} contributes only \textit{regular} local terms $w_{0} + d\,w_{1}(t) + d^{2}\,w_{2}(t) + \ldots$ to the propagator. The off-diagonal coefficients required below are
\cite{Decanini:2005eg, Decanini:2005gt}
\begin{align}
U_0(x,x')
={}&
1+\frac1{12}R_{\mu\nu}\sigma^{;\mu}\sigma^{;\nu}
-\frac1{24}\nabla_{(\rho}R_{\mu\nu)}
 \sigma^{;\mu}\sigma^{;\nu}\sigma^{;\rho}
+O(\sigma^2),
\label{U_0}
\\
U_1(x,x')
={}&
m^2+\left(\xi-\frac16\right)R
-\frac12\left(\xi-\frac16\right)
 \nabla_\mu R\,\sigma^{;\mu}
+O(\sigma),
\qquad D=3,
\label{U_1}
\\
V_0(x,x')
={}&
\frac12
\left[
m^2+\left(\xi-\frac16\right)R
\right]
-\frac14\left(\xi-\frac16\right)
 \nabla_\mu R\,\sigma^{;\mu}
+O(\sigma),
\qquad D=4.
\label{V_0}
\end{align}
The \(U_0\) term is common to the dimensions considered below. The
second line is the \(D=3\) coefficient entering the power singularity,
whereas the third line is the \(D=4\) coefficient multiplying the
logarithmic sector.

\subsubsection{Explicit computation for the conformal scalar in \(D=3\)}
For $D=3$, the singular part of \eqref{Hadamard_odd} is
\begin{equation}
G_H^{\mathrm{sing}}(x,x')
=
\frac{1}{4\pi d}
\left[
U_0(x,x')+\sigma U_1(x,x')+\bigo(d^4)
\right].
\end{equation}
Using \(\sigma^{;\mu}=-d\,t^\mu\) and \(\sigma=d^2/2\), the coefficients
in \eqref{U_0} and \eqref{U_1} give
    \begin{align}
U_0(x,x')
={}&
1+\frac{d^2}{12}R_{tt}
+\frac{d^3}{24}\nabla_tR_{tt}
+\bigo(d^4),
\\
\sigma U_1(x,x')
={}&
\frac{d^2}{2}
\left[
m^2+\left(\xi-\frac16\right)R
\right]
+\frac{d^3}{4}
\left(\xi-\frac16\right)\nabla_tR
+\bigo(d^4).
\end{align}
Therefore, for arbitrary \(m\), \(\xi\), and background,
\begin{align}
G_H^{\mathrm{sing}}(x,x')
=
\frac{1}{4\pi d}
\Bigg[
1
&+d^2\left\{
\frac1{12}R_{tt}
+\frac12
\left[
m^2+\left(\xi-\frac16\right)R
\right]
\right\} +d^3\left\{
\frac1{24}\nabla_tR_{tt}
+\frac14\left(\xi-\frac16\right)\nabla_tR
\right\}
+\bigo(d^4)
\Bigg].
\label{general_D3_Hadamard}
\end{align}
This result is local and assumes neither conformal flatness nor a
particular state.

We now specialise to the conformally coupled massless scalar,
\begin{equation}
m=0,
\qquad
\xi=\frac{D-2}{4(D-1)}\bigg|_{D=3}
=\frac18 .
\end{equation}
In three dimensions, \eqref{Schouten_1} becomes
\begin{equation}
P_{\mu\nu}=R_{\mu\nu}-\frac14Rg_{\mu\nu}.
\end{equation}
Consequently,
\begin{equation}
\frac1{12}R_{tt}-\frac1{48}R
=
\frac1{12}P_{tt},
\end{equation}
and, one order higher,
\begin{equation}
\frac1{24}\nabla_tR_{tt}
-\frac1{96}\nabla_tR
=
\frac1{24}\nabla_tP_{tt}.
\end{equation}
Thus, on an arbitrary three-dimensional background,
\begin{equation}
G_H^{\mathrm{sing}}(x,x')
=
\frac{1}{4\pi d}
\left[
1+\frac{d^2}{12}P_{tt}
+\frac{d^3}{24}\nabla_tP_{tt}
+O(d^4)
\right].
\label{D3_conformal_Hadamard}
\end{equation}
For the free scalar \(\Delta_\phi=(D-2)/2=1/2\), this agrees precisely
with the \(\alpha=1/2\) specialisation of
\eqref{X_inverse_alpha}. The trace pieces therefore cancel separately
at quadratic and cubic orders.

On a conformally flat patch one can make a stronger statement for the
locally Weyl-transformed flat vacuum. If
\(g_{\mu\nu}=e^{2\sigma}\delta_{\mu\nu}\), conformal covariance of the scalar
Green function gives
\begin{align}
\big\langle\phi(x)\phi(x')\big\rangle_g
&=
e^{-\sigma(x)/2}e^{-\sigma(x')/2}
\frac{1}{4\pi|x-x'|}
\nonumber\\
&=
\frac{1}{4\pi\,\widetilde X(x,x')^{\frac{1}{2}}}.
\label{D3_exact_conformal_propagator}
\end{align}
Thus the ambient expression is exact for this state on a conformally
flat patch. A different state or global image contributions can add a
smooth remainder without changing the universal singular coefficients
in \eqref{D3_conformal_Hadamard}.

Finally, because \(t^\mu t^\nu t^\rho\) projects onto the completely symmetric
part of \(\nabla_\mu P_{\nu\rho}\), the undifferentiated scalar propagator
cannot probe the antisymmetric Cotton projection:
\begin{equation}
C_{abc}t^at^bt^c=0.
\end{equation}
The free-field calculation therefore checks the universal symmetric
Schouten sector, but it does not determine the independent trilocal
Cotton contribution studied in section~\ref{sec:ansatz_OPE}.

\subsubsection{The check in \(D=4\) and the role of
\(\langle\phi^2\rangle\)}
In four dimensions, the Hadamard representation takes the form
\begin{equation}\label{eq:GH}
G_H(x,x')
=
\frac{1}{8\pi^2}
\left[
\frac{U_0(x,x')}{\sigma(x,x')}
+V(x,x')
 \log\!\left(\frac{2\sigma(x,x')}{\ell^2}\right)
+W(x,x')
\right].
\end{equation}
For the massless conformally coupled scalar,
\begin{equation}
m=0,
\qquad
\xi=\frac16,
\end{equation}
the coincidence value and the first off-diagonal derivative of
\(V_0\) in \eqref{V_0} vanish. The logarithmic sector therefore does
not contribute at linear order in the curvature and through the
three-derivative order retained here.

Using \eqref{U_0}, the pole part of \eqref{eq:GH} is
\begin{equation}
\frac{1}{8\pi^2}
\frac{U_0(x,x')}{\sigma(x,x')}
=
\frac{1}{4\pi^2d^2}
\left[
1+\frac{d^2}{12}R_{tt}
+\frac{d^3}{24}\nabla_tR_{tt}
+\bigo(d^4)
\right].
\label{D4_Hadamard_pole}
\end{equation}
This is the universal local singularity on an arbitrary
four-dimensional background. To determine the smooth term for a specified state, consider a
conformally flat patch \(g_{\mu\nu}=e^{2\sigma}\delta_{\mu\nu}\) and the
locally Weyl-transformed flat vacuum. Conformal covariance gives
\begin{align}
\big\langle\phi(x)\phi(x')\big\rangle_g
&=
e^{-\sigma(x)}e^{-\sigma(x')}
\frac{1}{4\pi^2|x-x'|^2}
\nonumber\\
&=
\frac{1}{4\pi^2\widetilde X(x,x')}
\nonumber\\
&=
\frac{1}{4\pi^2d^2}
\left[
1+\frac{d^2}{6}P_{tt}
+\frac{d^3}{12}\nabla_tP_{tt}
+\bigo(d^4)
\right].
\label{D4_exact_conformal_propagator}
\end{align}
In four dimensions,
\begin{equation}
P_{\mu\nu}
=
\frac12\left(R_{\mu\nu}-\frac16 R g_{\mu\nu}\right),
\end{equation}
and it follows that
\begin{align}
\big\langle\phi(x)\phi(x')\big\rangle_g = \frac{1}{4\pi^2d^2}
\Bigg[ 1 +d^2\left(
\frac1{12}R_{tt}-\frac1{72}R
\right) + d^3\left(
\frac1{24}\nabla_tR_{tt}
-\frac1{144}\nabla_tR
\right)
+O(d^4)
\Bigg].
\label{D4_Ricci_expansion}
\end{align}
Comparing \eqref{D4_Ricci_expansion} with
\eqref{D4_Hadamard_pole} gives
\begin{equation}
\big\langle\phi(x)\phi(x')\big\rangle_g
-
\left.G_H(x,x')\right|_{\frac{U_0}{\sigma}}
=
-\frac{R}{288\pi^2}
-\frac{d}{576\pi^2}\nabla_tR
+\bigo(d^2).
\label{D4_smooth_difference}
\end{equation}
Equivalently, in the normalisation above,
\begin{equation}
W(x,x)=-\frac{R}{36}.
\end{equation}
Since \(W(x,x')\) is symmetric, its first off-diagonal coefficient is
fixed by its coincidence value:
\begin{equation}
W(x,x')
=
-\frac{R}{36}
-\frac{d}{72}\nabla_tR
+\bigo(d^2).
\end{equation}

There are two equivalent operator bases in which to interpret this
result. In the minimally Hadamard-subtracted basis, the identity
coefficient is the pole \eqref{D4_Hadamard_pole}, while the
renormalised Wick square has, in the conformal vacuum,
\begin{equation}
\left\langle\phi^2\right\rangle_{\mathrm{ren}} =
-\frac{R}{288\pi^2}.
\label{D4_phi2_minimal}
\end{equation}
For the de Sitter space, this value of the $1$-point function is well known  (see e.g. \cite{birrell_davies_1982}, formula (7.43)). 
Furthermore, taking the derivative, we obtain 
\begin{equation}
\frac{d}{2}
\nabla_t
\left\langle\phi^2\right\rangle_{\mathrm{ren}}
=
-\frac{d}{576\pi^2}\nabla_tR,
\end{equation}
reproducing the second term in
\eqref{D4_smooth_difference}. 

Alternatively, we may pass to an operator basis in which the Wick square is a Weyl primary. The minimally subtracted $\phi^{2}$ is not, since the Hadamard parametrix is constructed from the chosen metric representative rather than from the conformal class, and under a Weyl transformation the subtracted operator picks up an inhomogeneous shift proportional to the identity operator.  Since $R\,\mathbb{1}$ is the only scalar of the correct dimension, the finite local ambiguity of the Hadamard scheme is exactly a multiple of $R\, \mathbb{1}$, and requiring the Wick square to transform as a primary of weight $2\Delta_{\phi} = 2$ fixes it uniquely. The required redefinition is
\begin{equation}
(\phi^2)_{\rm conf}
=
\phi^2+\frac{R}{288\pi^2}\,\mathds 1,
\label{D4_phi2_redefinition}
\end{equation}
which precisely compensates the anomalous shift. In this basis, $\langle(\phi^{2})_{\rm conf}\rangle$ vanishes in the conformal vacuum on every conformally flat background, as Weyl covariance of a non-trivial primary demands, and the identity contribution correspondingly becomes exactly $1/(4\pi^2\widetilde X)$.

The vanishing of \(V_0(x,x)\) at the
conformal point implies independence of the logarithmic Hadamard
length scale at the order considered here, but it does not remove
this independent finite curvature ambiguity. See \cite{Osborn_2000}, in particular section 11, for a prior discussion of $1$-point functions in free theories on manifolds of constant curvature. Requiring the Wick
square itself to transform conformally selects the distinguished
choice in \eqref{D4_phi2_redefinition}. We note that the choice of the $1$-point function at hand is independent of the  four-dimensional
stress-tensor anomaly. The latter also does not contribute in any way to the two-point function to the order of expansion we consider.

\section{The ansatz for correlation functions and OPE}
\label{sec:ansatz_OPE}
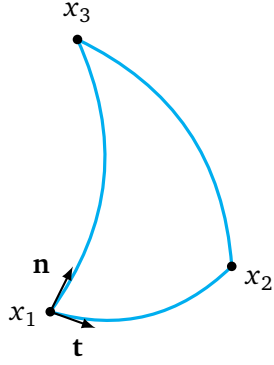
\begin{figure}[t]
    \centering
   \begin{tikzpicture}[>=latex,scale=1.2]

   \begin{scope}
     \draw[cyan, very thick] (0,0)  to[bend right] (2,0.5);
     \draw[cyan, very thick] (0,0)  to[bend right] (0.3,3);
     \draw[cyan, very thick] (2,0.5)  to[bend right] (0.3,3);
     \draw[black,thick,->] (0,0) --(0.5,-0.18);
     \draw[black,thick,->] (0,0) --(0.25,0.5);
   \end{scope}
   \fill[black] (0,0)circle (1.5pt);
   \fill[black] (2,0.5)circle (1.5pt);
   \fill[black] (0.3,3)circle (1.5pt);
   \draw (-0.3,-0.05) node{$x_1$};
   \draw (2.3,0.35) node{$x_2$};
   \draw (0.3,3.3) node{$x_3$};
   \draw (0.3,-0.4) node{${\bf t}$};
   \draw (-0.1,0.5) node{${\bf n}$};

   \end{tikzpicture}
\caption{The small geodesic triangle used in the
$x_2\rightarrow x_1$ expansion:
$x_2=\exp_{x_1}(dt)$ and $x_3=\exp_{x_1}(rn)$, with $d\ll r$ and both
distances small compared with the local curvature radius.}
\label{fig:geodesic_triangle1}
\end{figure}

We now focus on analysing the short-distance behaviour of three-point functions on a general, not conformally flat, background. 
We consider the three-point function 
\be
\langle {\cal O}_{1}(x_1){\cal O}_{2}(x_2){\cal O}_{3}(x_3)\rangle_{g_{\mu\nu}}
\ee
of three
scalar primary operators $\mathcal O_i$ that have generic dimensions $\Delta_i$,
with
\[
2\alpha_k=\Delta_i+\Delta_j-\Delta_k
\]
for $\{i,j,k\}$ a permutation of $\{1,2,3\}$. 
The ansatz for this correlator given in \eqref{corr_ansatz} depends on the pairwise invariants $\widetilde X_{ij}$, the short-distance expansion of which was obtained in section~\ref{ambient_sec}.  To use the corresponding expansions, we consider a small geodesic triangle with the vertices at the three-point function insertion points: $x_1,x_2,x_3$. More precisely, we choose a convex normal neighbourhood in which each pair of points $(x_{i},x_{j})$ is linked by a unique geodesic. We are interested in the limit $x_1\to x_2$ 
for which the geodesic distances satisfy 
\be
d(x_1,x_2) \ll d(x_1,x_3) 
\ee
Moreover, to be able to use the expansions (\ref{X(x,y)}) for each $\widetilde X_{ij}$ we need to have  
\[
d(x_1,x_2)\ll d(x_1,x_3)\ll R_{\rm curv} \, .
\]
This gives the precise meaning to the smallness of the geodesic triangle.
It is convenient to introduce a shorthand notation
\be \label{dandr}
d\equiv d(x_1,x_2)\, , \qquad r\equiv d(x_1,x_3) \, .
\ee
Using the covariant Taylor expansions along the corresponding geodesics we 
can write 
\begin{equation}
 x_2=\exp_{x_1}(dt),
 \qquad
 x_3=\exp_{x_1}(rn),
 \label{sec4_geo}
\end{equation}
where $t^\mu$ and $n^\mu$ are the unit length tangent vectors to the $(x_1,x_2)$ and $(x_1,x_3)$ geodesics respectively,  based at $x_1$ and oriented outward, see Figure \ref{fig:geodesic_triangle1}. 

In our expansions we are going to retain terms through cubic order in $d$ at fixed local $r$ and hence work
to linear order in curvature. All omitted orders are recorded explicitly
in the remainder of each truncated formula. We use
$\bigo(\partial^4)$ as shorthand for curvature jets of fourth derivative
order, including $\nabla^2P$, $\nabla C$, and $\nabla^2W$, while
$\bigo(\mathrm{curv}^2)$ includes all terms quadratic in the curvature
and its covariant derivatives, including mixed Schouten--Weyl and
Schouten--Cotton terms.
At these orders none of the corrections $\mathcal I_i^{(k)}$ in
\eqref{corr_ansatz} contributes:
$\mathcal I_i^{(1)}=0$ by the ambient identities, while
$\mathcal I_i^{(k)}$, $k\geq2$, is at least quadratic in curvature.
Thus the object to be tested is the  expression
\be \label{pairwise_part}
\frac{C_{123}}{\left(\widetilde{X}_{12}\right)^{\alpha_{3}}\left(\widetilde{X}_{13}\right)^{\alpha_{2}}\left(\widetilde{X}_{23}\right)^{\alpha_{1}}} \, .
\ee

\subsection{The local OPE test}
\label{sec4_match_sub}

We write the scalar contribution to the OPE in the same normalisation as
in section~\ref{confflat_sec}:
\begin{equation}
 \mathcal O_2(x_2)\mathcal O_1(x_1)
 \supset
 \frac{C_{123}}{\widetilde X_{12}^{\alpha_3}}
 C(g_{\mu\nu},t^\mu,d)\mathcal O_3(x_1).
 \label{sec4_OPE}
\end{equation}
Here $C(g_{\mu\nu},t^\mu,d)$ denotes the complete candidate local descendant
operator on a general background. We separate the part already fixed
on conformally flat geometries from possible terms that vanish on every
conformally flat background:
\be
C(g_{\mu\nu},t^\mu,d)
=
C_{\mathrm{cf}}(g_{\mu\nu},t^\mu,d)
+
C_{\mathrm{ncf}}(g_{\mu\nu},t^\mu,d).
\ee
Here $C_{\mathrm{cf}}$ is the complete Schouten-dressed operator
constructed in section~\ref{confflat_sec}, covariantly evaluated on the
chosen metric, whereas $C_{\mathrm{ncf}}$ is reserved for possible local terms involving the Weyl or Cotton tensors and therefore vanishes on every conformally flat background.

At linear order in curvature and within the parity-even
scalar-descendant sector considered here, constant Weyl rescalings and
the tensor symmetries show that no such non-conformally-flat correction
exists through order $d^3$. A two-point configuration supplies only the
tangent $t^\mu$, and the parity-even scalar contractions $W_{tttt}$ and
$C_{ttt}$ vanish identically. The first non-trivial contributions to  $C(g_{\mu\nu},t^\mu,d)$ involving the Cotton and Weyl tensors  are the following operators
\[
d^4t^\mu t^\rho W_{\mu\nu\rho\lambda}\nabla^\nu\nabla^\lambda,
\qquad
d^4t^\mu t^\rho C_{\mu\nu\rho}\nabla^\nu.
\]
Consequently, within this sector,
\[
C_{\mathrm{ncf}}(g_{\mu\nu},t^\mu,d)=\bigo(d^4),
\qquad
C(g_{\mu\nu},t^\mu,d)
=
C_{\mathrm{cf}}(g_{\mu\nu},t^\mu,d)+\bigo(d^4).
\]
Thus the OPE test through cubic order involves only the already
determined operator $C_{\mathrm{cf}}$.

We keep the unit two-point normalisation of \eqref{2pt_flat}. Since there
is no independent two-point scalar structure linear in the ambient
curvature, the two-point function needed at this order is
\begin{equation}
\left\langle
\mathcal O_3(x_1)\mathcal O_3(x_3)
\right\rangle_{g_{\mu\nu}}
=
\widetilde X_{13}^{-\Delta_3}
\left[1+\bigo(\mathrm{curv}^2)\right].
\label{sec4_2pt}
\end{equation}

Taking the correlator of \eqref{sec4_OPE} with
$\mathcal O_3(x_3)$ therefore gives, to the accuracy retained here,
\[
\langle {\cal O}_{1}(x_1){\cal O}_{2}(x_2){\cal O}_{3}(x_3)\rangle_{g_{\mu\nu}}=\frac{C_{123}}{\widetilde X_{12}^{\alpha_3}}
\left[
C(g_{\mu\nu},t^\mu,d)\widetilde X_{13}^{-\Delta_3}
+
\widetilde X_{13}^{-\Delta_3}
\bigo(\mathrm{curv}^2)
\right].
\]
On the other hand, after dropping the $\mathcal I_i^{(k)}$ corrections
at the retained order, the pairwise three-point ansatz (\ref{pairwise_part}) gives
\[
\langle {\cal O}_{1}(x_1){\cal O}_{2}(x_2){\cal O}_{3}(x_3)\rangle_{g_{\mu\nu}}=\frac{C_{123}}
{\widetilde X_{12}^{\alpha_3}
 \widetilde X_{13}^{\alpha_2}
 \widetilde X_{23}^{\alpha_1}}
\left[1+\bigo(\mathrm{curv}^2)\right].
\]
Cancelling the common factor
$C_{123}/\widetilde X_{12}^{\alpha_3}$, compatibility with a local OPE
would require
\begin{equation}
C(g_{\mu\nu},t^\mu,d)\widetilde X_{13}^{-\Delta_3}
\stackrel{?}{=}
\frac{1}
{\widetilde X_{13}^{\alpha_2}
 \widetilde X_{23}^{\alpha_1}}
+
\widetilde X_{13}^{-\Delta_3}
\bigo(\mathrm{curv}^2).
\label{sec4_OPE_test}
\end{equation}
Using $C=C_{\mathrm{cf}}+\bigo(d^4)$, the truncation of this condition
through the retained orders is equivalently
\be\label{eq:match?}
C_{\mathrm{cf}}(g_{\mu\nu},t^\mu,d)
\widetilde X_{13}^{-\Delta_3}
\stackrel{?}{=}
\frac{1}
{\widetilde X_{13}^{\alpha_2}
 \widetilde X_{23}^{\alpha_1}}
+
\widetilde X_{13}^{-\Delta_3}
\bigo\!\left(d^4,\partial^4,\mathrm{curv}^2\right).
\ee
We will refer to the above as the matching condition which asks the following question:
\textit{Does the short-distance expansion of the pairwise three-point ansatz agree with what is obtained by applying the local OPE descendant operator to the two-point function?}

To answer the above let us focus on the ratio $\widetilde X_{23}/\widetilde X_{13}$ which, as we shall see, arises directly from the matching condition~\eqref{eq:match?}. Indeed, since
$\alpha_1+\alpha_2=\Delta_3$,
\[
{1\over
 \widetilde X_{13}^{\alpha_2}
 \widetilde X_{23}^{\alpha_1}}
=
\widetilde X_{13}^{-\Delta_3}
\left(
{\widetilde X_{23}\over\widetilde X_{13}}
\right)^{-\alpha_1}.
\]
After the two-point factor $\widetilde X_{13}^{-\Delta_3}$ has been
extracted, all non-trivial dependence on the third side of the small
geodesic triangle is therefore contained in this ratio. The required covariant expansion of
$\widetilde X_{23}/\widetilde X_{13}$ is given in
\eqref{cos_full}. Substituting this expansion into
~\eqref{eq:match?} we find that  the terms coming from the descendants operator cancel all  Schouten tensor dependent terms. The remaining Weyl- and Cotton-dependent terms are  calculated
 in appendix~\ref{app:sec4_matching}.

Since $C_{\rm ncf}=\bigo(d^4)$ \footnote{In $D=3$, the parity-odd scalar structure
$d^3Y_{\mu\nu}t^\mu t^\nu$ is allowed, but it lies outside the
parity-even scalar-descendant sector considered here.}, it cannot modify any coefficient
in the matching condition through order $d^3$. Consequently, the
Weyl--Cotton discrepancy obtained using $C_{\rm cf}$ is also the
discrepancy for the complete candidate local operator
$C=C_{\rm cf}+C_{\rm ncf}$ at the retained order. Explicitly we find
\begin{equation}
\begin{split}
C(g_{\mu\nu},t^\mu,d)\widetilde X_{13}^{-\Delta_3}
-
\frac{1}
{\widetilde X_{13}^{\alpha_2}
 \widetilde X_{23}^{\alpha_1}}
&=-\widetilde X_{13}^{-\Delta_3}
\bigg\{
\frac{\alpha_1\alpha_2}{3(\Delta_3+1)}
d^2W_{tntn}
+
\frac{2\alpha_1(\alpha_1+1)\alpha_2}
{3(\Delta_3+2)}
\frac{d^3c}{r}W_{tntn}
\\
&\qquad
+
\frac{\alpha_1\alpha_2}{12(\Delta_3+1)}
d^2r\nabla_nW_{tntn}
+
\frac{\alpha_1\alpha_2(2\alpha_1+\alpha_2+3)}
{12(\Delta_3+1)(\Delta_3+2)}
d^3\nabla_tW_{tntn}
\\
&\qquad
+
\frac{\alpha_1(\alpha_1+1)\alpha_2}
{6(\Delta_3+2)}
d^3c\,\nabla_nW_{tntn}
+
\frac{\alpha_1\alpha_2}{6(\Delta_3+1)}
d^2r\bigl(C_{tnt}-cC_{ntn}\bigr)
\\
&\qquad
+
\frac{\alpha_1\alpha_2d^3}
{6(\Delta_3+1)(\Delta_3+2)}
\bigg[
\left\{
\alpha_2+1
-2(\alpha_1+1)(\Delta_3+1)c^2
\right\}C_{ntn}
\\
&
+
\left(
2\alpha_1^2+2\alpha_1\alpha_2
+2\alpha_1+\alpha_2-1
\right)cC_{tnt}
\bigg]
+\bigo\!\left(d^4,\partial^4,\mathrm{curv}^2\right)
\bigg\}
\end{split}
\label{sec4_mismatch}
\end{equation}
where 
\be
c = t \cdot n = g_{\mu\nu}(x_1)t^{\mu}n^{\nu} \, .
\ee

Several features of \eqref{sec4_mismatch} are important. First, after
all local scalar-descendant contributions through order $d^3$ have
been subtracted, the remaining difference is genuinely trilocal. In
the $x_2\to x_1$ channel, a local OPE coefficient may depend on $d$,
$t^\mu$, and background tensors at $x_1$, but not on the spectator
point $x_3$. By contrast, every displayed curvature structure in
\eqref{sec4_mismatch} depends on $x_3$ through $n^\mu$, and several
terms also contain $r=d(x_1,x_3)$ or $r^{-1}$. Thus the required
correction depends simultaneously on the OPE pair and on the
spectator point.

Second, the difference cannot be absorbed into a correction of the
two-point function. The ambient identities give
$\mathcal I_2^{(1)}=0$. Moreover, two-point data for the pair
$(x_1,x_3)$ may depend on $r$ and $n^\mu$, but not independently on
the second-channel data $d$ and $t^\mu$; it therefore cannot reproduce
the mixed $t^\mu$--$n^\mu$ structures in
\eqref{sec4_mismatch}.

Third, neither the corrections $\mathcal I_3^{(k)}$ already included in \eqref{corr_ansatz} nor the Fefferman--Graham datum $g_{(3)\mu\nu}$ suppressed in section~\ref{review_subsec} can remove the terms displayed in \eqref{sec4_mismatch}.

The strongest obstruction in $D=3$ is parity. Hodge-dualising the antisymmetric index pair of the Cotton tensor gives
\be
Y_{\mu\nu}
\equiv
\frac{1}{2}\epsilon_{\mu}{}^{\rho\lambda}
C_{\nu\rho\lambda}.
\ee
The Cotton--York tensor $Y_{\mu\nu}$ is symmetric, traceless and transverse, and therefore has the same algebraic and differential properties as $g_{(3)\mu\nu}$. Their parity properties, however, are different and hence they cannot mix. Consequently, a contribution proportional to $g_{(3)\mu\nu}$ cannot cancel a contribution proportional to $Y_{\mu\nu}$. The Hodge dualisation therefore does not provide a possible identification between the two tensors; instead, it makes the parity obstruction manifest.

Curvature counting provides a second, independent obstruction. The ambient identities imply: $\mathcal I_3^{(1)}=0$, while every $\mathcal I_3^{(k)}$ with $k\geq2$ is at least quadratic in the curvature. These corrections therefore cannot cancel the terms in \eqref{sec4_mismatch}, which are linear in $W$, $\nabla W$ and $C$. Moreover, as explained in section~\ref{review_subsec}, for $D\geq4$ the first contribution of $g_{(D)\mu\nu}$ occurs at relative order $d^D$ and hence lies beyond the cubic order retained here. Only in $D=3$ does $g_{(3)\mu\nu}$ enter a pairwise invariant at relative order $d^3$, which is why the three-dimensional case could not be discarded by separation counting alone. This does not alter the parity obstruction established above.

The third distinction is state dependence. The coefficient $g_{(3)\mu\nu}$ is independent Fefferman--Graham data: it can be varied by changing the state while keeping the background metric fixed. The Cotton tensor appearing in \eqref{sec4_mismatch}, by contrast, is fixed by the background geometry. Since the OPE matching condition must hold for arbitrary states, the state-dependent $g_{(3)\mu\nu}$ sector and the geometric Cotton sector must match separately. A cancellation obtained by tuning $g_{(3)\mu\nu}$ in one specially chosen state would not constitute a universal identity.
\Comment{
This distinction also has a direct OPE interpretation. The $g_{(3)\mu\nu}$ contribution enters through the stress-tensor conformal family: $\mathcal O_i\mathcal O_j
\supset
C_{ijT}\,
d^{D-\Delta_i-\Delta_j}\,
t^\mu t^\nu T_{\mu\nu}(x_1)$, with the state dependence appearing only after taking the expectation value, $\langle T_{\mu\nu}\rangle
\propto
g_{(3)\mu\nu}$. The stress-tensor family is certainly part of the complete OPE, but it is an operator family independent of the scalar family tested by the matching condition \eqref{eq:match?}. In a generic situation, this channel will produce different powers of $d$ than the ones provided by the $\tilde X_{ij}$ invariants.

In view of these distinctions, we conclude that while reinstating $g_{(3)\mu\nu}$ in our calculations restores the appropriate state-dependent stress-tensor contribution, it cannot cancel the Cotton-dependent geometric terms in \eqref{sec4_mismatch} or remove the need for the trilocal correction $J^{(1)}$.} This distinction also has a direct OPE interpretation. The matching condition \eqref{eq:match?} isolates the \(\mathcal{O}_1 \mathcal{O}_2\to \mathcal{O}_3\) conformal family:
$$
 \big\langle \mathcal{O}_1(x_1)\mathcal{O}_2(x_2)\mathcal{O}_3(x_3)\big\rangle_{g_{\mu\nu}}
 \supset
 \frac{C_{123}}{\widetilde X_{12}^{\alpha_3}}\, C(g_{\mu\nu},t^\mu,d)\,
 \big\langle \mathcal{O}_3(x_1)\mathcal{O}_3(x_3)\big\rangle_{g_{\mu\nu}} .
$$

Although this is the \(\mathcal{O}_3\) family in the \(x_2\to x_1\) channel, the two-point function on the right-hand side, together with its derivatives generated by \( C(g_{\mu\nu},t^\mu,d)\), can itself contain VEVs of the stress tensor and its composites. For example, its own short-distance OPE contains

$$
 \mathcal{O}_3(x_1)\mathcal{O}_3(x_3)
 \supset
 C_{33T}\,r^{D-2\Delta_3}
 n^\mu n^\nu T_{\mu\nu}(x_1)+\cdots ,
$$
so that taking the expectation value introduces
\(\langle T_{\mu\nu}\rangle\propto g_{(D)\mu\nu}\). The analogous multi-stress-tensor contributions are organised by the two-point invariants \(\mathcal{I}_2^{(k)}\). Thus stress-tensor VEVs may enter within the \(\mathcal{O}_3\)-family contribution tested by \eqref{eq:match?}. For dimensional reasons, the local part of such  VEVs would contain at least two factors of the Riemann tensor.
Such contributions are distinct from a possible direct \(\mathcal{O}_1 \mathcal{O}_2\to[T^k]\) channel, which would involve a mixed correlator of the form
\(\langle[T^k](x_1)\mathcal{O}_3(x_3)\rangle_{g_{\mu\nu}}\). Such direct operator families are not included in the matching condition \eqref{eq:match?}.

Another point worth mentioning regarding formula \eqref{sec4_mismatch} is that at the accuracy considered here, the additional two-point invariants do not modify the calculation: the ambient identities give \(\mathcal{I}_2^{(1)}=0\), while \(\mathcal{I}_2^{(k)}\), \(k\geq2\), is at least quadratic in the ambient curvature. The single-stress-tensor state dependence is instead contained in the pairwise invariant itself. In \(D=3\), this includes the freely specifiable datum \(g_{(3)\mu\nu}\), which was set to zero in obtaining \eqref{sec4_mismatch}. Equation \eqref{sec4_mismatch} therefore establishes a Cotton-dependent discrepancy already in the \(g_{(3)}\)-independent sector. Reinstating \(g_{(3)\mu\nu}\) adds further state-dependent terms, but it cannot erase this background-dependent discrepancy as a universal identity: the coefficients independent of \(g_{(3)\mu\nu}\) must match separately, and the parity distinction established above provides an additional obstruction. Thus the pairwise \(\mathcal{O}_3\)-family contribution still requires the trilocal correction \(J^{(1)}\).

As a final remark about \eqref{sec4_mismatch}, we note that  every term in that expression contains the factor
$\alpha_1\alpha_2$. The mismatch therefore vanishes when either of the two relevant external operators is the identity, as required by the absence of a linear conformal-curvature correction to the two-point function.

\subsection{The required trilocal correction}
\label{sec4_J_sub}

If the theory is to possess a local covariant OPE, the pairwise
three-point ansatz must therefore be corrected. We now introduce
$J^{(1)}$ by writing
\begin{equation}
\begin{split}
\left\langle
\mathcal O_1(x_1)\mathcal O_2(x_2)\mathcal O_3(x_3)
\right\rangle_{g_{\mu\nu}}
={}&
\frac{C_{123}}
{\widetilde X_{12}^{\alpha_3}
 \widetilde X_{13}^{\alpha_2}
 \widetilde X_{23}^{\alpha_1}}
\left[
1+J^{(1)}+\mathcal I_3^{(2)}+\cdots
\right].
\end{split}
\label{sec4_corrected_3pt}
\end{equation}
At linear order in conformal curvature, matching
\eqref{sec4_corrected_3pt} to the local OPE gives
\begin{equation}
\begin{aligned}
J^{(1)}
={}&
\widetilde X_{13}^{\alpha_2}
\widetilde X_{23}^{\alpha_1}
C(g_{\mu\nu},t^\mu,d)
\widetilde X_{13}^{-\Delta_3}
-1
+\bigo(\mathrm{curv}^2)
\\
={}&
\widetilde X_{13}^{\alpha_2}
\widetilde X_{23}^{\alpha_1}
C_{\mathrm{cf}}(g_{\mu\nu},t^\mu,d)
\widetilde X_{13}^{-\Delta_3}
-1
+
\bigo\!\left(d^4,\partial^4,\mathrm{curv}^2\right).
\end{aligned}
\label{sec4_J_definition}
\end{equation}
Using \eqref{sec4_mismatch}, we obtain
\begin{equation}
\begin{aligned}
 J^{(1)}={}&
 -\frac{\alpha_1\alpha_2}
 {3(\Delta_3+1)}
 d^2W_{tntn}
 -
 \frac{2\alpha_1\alpha_2(\alpha_2+1)}
 {3(\Delta_3+1)(\Delta_3+2)}
 \frac{d^3c}{r}W_{tntn}
 \\
 &-
 \frac{\alpha_1\alpha_2}
 {12(\Delta_3+1)}
 d^2r\nabla_nW_{tntn}
 -
 \frac{\alpha_1\alpha_2(2\alpha_1+\alpha_2+3)}
 {12(\Delta_3+1)(\Delta_3+2)}
 d^3\nabla_tW_{tntn}
 \\
 &-
 \frac{\alpha_1\alpha_2(\alpha_2+1)}
 {6(\Delta_3+1)(\Delta_3+2)}
 d^3c\,\nabla_nW_{tntn}
 +
 \frac{\alpha_1\alpha_2}
 {6(\Delta_3+1)}
 d^2r\bigl(cC_{ntn}-C_{tnt}\bigr)
 \\
 &+
 \frac{\alpha_1\alpha_2d^3}
 {6(\Delta_3+1)(\Delta_3+2)}
 \left[
 (\alpha_2+1)(2c^2-1)C_{ntn}
 +(2\alpha_1-\alpha_2+1)cC_{tnt}
 \right]
 \\
 &+
 \bigo\!\left(d^4,\partial^4,\mathrm{curv}^2\right).
\end{aligned}
\label{sec4_J}
\end{equation}

Equation~\eqref{sec4_J} fixes the necessary terms in the
$x_2\to x_1$ short-distance expansion through order $d^3$, at linear
order in curvature and through the retained curvature-derivative
order; it does not determine a unique function of three separated
points. Terms beginning at order $d^4$, the omitted
$\bigo(\partial^4,\mathrm{curv}^2)$ contributions, and the
compatible short-distance expansions in the other two OPE channels
remain to be determined. On a conformally flat background,
$W_{\mu\nu\rho\lambda}=C_{\mu\nu\rho}=0$, so the displayed correction
vanishes and the result governed by $C_{\rm cf}$ is recovered.

The Weyl covariance of \eqref{sec4_J} is not manifest term-by-term.
Appendix~\ref{app:sec4_Weyl_covariance} provides a non-trivial sanity check: the inhomogeneous variation of $J^{(1)}$ vanishes to the appropriate number of derivatives. (In $D\geq4$, this happens due to a cancellation between the variation of the Weyl tensor terms and the Cotton ones. In $D=3$, where the Weyl tensor vanishes identically, the displayed Cotton contribution is separately covariant at the retained curvature-jet order.)

On conformally flat spaces, the curvature dressing of the descendant tower
is completely fixed by flat-space OPE data and is therefore kinematic. On
a general background, the existence of a local OPE instead imposes a
non-trivial constraint on the three-point function. The calculation above
fixes the short-distance trilocal correction required by that constraint, but
whether its global completion remains universal kinematics or constitutes
new dynamical input is left open.

\section{Discussion} \label{discussion_sec}
We have argued that on a general background, the existence of a local covariant OPE turns from a consequence of symmetry into a non-trivial constraint on the correlation functions. On conformally flat spaces, the payoff is completeness. The conformally flat operator $C_{\mathrm{cf}}$ of
section~\ref{confflat_sec} is generated to arbitrary order by flat space alone. It factorises into three geometric maps: the endpoint Weyl factor, the covariant expansion of the displacement vector, and the local map $\mathbb{T}$ that trades flat derivatives for covariant ones, and every term it produces arrives multiplied by a flat-space descendant coefficient. The map $\mathbb{T}$ has, moreover, a life of its own. It can be reconstructed from the tractor tensors generated by successive applications of the Thomas operator.
This construction explains why combinations such as
$\nabla_\mu\nabla_\nu-\Delta_3P_{\mu\nu}$ appear naturally in the curved descendant tower -- they are components of the tractors generated by the Thomas operator acting on the primary. The construction of section \ref{confflat_sec} is essentially algorithmic, thus, giving an example of how to construct OPE on a curved space to all orders.

\Comment{
On general backgrounds we showed that locality turns from a consequence into a constraint. Parisini, Skenderis, and Withers \cite{Parisini:2022wkb, Parisini:2023nbd} proposed a natural ansatz for correlation functions on general backgrounds, built in the ambient space from pairwise invariants $\widetilde{X}_{ij}$ and  ${\cal I}_{i}^{(k)}$. When we take the short-distance limit of the three-point ansatz, we find that it does not automatically admit a local covariant OPE. Corrections must be added, and they are controlled by exactly the tensors that measure the failure of conformal flatness, the Cotton tensor in three dimensions, joined in four dimensions and above by the Weyl tensor and its first derivatives. Demanding that the OPE exists is therefore doing real work. It tells us something the ansatz alone did not know. 
}

On general backgrounds we showed that locality turns from a consequence into a constraint. Parisini, Skenderis, and Withers \cite{Parisini:2022wkb, Parisini:2023nbd} proposed a natural ansatz for correlation functions on general backgrounds, built in the ambient space from the pairwise invariants \(\widetilde X_{ij}\) and \(\mathcal{I}_i^{(k)}\). As emphasised in those works, the ambient expressions constructed in this way provide a universal sector rather than the most general solution of the separated-point Weyl-covariance constraints; contributions from other operator families may require additional ambient data and invariants. More work needs to be done to see whether  some modifications of the ambient formalism may be needed to reproduce the trilocal correction we found in a natural way \footnote{We thank Kostas Skenderis for helpful discussions of this
particular point.}. It is hard to imagine how that may work given that $J^{(1)}$ 
has no free parameters in it which could account for choosing a state. 

Taking the short-distance limit of the pairwise \(\mathcal{O}_3\)-family contribution, we find that, at linear order in conformal curvature, it does not by itself reproduce the corresponding local covariant descendant expansion. Its completion requires the trilocal term \(J^{(1)}\), controlled by precisely the tensors that measure the failure of conformal flatness: the Cotton tensor in three dimensions, joined in four dimensions and above by the Weyl tensor and its first derivatives. Thus our result constrains the pairwise continuation of the \(\mathcal{O}_3\) conformal family beyond what Weyl covariance alone fixes; it does not assert that the construction of \cite{Parisini:2022wkb,Parisini:2023nbd} was intended to represent the complete three-point function.

There are several directions in which our work can be extended. Perhaps the most obvious one concerns constructing a global completion of the pairwise \(C_{123}\) contribution tested above and understanding how it is embedded in a full correlator containing the additional operator sectors not captured by the universal ambient construction of~\cite{Parisini:2022wkb,Parisini:2023nbd}. The quantity \(J^{(1)}\) determined here contains only the first two terms in its short-distance expansion and must therefore be completed to a global correction term. It is hard to imagine what could be the geometrical meaning of this correction. While the building blocks proposed in \cite{Parisini:2022wkb, Parisini:2023nbd} have a clear provenance from parallel transport in the ambient space this doesn't seem to work for $J^{(1)}$. Given the trilocal nature of $J^{(1)}$ one can try to relate it to holonomy of $T$ around the 
geodesic triangle that does indeed pick up the ambient sectional curvature. 
However this clashes with  the $r$-dependence of $J^{(1)}$ which  is non-homogeneous with different powers of $r$ present in the terms entering $J^{(1)}$. More generally, it is not clear that the global correction that would complete $J^{(1)}$ can be fixed by locality and Weyl covariance alone.
In even dimensions the conformal anomaly, which did not enter the analysis of this paper, may contribute as well.

The second direction concerns operators with spin and operators which are odd under parity transformations. We plan to incorporate such operators into our analysis in the forthcoming work \cite{WIP}. There we extend the covariant OPE of section \ref{confflat_sec} to spinning primaries and show that the parity-odd conformal-curvature structures already appear in free-theory correlators once spin is present. In three dimensions, the Cotton--York tensor $Y_{\mu\nu}$ enters the scalar channel through $Y_{\mu\nu}t^{\mu}t^{\nu}$ at order $d^{3}$, the vector channel through $Y_{\mu\nu}t^{\nu}$, and the spin-two channel through $Y_{\mu\nu}$ itself, while in $D\geq 4$ the spin-two channel admits the unique Weyl structure $W_{\mu \rho\nu \lambda}t^{\rho}t^{\lambda}$ already at order $d^{3}$, with the Bach tensor first appearing at order $d^{4}$. Selection rules follow, for instance, the absence of a parity-odd vector structure at order $d^{2}$, and in $D = 3$ the unitarity bound appears as a resonance with the conformal-Laplacian descendant. Large-$N$ Chern--Simons--matter theories furnish concrete examples in which the parity-odd sector is non-vanishing and computable. The gravitational Chern--Simons term, whose metric variation is the Cotton tensor, is the natural source for the parity-odd Cotton tensor structures in curved-space correlators, and in these theories it is accompanied by separated-point parity-odd OPE coefficients that can be computed explicitly. We plan to investigate such theories in more detail  in future work.

The third future direction carries the OPE construction into Lorentzian signature, where the observables that motivate much of this subject, the energy correlators built from light-ray operators, actually live. On conformally flat backgrounds, the continuation is remarkably benign. The curvature dressing is signature blind; only the kinematic power of the geodesic interval continues through the Wightman prescription, and the conformal anomaly is isolated as the unique carrier of a light-cone branch cut. The light transform respects the Weyl map up to a positive redshift weight, the small-angle detector OPE becomes a transverse coincidence limit dressed by the Schouten tensor, and on de Sitter the small-angle energy-energy correlator can be written down and compared, term-by-term, with its flat-space limit. Averaged null energy positivity remains the delicate point since positivity is not inherited through the Weyl map, although on maximally symmetric backgrounds the anomaly null-projects to zero; the question becomes an intrinsic achronal one.

\section*{Acknowledgements}
We want to thank Kostas Skenderis, Sergey Solodukhin and Omar Zanusso for useful discussions.
AD acknowledges support from a recently concluded STFC Consolidated Grant ST/T000600/1 -- ``Particle Theory at the Higgs Centre''. The work of AD is supported by the research grant ANRF/ECRG/2024/000247/PMS from the Anusandhan National Research Foundation (ANRF), India. NU is supported by DOE grant DE-SC0017660.

\appendix

\section{Cosine law and geodesic triangle expansion} \label{app:cosine-law}

This appendix collects the details underlying the covariant cosine law
expansion used in sections~\ref{ambient_sec} and \ref{sec:ansatz_OPE}.  We keep the orientation and notation of \eqref{dandr}, \eqref{sec4_geo}. In particular, $t^\mu$ and $n^\mu$ point away from $x_1$, towards $x_2$ and $x_3$, respectively. All tensors below are evaluated at $x_1$.

Throughout this appendix we work to linear order in curvature, retain
curvature jets through three-derivative order, and expand through cubic
order in $d$. The formulae defining $\chi_S$, $\chi_W$, and $\chi_C$
below are understood at this retained order; the combined remainder is
displayed explicitly in \eqref{cos_full}.

\subsection{Covariant Taylor expansion}

For fixed $x_3$, the pairwise invariant $\widetilde X(x,x_3)$ is a scalar in
its first argument.  Its Taylor expansion along the geodesic from $x_1$ to
$x_2$ is therefore
\begin{equation}
 \widetilde X_{23}
 =\widetilde X_{13}+dY_t+\frac{d^2}{2}Y_{tt}
 +\frac{d^3}{6}Y_{ttt}+O(d^4),
 \label{cos_Taylor}
\end{equation}
where
\[
Y_{t\cdots t}
\equiv
\left.
t^{\mu_1}\cdots t^{\mu_k}
\nabla_{\mu_1}\cdots\nabla_{\mu_k}
\widetilde X(x,x_3)
\right|_{x=x_1}.
\]
In the locally flat limit,
\begin{equation}
 \widetilde X_{13}=r^2,
 \qquad
 Y_t=-2rc,
 \qquad
 Y_{tt}=2,
 \qquad
 Y_{ttt}=0.
 \label{cos_flat}
\end{equation}
Dividing \eqref{cos_Taylor} by \(\widetilde X_{13}=r^{2}\), we recover the
ordinary flat-space cosine law,
\be
\frac{\widetilde X_{23}}{\widetilde X_{13}}
=
1-2\frac{d}{r}c+\frac{d^{2}}{r^{2}},
\qquad
c=t\cdot n=\cos\theta .
\ee
Equivalently,
\be
\widetilde X_{23}
=
r^{2}+d^{2}-2rd\cos\theta .
\ee

For the curvature calculation it is useful to use Synge's world function
$\sigma(x_1,x_3)=r^2/2$.  With our orientation,
\begin{equation}
 \sigma^\mu\equiv\nabla^\mu\sigma=-rn^\mu.
 \label{cos_sigma}
\end{equation}
The invariant \eqref{X(x,y)} can equivalently be written, to first order in
the Schouten tensor and its first derivative, as
\begin{equation}
 \widetilde X_{13}
 =2\sigma-\frac{\sigma}{3}\sigma^\mu\sigma^\nu P_{\mu\nu}
 +\frac{\sigma}{6}\sigma^\mu\sigma^\nu\sigma^\rho\nabla_\rho P_{\mu\nu}
 +O(\partial^4,\mathrm{curv}^2).
 \label{cos_X}
\end{equation}
Using $\sigma=r^2/2$ and $\sigma^\mu=-rn^\mu$, equation~\eqref{cos_X}
becomes
\be
\widetilde X_{13}
=
r^2\left[
1-\frac{r^2}{6}P_{nn}
-\frac{r^3}{12}\nabla_nP_{nn}
+\bigo(\partial^4,\mathrm{curv}^2)
\right].
\ee
Thus the plus sign of the last term in \eqref{cos_X} is consistent
with the negative radial cubic term, because each factor $\sigma^\mu$
contributes $-rn^\mu$.

The completely symmetric Schouten pieces obtained by differentiating
\eqref{cos_X} assemble into
\begin{equation}
 \begin{split}
 \chi_S={}&\frac{dr}{3}P_{tn}-\frac{d^2}{3}P_{nn}
 -\frac{d^2}{2}P_{tt}+\frac{d^3c}{3r}P_{tt}
 +\frac{d^3}{3r}P_{tn}
 \\
 &+\frac{dr^2}{12}\nabla_{(t}P_{nn)}
 -\frac{d^2r}{12}\nabla_nP_{nn}
 -\frac{d^3}{6}\nabla_tP_{tt},
 \end{split}
 \label{cos_S}
\end{equation}
The mixed-derivative term is explicitly the totally symmetric projection.  Replacing it by the raw derivative $\nabla_tP_{nn}$ would incorrectly move a Cotton term into the Schouten sector.

For completeness, the full first derivative can be obtained using
$\nabla_\mu r=-n_\mu$ and
\be
\nabla_\mu n_\nu
=
-\frac{1}{r}
\left(g_{\mu\nu}-n_\mu n_\nu\right)
+\bigo(rR).
\ee
To linear order in curvature one finds
\begin{equation}
\begin{split}
\nabla_\mu\widetilde X_{13}
={}&
-2rn_\mu
+\frac{r^3}{3}
\left(n_\mu P_{nn}+P_{n\mu}\right)
\\
&-r^4\left[
\frac{1}{12}n^\alpha n^\beta
\nabla_\mu P_{\alpha\beta}
-\frac{1}{6}n^\alpha\nabla_nP_{\alpha\mu}
-\frac{1}{6}n_\mu\nabla_nP_{nn}
\right]
+\bigo(\partial^4,\mathrm{curv}^2).
\end{split}
\end{equation}
Contracting with $t^\mu$ gives
\begin{equation}
\begin{split}
Y_t
={}&
-2rc
+\frac{r^3}{3}
\left(cP_{nn}+P_{tn}\right)
\\
&-r^4\left[
\frac{1}{12}\nabla_tP_{nn}
-\frac{1}{6}\nabla_nP_{tn}
-\frac{c}{6}\nabla_nP_{nn}
\right]
+\bigo(\partial^4,\mathrm{curv}^2).
\end{split}
\end{equation}

\subsection{Weyl part of the cosine law}

The off-diagonal Hessian of the world function has the expansion (see e.g. \cite{Decanini:2005eg})
\begin{equation}
 \nabla_\nu\nabla_\mu\sigma
 =g_{\mu\nu}-\frac13R_{\mu\rho\nu\lambda}\sigma^\rho\sigma^\lambda
 +\frac1{12}(\nabla_\eta R_{\mu\rho\nu\lambda})
 \sigma^\rho\sigma^\lambda\sigma^\eta+\ldots.
 \label{cos_Hessian}
\end{equation}
After the Schouten terms have been separated as in
\eqref{cos_S}, its trace-free curvature projection gives
\begin{equation}
 \left.Y_{tt}\right|_W
 =-\frac{2r^2}{3}W_{tntn}
 -\frac{r^3}{6}\nabla_nW_{tntn},
 \qquad
 \left.Y_{ttt}\right|_W
 =-\frac{r^2}{2}\nabla_tW_{tntn}.
 \label{cos_Wproj}
\end{equation}
The first term is the sectional-curvature correction associated with the
two-plane spanned by $t^\mu$ and $n^\mu$.  The remaining two terms measure the
variation of that sectional curvature along the two sides of the triangle.

Substituting \eqref{cos_Wproj} into
\eqref{cos_Taylor}, and using
$\widetilde X_{13}=r^2$ in a term which is already linear in the Weyl tensor,
we obtain
\begin{equation}
 \chi_W
 =-\frac{d^2}{3}W_{tntn}
 -\frac{d^2r}{12}\nabla_nW_{tntn}
 -\frac{d^3}{12}\nabla_tW_{tntn}.
 \label{cos_W}
\end{equation}

\subsection{Cotton part of the cosine law}

We next isolate the Cotton part of the first derivative of the
Schouten tensor. With the
convention $C_{\mu\nu\rho}=\nabla_\nu P_{\rho\mu}-\nabla_\rho P_{\nu\mu}$ used in
section~\ref{ambient_sec},
\begin{equation}
 \nabla_\mu P_{\nu\rho}
 =\nabla_{(\mu}P_{\nu\rho)}+\frac13C_{\rho\mu\nu}+\frac13C_{\nu\mu\rho}.
 \label{cos_dP}
\end{equation}
The contractions needed below are therefore
\begin{equation}
 \begin{aligned}
 \left.\nabla_tP_{nn}\right|_C&=\frac23C_{ntn},
 &\qquad
 \left.\nabla_nP_{tn}\right|_C&=-\frac13C_{ntn},
 \\
 \left.\nabla_nP_{tt}\right|_C&=\frac23C_{tnt},
 &
 \left.\nabla_tP_{tn}\right|_C&=-\frac13C_{tnt}.
 \end{aligned}
 \label{cos_Cproj}
\end{equation}
The last two equations follow in the same way as the first two.  They also
make it manifest that each Cotton contraction vanishes when $n^\mu=t^\mu$.

Differentiating \eqref{cos_X}, and then using
\eqref{cos_Cproj}, immediately gives
\begin{equation}
 \left.Y_t\right|_C=-\frac{r^4}{9}C_{ntn}.
 \label{cos_YtC}
\end{equation}
The Hessian requires a little more care.  There are three sources of terms
containing one derivative of the Schouten tensor in
$Y_{tt}=t^\mu t^\nu\nabla_\mu\nabla_\nu\widetilde X_{13}$.

First, the curvature-derivative term in the Synge Hessian contributes
\begin{equation}
 \left.(2\sigma)_{tt}\right|_{\nabla P}
 =-\frac{r^3}{6}
 \left(\nabla_nP_{tt}+\nabla_nP_{nn}-2c\nabla_nP_{tn}\right).
 \label{cos_src1}
\end{equation}
Using \eqref{cos_Cproj}, its Cotton part is
\begin{equation}
 \left.(2\sigma)_{tt}\right|_C
 =-\frac{r^3c}{9}C_{ntn}-\frac{r^3}{9}C_{tnt}.
 \label{cos_src1-C}
\end{equation}

Second, differentiating the quadratic Schouten term in
\eqref{cos_X} gives
\begin{equation}
 \left.\left(-\frac{\sigma}{3}\sigma^\mu\sigma^\nu P_{\mu\nu}\right)_{tt}
 \right|_{\nabla P}
 =\frac{2r^3}{3}
 \left(c\nabla_tP_{nn}+\nabla_tP_{tn}\right).
 \label{cos_src2}
\end{equation}
Its Cotton projection is
\begin{equation}
 \left.\left(-\frac{\sigma}{3}\sigma^\mu\sigma^\nu P_{\mu\nu}\right)_{tt}
 \right|_C
 =\frac{4r^3c}{9}C_{ntn}-\frac{2r^3}{9}C_{tnt}.
 \label{cos_src2-C}
\end{equation}

Finally, differentiating the cubic term in
\eqref{cos_X} produces a combination proportional to
\begin{equation}
 2c\left(\nabla_tP_{nn}+2\nabla_nP_{tn}\right)
 +\nabla_nP_{tt}+2\nabla_tP_{tn}.
 \label{cos_src3}
\end{equation}
Each of the two combinations has zero Cotton projection by
\eqref{cos_Cproj}.  Thus the cubic term contributes
only to the totally symmetric Schouten sector in this Hessian calculation.
Adding \eqref{cos_src1-C} and
\eqref{cos_src2-C} gives
\begin{equation}
 \left.Y_{tt}\right|_C
 =\frac{r^3}{3}(cC_{ntn}-C_{tnt}).
 \label{cos_YttC}
\end{equation}

It is important that the third derivative is not obtained by differentiating
only the Cotton projection in \eqref{cos_YttC}.  The part of the
full Hessian which is linear in the undifferentiated Schouten tensor is
\begin{equation}
 \left.Y_{tt}\right|_P=-r^2(P_{tt}+P_{nn}).
 \label{cos_YttP}
\end{equation}
Indeed, the world-function term and the quadratic Schouten term give,
respectively,
\begin{equation}
 \begin{split}
 &\left.(2\sigma)_{tt}\right|_P
 =-\frac{2r^2}{3}(P_{tt}+P_{nn}-2cP_{tn}),
 \\
 &\left.\left(-\frac{\sigma}{3}\sigma^\mu\sigma^\nu P_{\mu\nu}\right)_{tt}
 \right|_P
 =-\frac{r^2}{3}(P_{tt}+P_{nn}+4cP_{tn}),
 \end{split}
 \label{cos_YttP-sources}
\end{equation}
so that the mixed $P_{tn}$ terms cancel.  Differentiating
\eqref{cos_YttP} once more generates the Cotton term
\begin{equation}
 \left.\nabla_t\bigl(Y_{tt}|_P\bigr)\right|_C
 =-r^2\left[
 \left.\nabla_tP_{tt}\right|_C+
 \left.\nabla_tP_{nn}\right|_C\right]
 =-\frac{2r^2}{3}C_{ntn},
 \label{cos_YttP-to-C}
\end{equation}
because $\left.\nabla_tP_{tt}\right|_C=0$ and
$\left.\nabla_tP_{nn}\right|_C=2C_{ntn}/3$.  On the other hand, using
$\nabla_t r=-c$ and $\nabla_tn^\mu=-(t^\mu-cn^\mu)/r$, differentiation of
\eqref{cos_YttC} gives
\begin{equation}
 \nabla_t\left[\frac{r^3}{3}(cC_{ntn}-C_{tnt})\right]
 =r^2\left(-\frac13C_{ntn}+cC_{tnt}\right)+\bigo\!\left(r^3\nabla C,\mathrm{curv}^2\right).
 \label{cos_dYttC}
\end{equation}
Adding \eqref{cos_YttP-to-C} and
\eqref{cos_dYttC} therefore yields
\begin{equation}
 \left.Y_{ttt}\right|_C
 =r^2(-C_{ntn}+cC_{tnt})+\bigo\!\left(r^3\nabla C,\mathrm{curv}^2\right).
 \label{cos_YtttC}
\end{equation}

We finally record the mixed traces needed for the trace descendant in
section~\ref{sec:ansatz_OPE}. Let \(v^\mu\in T_{x_1}M\) be arbitrary.  For
\(v^\mu\neq0\), write \(v^\mu=\rho u^\mu\), where
\(u^\mu u_\mu=1\).  Applying equation~\eqref{cos_YttC} with
\(t^\mu\) replaced by \(u^\mu\), and using the bilinearity of the
Hessian, gives
\begin{equation}
\begin{aligned}
 \left.Y_{vv}\right|_C
 &=\rho^2\left.Y_{uu}\right|_C =\frac{r^3}{3}\rho^2
   \left[(u\mathbin{\cdot}n)C_{nun}-C_{unu}\right] \\
 &=\frac{r^3}{3}
   \left[(v\mathbin{\cdot}n)C_{nvn}-C_{vnv}\right].
\end{aligned}
\label{cos_YvvC}
\end{equation}
The final expression also holds trivially for \(v^\mu=0\).

To extract the mixed contraction, define
\(Q(v):=\left.Y_{vv}\right|_C\).  Symmetry and bilinearity of the
Hessian imply
\[
 Q(n+t)=Q(n)+2\left.Y_{nt}\right|_C+Q(t).
\]
Using \(n^2=1\), \(c=t\mathbin{\cdot}n\), and
\(C_{abc}=-C_{acb}\), equation~\eqref{cos_YvvC} gives
\[
 Q(n)=0,\qquad
 Q(t)=\frac{r^3}{3}\left(cC_{ntn}-C_{tnt}\right),
\]
and
\[
 Q(n+t)
 =\frac{r^3}{3}\left[(2+c)C_{ntn}-C_{tnt}\right].
\]
Therefore,
\be
 \left.Y_{nt}\right|_C
 =\frac12\bigl[Q(n+t)-Q(n)-Q(t)\bigr]
 =\frac{r^3}{3}C_{ntn}.
\ee
Together with
\be
\begin{split}
    (\nabla^\mu\widetilde X_{13})^{(0)}=-2rn^\mu, \quad (\nabla_\mu\nabla_t\widetilde X_{13})^{(0)}=2t_\mu, \quad t_\mu\left.\nabla^\mu\widetilde X_{13}\right|_C
=\left.Y_t\right|_C=-r^4C_{ntn}/9
\end{split}
\ee
this implies
\begin{equation}
\begin{aligned}
\left.
\nabla^\mu\widetilde X_{13}\,
\nabla_t\nabla_\mu\widetilde X_{13}
\right|_C
&=
-2r\left.Y_{nt}\right|_C
+2\left.Y_t\right|_C
=
-\frac{8r^4}{9}C_{ntn}.
\end{aligned}
\label{cos_mixed_trace}
\end{equation}

For the second trace, the required diagonal cubic contractions are
\begin{equation}
 \begin{split}
 \left.Y_{vvv}\right|_C
 &=r^2\left[-v^2C_{nvn}+(v\mathbin{\cdot}n)C_{vnv}\right],
 \\
 \left.Y_{vvv}\right|_{\nabla W}
 &=-\frac{r^2}{2}\nabla_vW_{vnvn}.
 \end{split}
 \label{cos_cubic}
\end{equation}
For a symmetric cubic tensor, taking two derivatives of its diagonal
polynomial with respect to $v^\mu$ gives six times its trace.  Applying this
identity to
\eqref{cos_cubic} gives
\begin{equation}
 \begin{split}
 g^{\mu\nu}t^\rho\left.Y_{(\mu\nu\rho)}\right|_C
 &=-\frac{D+3}{3}r^2C_{ntn},
 \\
 g^{\mu\nu}t^\rho\left.Y_{(\mu\nu\rho)}\right|_{\nabla W}
 &=-\frac{D-3}{3}r^2C_{ntn},
 \end{split}
 \label{cos_trace_parts}
\end{equation}
where we used $\nabla^\mu W_{\mu\nu\rho\lambda}=(D-3)C_{\nu\rho\lambda}$.  The trace of the
symmetrised third derivative differs from $\nabla_t\Box\widetilde X_{13}$ only by a
Ricci-tensor multiple of $\nabla Y$; this is a Schouten contribution and has
no Cotton projection at the order retained.  Consequently,
\begin{equation}
 \left.\nabla_t\Box\widetilde X_{13}\right|_C
 =-\frac{2D}{3}r^2C_{ntn}.
 \label{cos_trace3}
\end{equation}

Substitution of \eqref{cos_YtC},
\eqref{cos_YttC}, and \eqref{cos_YtttC} into
\eqref{cos_Taylor} gives
\begin{equation}
 \begin{split}
 \chi_C={}&
 \frac{d\left.Y_t\right|_C}{r^2}
 +\frac{d^2\left.Y_{tt}\right|_C}{2r^2}
 +\frac{d^3\left.Y_{ttt}\right|_C}{6r^2}
 \\
 ={}&-\frac19dr^2C_{ntn}
 +\frac16d^2r(cC_{ntn}-C_{tnt})
 +\frac16d^3(-C_{ntn}+cC_{tnt}).
 \end{split}
 \label{cos_C}
\end{equation}

Combining \eqref{cos_S},
\eqref{cos_W}, and
\eqref{cos_C}, we recover
\begin{equation}
\frac{\widetilde X_{23}}{\widetilde X_{13}}
=
1-2\frac{dc}{r}+\frac{d^2}{r^2}
+\chi_S+\chi_C+\chi_W
+\bigo(d^4,\partial^4,\mathrm{curv}^2).
\label{cos_full}
\end{equation}
Both $\chi_C$ and $\chi_W$ vanish for a degenerate triangle with
$n^\mu=t^\mu$: the Weyl contraction $W_{tttt}$ vanishes by antisymmetry, and the
derivative terms vanish as well because
$\nabla_tW_{tttt}=0$ when the tangent slots are parallel transported.  The
Cotton contractions $C_{ttt}$ vanish for the same reason.  This provides a
simple geometrical check on the result. 

\section{Details of the general-background OPE matching}
\label{app:sec4_matching}

This appendix gives the intermediate steps leading to the mismatch
\eqref{sec4_mismatch}. We first record the flat-space normalisation and the
action of the conformally flat descendant operator $C_{\rm cf}$, and then treat the Weyl and Cotton sectors separately.

\paragraph{Sector notation.}
Throughout this appendix, if \(F\) has been expanded to the curvature
and short-distance order under consideration, then
\(\left.F\right|_{\mathcal S}\) denotes the sum of the terms of \(F\)
belonging to the labelled sector \(\mathcal S\). 
In particular, \(\left.F\right|_W\) and \(\left.F\right|_C\) retain,
respectively, the terms linear in the independent Weyl and Cotton
curvature jets, after the flat and Schouten sectors have been removed,
and
\[
 \left.F\right|_{W,C}
 \equiv\left.F\right|_W+\left.F\right|_C
\]
at the linear-curvature order used here.  In \(D\geq4\), traced
derivatives of the Weyl tensor are assigned to the Cotton sector using
\(\nabla^\mu W_{\mu\nu\rho\lambda}=(D-3)C_{\nu\rho\lambda}\).  A refined label such as
\(d^2r\,C\) or \(d^3C\) additionally selects the indicated
short-distance order within the Cotton sector, whereas
\(\left.F\right|_{\Upsilon W}\) retains terms linear in both
\(\Upsilon_\mu=\nabla_
\mu\sigma\) and an undifferentiated Weyl tensor.
The selector acts on the entire delimited expression immediately to
its left.  When a Weyl variation is involved,
\(\delta_\sigma(\left.F\right|_W)\) means that the sector is selected
before varying, whereas
\(\left.\delta_\sigma F\right|_{\Upsilon W}\) means that the indicated
component is extracted after varying.

\subsection{Matching preliminaries}
\label{app:sec4_prelim}

Before turning on curvature, we have:
\[
\frac{\widetilde X_{23}}{\widetilde X_{13}}
=
1-2\frac{dc}{r}+\frac{d^2}{r^2}.
\]
Hence, after factoring
$\widetilde X_{13}^{-\Delta_3}$ from the right-hand side of
\eqref{sec4_OPE_test}, the remaining dimensionless factor expands as
\begin{equation}
\begin{split}
\left(
1-2\frac{dc}{r}+\frac{d^2}{r^2}
\right)^{-\alpha_1}
={}&
1+2\alpha_1\frac{dc}{r}
+2\alpha_1(\alpha_1+1)\frac{d^2c^2}{r^2}
-\alpha_1\frac{d^2}{r^2}
\\
&+
\frac{4}{3}\alpha_1(\alpha_1+1)(\alpha_1+2)
\frac{d^3c^3}{r^3}
-2\alpha_1(\alpha_1+1)\frac{d^3c}{r^3}
+\bigo(d^4).
\end{split}
\label{sec4_flat}
\end{equation}
The derivative part of the local differential operator that reproduces
this expansion follows directly from \eqref{C_flat}; we denote it by
$C_{\mathrm{der}}$.

For bookkeeping through cubic order, we decompose the conformally flat
operator as
\be\label{eq:C_der}
C_{\rm cf}(g_{\mu\nu},t^\mu,d)
=
C_{\rm der}(g_{\mu\nu},t^\mu,d)
+
C_{\rm Sch}(g_{\mu\nu},t^\mu,d)
+
\bigo(d^4).
\ee
This is not a division into differential and non-differential terms.
In the ordered covariant-derivative basis of \eqref{C_new_exp},
$C_{\rm der}$ contains those monomials whose coefficients contain no
explicit Schouten tensor or its covariant derivatives, whereas
$C_{\rm Sch}$ contains every monomial with an explicit
$P_{\mu\nu}$ or $\nabla_\rho P_{\mu\nu}$ coefficient. Thus a mixed
term such as $P_{\mu\nu}\nabla_\rho$ belongs to $C_{\rm Sch}$.
This decomposition is only a bookkeeping device; only the sum
$C_{\rm cf}=C_{\rm der}+C_{\rm Sch}$ has intrinsic significance.

For the Weyl and Cotton projections below, the required derivative
part is
\begin{equation}
 \begin{split}
 C_{\mathrm{der}}(g_{\mu\nu},t^\mu,d)={}1
 &+\frac{\alpha_1}{\Delta_3}d\nabla_t
 +\frac{\alpha_1(\alpha_1+1)}
 {2\Delta_3(\Delta_3+1)}d^2\nabla_t^2 +\frac{\alpha_1\alpha_2}
 {2\Delta_3(\Delta_3+1)(D-2-2\Delta_3)}d^2\Box
 \\
 &+\frac{\alpha_1(\alpha_1+1)(\alpha_1+2)}
 {6\Delta_3(\Delta_3+1)(\Delta_3+2)}d^3\nabla_t^3 +\frac{\alpha_1(\alpha_1+1)\alpha_2}
 {2\Delta_3(\Delta_3+1)(\Delta_3+2)(D-2-2\Delta_3)}
 d^3\nabla_t\Box
 \\
 &+\bigo(d^4).
 \end{split}
 \label{sec4_Cop}
\end{equation}
Here $\nabla_t=t^\mu\nabla_\mu$, and in the last term the ordering means
$\nabla_t(\Box\mathcal O_3)$. Equation~\eqref{sec4_Cop} is not a
second definition of the complete operator in \eqref{CC}. It displays
only the component $C_{\mathrm{der}}$ required for the independent
Weyl and Cotton projections below. The complementary component
$C_{\mathrm{Sch}}$ consists of the explicit Schouten terms in
\eqref{C_new_exp}, and their sum agrees with $C_{\mathrm{cf}}$ through cubic order.

The complete operator $C_{\mathrm{cf}}$ is the covariant operator
constructed in section~\ref{confflat_sec}. It is not obtained by the
naive replacement $\partial_\mu\mapsto\nabla_\mu$. Rather, its construction
combines the endpoint factor, the covariant expansion of $\Delta x^\mu$,
and the local map $\mathbb T$ from flat derivatives to their curved
expressions. The map $\mathbb T$ is not itself the Thomas operator; as
explained in subsection~\ref{sec:tractor_bundles}, it is extracted from
the tractor tensors generated by successive applications of the Thomas
operator and by matching their components. For example, equation \eqref{T3derivatives2} gives the fully symmetrised cubic
image,
\begin{equation}
 \begin{split}
 {\mathbb T}\partial_{(\mu}\partial_\nu\partial_{\rho)}\mathcal O_3
 ={}&\nabla_{(\mu}\nabla_\nu\nabla_{\rho)}\mathcal O_3
 -\Delta_3\bigl(\nabla_{(\mu}P_{\nu\rho)}\bigr)\mathcal O_3 -(3\Delta_3+2)P_{(\mu\nu}\nabla_{\rho)}\mathcal O_3
 +g_{(\mu\nu}P_{\rho)}{}^{\lambda}\nabla_\lambda\mathcal O_3.
 \end{split}
 \label{sec4_T3}
\end{equation}
Likewise, the second identity in \eqref{Tsecond_der} gives
\begin{equation}
 \begin{split}
 t^\nu{\mathbb T}(\partial_\nu\partial^2\mathcal O_3)
 =\bigl[&\nabla_t\Box-\Delta_3(\nabla_tP)-\Delta_3P\nabla_t
 +(D-2-2\Delta_3)P_t{}^\mu\nabla_\mu\bigr]\mathcal O_3,
 \end{split}
 \label{sec4_Ttrace}
\end{equation}
where $P=P^\mu{}_\mu$, in the notation of section~\ref{confflat_sec}. The explicit Schouten terms in these expressions are retained in
$C_{\mathrm{Sch}}$, and hence in $C_{\mathrm{cf}}$. Together with the
Schouten structures produced when $C_{\mathrm{der}}$ acts on
$\widetilde X_{13}^{-\Delta_3}$, they reproduce the complete
$\chi_S$ sector. In the calculation below we organise the final answer into its Schouten, Weyl, and Cotton projections. Thus no term is dropped from the operator itself; after the Schouten sector has been matched, we simply isolate the remaining independent Weyl and Cotton structures.

At linear order in curvature and through cubic order in $d$, the
explicit Schouten complement has no independent Weyl or Cotton
projection:
\[
\left.
\left[
C_{\mathrm{Sch}}(g_{\mu\nu},t^\mu,d)
\widetilde X_{13}^{-\Delta_3}
\right]\right|_{W,C}
=
\widetilde X_{13}^{-\Delta_3}
\bigo\!\left(d^4,\partial^4,\mathrm{curv}^2\right).
\]
Indeed, $C_{\mathrm{Sch}}$ acting on the curvature-independent part of
$\widetilde X_{13}^{-\Delta_3}$ produces only Schouten structures,
whereas its action on a Weyl or Cotton contribution is already
quadratic in curvature. The derivatives of $P_{\mu\nu}$ appearing in $C_{\mathrm{Sch}}$ occur only in symmetrised or traced combinations. They therefore do not produce the antisymmetric combination: $C_{\mu\nu\rho}=\nabla_\nu P_{\rho\mu}-\nabla_\rho P_{\nu\mu}$. 

Consequently,
\[
\left.
\left[
C_{\mathrm{cf}}(g_{\mu\nu},t^\mu,d)
\widetilde X_{13}^{-\Delta_3}
\right]\right|_{W,C}
=
\left.
\left[
C_{\mathrm{der}}(g_{\mu\nu},t^\mu,d)
\widetilde X_{13}^{-\Delta_3}
\right]\right|_{W,C}
+
\widetilde X_{13}^{-\Delta_3}
\bigo\!\left(d^4,\partial^4,\mathrm{curv}^2\right).
\]

For later comparison, the five derivative structures displayed above act in
flat space as follows:
\begin{equation}
\begin{aligned}
d\nabla_t\widetilde X_{13}^{-\Delta_3}
&=
2\Delta_3\frac{dc}{r}\,
\widetilde X_{13}^{-\Delta_3},
\\
d^2\nabla_t^2\widetilde X_{13}^{-\Delta_3}
&=
\left[
4\Delta_3(\Delta_3+1)\frac{d^2c^2}{r^2}
-2\Delta_3\frac{d^2}{r^2}
\right]
\widetilde X_{13}^{-\Delta_3},
\\
d^2\Box\widetilde X_{13}^{-\Delta_3}
&=
-2\Delta_3(D-2-2\Delta_3)
\frac{d^2}{r^2}\,
\widetilde X_{13}^{-\Delta_3},
\\
d^3\nabla_t^3\widetilde X_{13}^{-\Delta_3}
&=
\left[
8\Delta_3(\Delta_3+1)(\Delta_3+2)
\frac{d^3c^3}{r^3}
-12\Delta_3(\Delta_3+1)
\frac{d^3c}{r^3}
\right]
\widetilde X_{13}^{-\Delta_3},
\\
d^3\nabla_t\Box\widetilde X_{13}^{-\Delta_3}
&=
-4\Delta_3(\Delta_3+1)(D-2-2\Delta_3)
\frac{d^3c}{r^3}\,
\widetilde X_{13}^{-\Delta_3}.
\end{aligned}
\label{sec4_flat_act}
\end{equation}

Substitution into \eqref{sec4_Cop} reproduces \eqref{sec4_flat} term-by-term. This gives a useful normalisation and sign check before the curvature calculation.

At linear order in curvature, using the geodesic-triangle expansion \eqref{cos_full}, with the three curvature sectors given in \eqref{cos_S}, \eqref{cos_C}, and \eqref{cos_W}, we have:
\begin{equation}
\begin{split}
\left(
\frac{\widetilde X_{23}}{\widetilde X_{13}}
\right)^{-\alpha_1}
={}&
\left(
1-2\frac{dc}{r}+\frac{d^2}{r^2}
\right)^{-\alpha_1}
\\
&-
\alpha_1(\chi_S+\chi_C+\chi_W)
\left(
1-2\frac{dc}{r}+\frac{d^2}{r^2}
\right)^{-\alpha_1-1}
+\bigo(d^4,\partial^4,\mathrm{curv}^2).
\end{split}
\label{sec4_curv_exp}
\end{equation}
The combined Schouten projection of
$C_{\mathrm{der}}+C_{\mathrm{Sch}}$, whose sum agrees with
$C_{\mathrm{cf}}$ through cubic order, reproduces the $\chi_S$
contribution. This is the conformally flat result of section~\ref{confflat_sec}, now written in geodesic variables. Consequently, there is no Schouten residual. The non-trivial comparison is between the $\chi_W$ and $\chi_C$ terms on the right-hand side and the Weyl and Cotton responses of $C_{\rm der}$ on the left-hand side.

Now using \eqref{sec4_curv_exp}, we obtain
\begin{equation}
\begin{aligned}
\left.
\left[
 C_{\mathrm{cf}}(g_{\mu\nu},t^\mu,d)
 \widetilde X_{13}^{-\Delta_3}
 -
 \frac{1}{
   \widetilde X_{13}^{\alpha_2}
   \widetilde X_{23}^{\alpha_1}}
\right]
\right|_{W,C}
={}&
\left.
\left[
 C_{\mathrm{cf}}(g_{\mu\nu},t^\mu,d)
 \widetilde X_{13}^{-\Delta_3}
\right]
\right|_{W,C}
\\
&+\alpha_1\widetilde X_{13}^{-\Delta_3}
 (\chi_C+\chi_W)
 \left(
   1-2\frac{dc}{r}+\frac{d^2}{r^2}
 \right)^{-\alpha_1-1}
\\
&+\widetilde X_{13}^{-\Delta_3}
 O(d^4,\partial^4,\mathrm{curv}^2).
\end{aligned}
\label{sec4_projected_comparison}
\end{equation}

\subsection{The Weyl-tensor sector}
\label{app:sec4_W_matching}

We now evaluate the two terms on the right-hand side of
\eqref{sec4_projected_comparison}, first in the Weyl sector and then
in the Cotton sector.

From \eqref{cos_W},
\begin{equation}
 \chi_W=-\frac{d^2}{3}W_{tntn}
 -\frac{d^2r}{12}\nabla_nW_{tntn}
 -\frac{d^3}{12}\nabla_tW_{tntn}.
 \label{sec4_chiW}
\end{equation}
Since this expression starts at order $d^2$, it is sufficient to use
\begin{equation}
\left(
1-2\frac{dc}{r}+\frac{d^2}{r^2}
\right)^{-\alpha_1-1}
=
1+2(\alpha_1+1)\frac{dc}{r}
+\bigo(d^2).
\label{sec4_binW}
\end{equation}
in \eqref{sec4_curv_exp}. We then find
\begin{equation}
 \begin{split}
 \left.
 \frac{1}{\widetilde X_{13}^{\alpha_2}
              \widetilde X_{23}^{\alpha_1}}
 \right|_W
 =\,&\widetilde X_{13}^{-\Delta_3}\bigg[{}
 \frac{\alpha_1}{3}d^2W_{tntn}
 +\frac{2\alpha_1(\alpha_1+1)}{3}\frac{d^3c}{r}W_{tntn} +\frac{\alpha_1}{12}d^2r\nabla_nW_{tntn}
 +\frac{\alpha_1}{12}d^3\nabla_tW_{tntn}
 \\
 &+\frac{\alpha_1(\alpha_1+1)}{6}
 d^3c\,\nabla_nW_{tntn} + \bigo\!\left(d^4,\partial^4,\mathrm{curv}^2\right)\bigg].
 \end{split}
 \label{sec4_pairW}
\end{equation}

We next calculate the part generated by $C_{\rm der}$.  The
chain rule, together with the Weyl projections of the derivatives of
$\widetilde X_{13}$ given in appendix~\ref{app:cosine-law}, gives
\begin{equation}
 \begin{split}
 \frac{\left.d^2\nabla_t^2
 \widetilde X_{13}^{-\Delta_3}\right|_W}
 {\widetilde X_{13}^{-\Delta_3}}
 ={}&\frac{2\Delta_3}{3}d^2W_{tntn}
 +\frac{\Delta_3}{6}d^2r\nabla_nW_{tntn},
 \\
 \frac{\left.d^3\nabla_t^3
 \widetilde X_{13}^{-\Delta_3}\right|_W}
 {\widetilde X_{13}^{-\Delta_3}}
 ={}&4\Delta_3(\Delta_3+1)\frac{d^3c}{r}W_{tntn}
 +\frac{\Delta_3}{2}d^3\nabla_tW_{tntn} +\Delta_3(\Delta_3+1)d^3c\,\nabla_nW_{tntn}.
 \end{split}
 \label{sec4_descW}
\end{equation}
The remaining derivative structures in \eqref{sec4_Cop} do not contribute to the independent Weyl projections displayed above. The trace of $\nabla W$ is converted by $\nabla^\mu W_{\mu\nu\rho\lambda}=(D-3)C_{\nu\rho\lambda}$ into the Cotton response treated below. Multiplying
\eqref{sec4_descW} by the corresponding coefficients in
\eqref{sec4_Cop}, we obtain
\begin{equation}
 \begin{split}
\left.
C_{\mathrm{der}}(g_{\mu\nu},t^\mu,d)
\widetilde X_{13}^{-\Delta_3}
\right|_W
 &=\widetilde X_{13}^{-\Delta_3}\bigg[{}
 \frac{\alpha_1(\alpha_1+1)}{3(\Delta_3+1)}d^2W_{tntn} +\frac{2\alpha_1(\alpha_1+1)(\alpha_1+2)}
 {3(\Delta_3+2)}\frac{d^3c}{r}W_{tntn}\\ &+\frac{\alpha_1(\alpha_1+1)}{12(\Delta_3+1)}
 d^2r\nabla_nW_{tntn} +\frac{\alpha_1(\alpha_1+1)(\alpha_1+2)}
 {12(\Delta_3+1)(\Delta_3+2)}d^3\nabla_tW_{tntn}\\ &+\frac{\alpha_1(\alpha_1+1)(\alpha_1+2)}
 {6(\Delta_3+2)}d^3c\,\nabla_nW_{tntn} + \bigo\!\left(d^4,\partial^4,\mathrm{curv}^2\right)\bigg].
 \end{split}
 \label{sec4_localW}
\end{equation}
Since $C_{\mathrm{Sch}}$ has no independent Weyl projection at this
order, equation~\eqref{sec4_localW} is also the Weyl response of
$C_{\mathrm{cf}}$ through cubic order, modulo the displayed remainder.

Subtracting \eqref{sec4_localW} from
\eqref{sec4_pairW}, and repeatedly using
$\Delta_3=\alpha_1+\alpha_2$, gives
\begin{equation}
 \begin{split}
 \left.
 \left[
 \frac{1}{\widetilde X_{13}^{\alpha_2}
              \widetilde X_{23}^{\alpha_1}}
 -C_{\mathrm{cf}}(g_{\mu\nu},t^\mu,d)\widetilde X_{13}^{-\Delta_3}
 \right]\right|_W
 =\widetilde X_{13}^{-\Delta_3}\bigg[{}
 &\frac{\alpha_1\alpha_2}{3(\Delta_3+1)}d^2W_{tntn} +\frac{2\alpha_1(\alpha_1+1)\alpha_2}
 {3(\Delta_3+2)}\frac{d^3c}{r}W_{tntn}
 \\
 &+\frac{\alpha_1\alpha_2}{12(\Delta_3+1)}
 d^2r\nabla_nW_{tntn} +\frac{\alpha_1\alpha_2(2\alpha_1+\alpha_2+3)}
 {12(\Delta_3+1)(\Delta_3+2)}d^3\nabla_tW_{tntn}
 \\
 &+\frac{\alpha_1(\alpha_1+1)\alpha_2}
 {6(\Delta_3+2)}d^3c\,\nabla_nW_{tntn} + \bigo\!\left(d^4,\partial^4,\mathrm{curv}^2\right)\bigg].
 \end{split}
 \label{sec4_resW}
\end{equation}
Thus the pairwise ambient expression alone fails the local scalar-family OPE test already at order $d^2W_{tntn}$.

\subsection{The Cotton-tensor sector}
\label{app:sec4_C_matching}

We now repeat the comparison for the Cotton tensor. Using the convention
$C_{\mu\nu\rho}=\nabla_\nu P_{\rho\mu}-\nabla_\rho P_{\nu\mu}$ of section~\ref{ambient_sec}, the Cotton correction to the cosine law is given by \eqref{cos_C}:
\begin{equation}
 \chi_C=-\frac{1}{9}dr^2C_{ntn}
 +\frac{1}{6}d^2r(cC_{ntn}-C_{tnt})
 +\frac{1}{6}d^3(-C_{ntn}+cC_{tnt}).
 \label{sec4_chiC}
\end{equation}
Because the first term is linear in $d$, the required binomial expansion is
now
\begin{equation}
\begin{split}
\left(
1-2\frac{dc}{r}+\frac{d^2}{r^2}
\right)^{-\alpha_1-1}
={}&
1+2(\alpha_1+1)\frac{dc}{r}
+2(\alpha_1+1)(\alpha_1+2)
\frac{d^2c^2}{r^2}
-
(\alpha_1+1)\frac{d^2}{r^2}
+\bigo(d^3).
\end{split}
\label{sec4_binC}
\end{equation}
Substituting \eqref{sec4_chiC} and
\eqref{sec4_binC} into
\eqref{sec4_curv_exp}, we find
\begin{equation}
 \begin{split}
 \left.
 \frac{1}{\widetilde X_{13}^{\alpha_2}
              \widetilde X_{23}^{\alpha_1}}
 \right|_C
 =\widetilde X_{13}^{-\Delta_3}\bigg\{{}
 &\frac{\alpha_1}{9}dr^2C_{ntn} +d^2r\left[
 \frac{\alpha_1(4\alpha_1+1)}{18}cC_{ntn}
 +\frac{\alpha_1}{6}C_{tnt}\right]
 \\
 &+d^3\left[
 \left(
 \frac{\alpha_1(\alpha_1+1)(2\alpha_1+1)}{9}c^2
 +\frac{\alpha_1(1-2\alpha_1)}{18}
 \right)C_{ntn}
 +\frac{\alpha_1(2\alpha_1+1)}{6}cC_{tnt}
 \right]\bigg\}\\
 &+\widetilde X_{13}^{-\Delta_3}\bigo\!\left(d^4,\partial^4,\mathrm{curv}^2\right).
 \end{split}
 \label{sec4_pairC}
\end{equation}

The corresponding Cotton pieces generated by $C_{\rm der}$ in
\eqref{sec4_Cop} are
\begin{equation}
 \begin{aligned}
 \frac{\left.d\nabla_t\widetilde X_{13}^{-\Delta_3}\right|_C}
 {\widetilde X_{13}^{-\Delta_3}}
 & =\frac{\Delta_3}{9}dr^2C_{ntn},
 \\
 \frac{\left.d^2\nabla_t^2\widetilde X_{13}^{-\Delta_3}\right|_C}
 {\widetilde X_{13}^{-\Delta_3}}
 & =\Delta_3d^2r\left[
 \frac{4\Delta_3+1}{9}cC_{ntn}+\frac{1}{3}C_{tnt}\right],
 \\
 \frac{\left.d^3\nabla_t^3\widetilde X_{13}^{-\Delta_3}\right|_C}
 {\widetilde X_{13}^{-\Delta_3}}
 & =\Delta_3d^3\bigg\{
 \left[\frac{2(\Delta_3+1)(2\Delta_3+1)}{3}c^2
 +\frac{1-2\Delta_3}{3}\right]C_{ntn}
 \\
 &\hspace{3.5cm} +(2\Delta_3+1)cC_{tnt}\bigg\},
 \\
 \frac{\left.d^3\nabla_t\Box\widetilde X_{13}^{-\Delta_3}\right|_C}
 {\widetilde X_{13}^{-\Delta_3}}
 & =\frac{2\Delta_3(2-\Delta_3)(D-2-2\Delta_3)}{9}
 d^3C_{ntn}.
\end{aligned}
\label{sec4_descC}
\end{equation}
There is no Cotton term in $d^2\Box\widetilde X_{13}^{-\Delta_3}$ at this
order.  The last line of \eqref{sec4_descC} is worth
deriving explicitly.  The chain rule gives
\begin{equation}
 \begin{split}
 \frac{\nabla_t\Box\widetilde X_{13}^{-\Delta_3}}
 {\widetilde X_{13}^{-\Delta_3}}
 ={}&-\Delta_3(\Delta_3+1)(\Delta_3+2)
 \frac{(\nabla_t\widetilde X_{13})
       (\nabla^\mu\widetilde X_{13}\nabla_\mu\widetilde X_{13})}
      {\widetilde X_{13}^{3}}
 \\
 &+\Delta_3(\Delta_3+1)
 \frac{2\nabla^\mu\widetilde X_{13}
          \nabla_t\nabla_\mu\widetilde X_{13}
       +(\nabla_t\widetilde X_{13})\Box\widetilde X_{13}}
      {\widetilde X_{13}^{2}}
 -\Delta_3\frac{\nabla_t\Box\widetilde X_{13}}
                    {\widetilde X_{13}}.
 \end{split}
 \label{sec4_trace_chain}
\end{equation}
At the order under consideration,
\begin{equation}
 \begin{gathered}
 \left.\nabla_t\widetilde X_{13}\right|_C
 =-\frac{r^4}{9}C_{ntn},
 \qquad
 \left.\nabla^\mu\widetilde X_{13}
       \nabla_t\nabla_\mu\widetilde X_{13}\right|_C
 =-\frac{8r^4}{9}C_{ntn},
 \\
 \left.\nabla_t\Box\widetilde X_{13}\right|_C
 =-\frac{2D}{3}r^2C_{ntn},
 \qquad
 \left.\nabla^\mu\widetilde X_{13}\nabla_\mu\widetilde X_{13}\right|_C
 =\left.\Box\widetilde X_{13}\right|_C=0,
 \end{gathered}
 \label{sec4_trace_data}
\end{equation}
The two mixed traces in the second and third expressions are derived in
\eqref{cos_mixed_trace} and
\eqref{cos_trace3}.  In $D\geq4$, the Cotton
projection of $\nabla_t\Box\widetilde X_{13}$ includes the divergence of the
$\nabla W$ term through $\nabla^\mu W_{\mu\nu\rho\lambda}=(D-3)C_{\nu\rho\lambda}$.
The corresponding flat values are
$\nabla^\mu\widetilde X_{13}\nabla_\mu\widetilde X_{13}=4r^2$ and
$\Box\widetilde X_{13}=2D$.  Substitution into
\eqref{sec4_trace_chain} gives
\begin{equation}
 \begin{split}
 \frac{\left.d^3\nabla_t\Box
 \widetilde X_{13}^{-\Delta_3}\right|_C}
 {\widetilde X_{13}^{-\Delta_3}}
 &=d^3\bigg[{}
 \frac{4}{9}\Delta_3(\Delta_3+1)(\Delta_3+2) -\frac{2}{9}\Delta_3(\Delta_3+1)(D+8)
 +\frac{2D}{3}\Delta_3\bigg]C_{ntn}
 \\
 &=\frac{2\Delta_3(2-\Delta_3)(D-2-2\Delta_3)}{9}
 d^3C_{ntn},
 \end{split}
 \label{sec4_traceC}
\end{equation}
which proves the last line of
\eqref{sec4_descC}. This shows that the trace descendant receives contributions from every term in the chain rule.  In $D\geq4$, the same trace descendant also combines the traced $\nabla W$ projection with the
Cotton contribution.

Multiplying \eqref{sec4_descC} by the
coefficients in \eqref{sec4_Cop} gives
\begin{equation}
\begin{aligned}
\left.
\left[
C_{\mathrm{der}}(g_{\mu\nu},t^\mu,d)
\widetilde X_{13}^{-\Delta_3}
\right]
\right|_C
={}&
\widetilde X_{13}^{-\Delta_3}
\Bigg\{
\frac{\alpha_1}{9}\,d r^2 C_{ntn}
+d^2r
\left[
\frac{\alpha_1(\alpha_1+1)(4\Delta_3+1)}
     {18(\Delta_3+1)}
cC_{ntn}
+
\frac{\alpha_1(\alpha_1+1)}
     {6(\Delta_3+1)}
C_{tnt}
\right]
\\
&
+d^3
\Bigg[
\frac{\alpha_1(\alpha_1+1)(\alpha_1+2)(2\Delta_3+1)}
     {9(\Delta_3+2)}
c^2C_{ntn}
\\
&
+
\left(
\frac{\alpha_1(\alpha_1+1)(\alpha_1+2)(1-2\Delta_3)}
     {18(\Delta_3+1)(\Delta_3+2)}
+
\frac{\alpha_1(\alpha_1+1)\alpha_2(2-\Delta_3)}
     {9(\Delta_3+1)(\Delta_3+2)}
\right)C_{ntn}
\\
&
+
\frac{\alpha_1(\alpha_1+1)(\alpha_1+2)(2\Delta_3+1)}
     {6(\Delta_3+1)(\Delta_3+2)}
cC_{tnt}
\Bigg]
+\bigo\!\left(d^4,\partial^4,\mathrm{curv}^2\right)
\Bigg\}.
\end{aligned}
\label{sec4_localC}
\end{equation}
Since $C_{\mathrm{Sch}}$ has no independent Cotton projection at this
order, equation~\eqref{sec4_localC} is also the Cotton response of
$C_{\mathrm{cf}}$ through cubic order, modulo the displayed remainder.

After factoring out the common
$\widetilde X_{13}^{-\Delta_3}$, the first term in
\eqref{sec4_pairC} is reproduced completely by the first descendant:
\begin{equation}
 \frac{\alpha_1}{9}dr^2C_{ntn}
 -\frac{\alpha_1}{\Delta_3}
 \frac{\Delta_3}{9}dr^2C_{ntn}=0.
 \label{sec4_leadC}
\end{equation}
Thus, the raw $dr^2C_{ntn}$ term in the conformal cosine law is not itself a residual. At the next order, subtracting
\eqref{sec4_localC} from \eqref{sec4_pairC} gives
\begin{equation}
 \left.
 \left[
 \frac{1}{\widetilde X_{13}^{\alpha_2}
              \widetilde X_{23}^{\alpha_1}}
 -C_{\mathrm{cf}}(g_{\mu\nu},t^\mu,d)\widetilde X_{13}^{-\Delta_3}
 \right]\right|_{d^2rC}
 =\widetilde X_{13}^{-\Delta_3}
 \frac{\alpha_1\alpha_2}{6(\Delta_3+1)}
 d^2r(C_{tnt}-cC_{ntn}).
 \label{sec4_resC2}
\end{equation}
At cubic order, the same subtraction yields
\begin{equation}
 \begin{split}
 \left.
 \left[
 \frac{1}{\widetilde X_{13}^{\alpha_2}
              \widetilde X_{23}^{\alpha_1}}
 -C_{\mathrm{cf}}(g_{\mu\nu},t^\mu,d)\widetilde X_{13}^{-\Delta_3}
 \right]\right|_{d^3C}
 ={}&\widetilde X_{13}^{-\Delta_3}
 \frac{\alpha_1\alpha_2d^3}
 {6(\Delta_3+1)(\Delta_3+2)}
 \bigg[\left\{\alpha_2+1
 -2(\alpha_1+1)(\Delta_3+1)c^2\right\}C_{ntn}
 \\
 &+\left(2\alpha_1^2+2\alpha_1\alpha_2
 +2\alpha_1+\alpha_2-1\right)cC_{tnt}
 \bigg].
 \end{split}
 \label{sec4_resC3}
\end{equation}
Equations \eqref{sec4_resC2} and
\eqref{sec4_resC3} are the Cotton analogue of
\eqref{sec4_resW}.

\subsection{Weyl covariance of the curvature correction}
\label{app:sec4_Weyl_covariance}

Now let us provide a Weyl-covariance check referred to in
section~\ref{sec:ansatz_OPE}. It is a consistency check on the coefficients of the short-distance correction \eqref{sec4_J}, rather than an independent derivation of that correction.

Under a constant rescaling
$g_{\mu\nu}\mapsto\Lambda^2g_{\mu\nu}$,
\begin{equation}
 d\mapsto\Lambda d,
 \qquad r\mapsto\Lambda r,
 \qquad t^\mu\mapsto\Lambda^{-1}t^\mu,
 \qquad n^\mu\mapsto\Lambda^{-1}n^\mu.
 \label{sec4_const_data}
\end{equation}
The quantities $c$ and $d/r$ are invariant. The all-lower Weyl tensor has weight two, while the all-lower Cotton tensor is invariant under a constant rescaling. It follows that
\begin{equation}
 W_{tntn}\mapsto\Lambda^{-2}W_{tntn},
 \qquad
 \nabla_tW_{tntn},\nabla_nW_{tntn}\mapsto\Lambda^{-3}(\cdots),
 \qquad
 C_{ntn},C_{tnt}\mapsto\Lambda^{-3}(\cdots).
 \label{sec4_weights}
\end{equation}
Every term in \eqref{sec4_J} therefore has Weyl weight zero. The more restrictive test is a non-constant infinitesimal rescaling.  Following
\eqref{gen_Weyl_metric}, we write
\begin{equation}
 \hat g_{\mu\nu}=e^{2\sigma(x)}g_{\mu\nu},
 \qquad
 \Upsilon_\mu=\nabla_\mu\sigma,
 \qquad
 \Upsilon_t=\Upsilon_\mu t^\mu,
 \qquad
 \Upsilon_n=\Upsilon_\mu n^\mu.
 \label{sec4_Weyl_data}
\end{equation}
We keep the inhomogeneous terms linear in $\Upsilon_\mu$, linear in
conformal curvature, and through cubic order in $d$. The
derivative-of-$\sigma$ parts of the geodesic data are
\begin{align}
 \begin{split}
 \delta_\sigma d&=\frac{d^2}{2}\Upsilon_t+\cdots,
 \qquad\qquad \qquad\ \ \ 
 \delta_\sigma r=\frac{r^2}{2}\Upsilon_n+\cdots,
 \\
 \delta_\sigma t^\mu&=-\frac d2(\Upsilon^\mu-t^\mu\Upsilon_t)+\cdots,
 \qquad
 \delta_\sigma n^\mu=-\frac r2(\Upsilon^\mu-n^\mu\Upsilon_n)+\cdots.
 \end{split}
 \label{sec4_geo_Weyl}
\end{align}
With our Cotton convention,
\begin{equation}
 \delta_\sigma C_{\mu\nu\rho}=\Upsilon^\lambda W_{\lambda \mu\nu\rho}.
 \label{sec4_deltaC}
\end{equation}
Consequently
\begin{equation}
 \delta_\sigma C_{ntn}=\Upsilon^\mu n^\nu t^\rho n^\lambda W_{\mu\nu\rho\lambda},
 \qquad
 \delta_\sigma C_{tnt}=t^\mu\Upsilon^\nu t^\rho n^\lambda W_{\mu\nu\rho\lambda},
 \label{sec4_deltaCproj}
\end{equation}

The inhomogeneous connection variations of the first derivatives of the
Weyl tensor are
\begin{equation}
 \begin{split}
 \left.\delta_\sigma\!\left(\nabla_tW_{tntn}\right)
\right|_{\mathrm{inhom}}
 &=-4\Upsilon_tW_{tntn}+2\Upsilon^\mu n^\nu t^\rho n^\lambda W_{\mu\nu\rho\lambda}+2ct^\mu\Upsilon^\nu t^\rho n^\lambda W_{\mu\nu\rho\lambda},
 \\
 \left.\delta_\sigma\!\left(\nabla_nW_{tntn}\right)
\right|_{\mathrm{inhom}}
 &=-4\Upsilon_nW_{tntn}+2c\Upsilon^\mu n^\nu t^\rho n^\lambda W_{\mu\nu\rho\lambda}+2t^\mu\Upsilon^\nu t^\rho n^\lambda W_{\mu\nu\rho\lambda}.
 \end{split}
 \label{sec4_delta_dW}
\end{equation}
Together with \eqref{sec4_geo_Weyl}, these formulae imply
the following variations.  We display only the terms proportional to
$\Upsilon W$ which participate in the inhomogeneous $\nabla W$--Cotton
mixing. Variations proportional to $\Upsilon\nabla W$ or $\Upsilon C$ belong
to the higher curvature-jet completion omitted here and are included
in the general remainder.
\begin{equation}
 \begin{split}
 \left.\delta_\sigma(d^2W_{tntn})\right|_{\Upsilon W}={}&
 2d^3\Upsilon_tW_{tntn}+d^2r\Upsilon_nW_{tntn} -d^3\Upsilon^\mu n^\nu t^\rho n^\lambda W_{\mu\nu\rho\lambda}
 -d^2rt^\mu\Upsilon^\nu t^\rho n^\lambda W_{\mu\nu\rho\lambda}+\bigo(d^4),
 \\
 \left.\delta_\sigma(d^2r\nabla_nW_{tntn})\right|_{\Upsilon W}={}&
 d^2r\bigl(2c\Upsilon^\mu n^\nu t^\rho n^\lambda W_{\mu\nu\rho\lambda} + 2t^\mu\Upsilon^\nu t^\rho n^\lambda W_{\mu\nu\rho\lambda}
 -4\Upsilon_nW_{tntn}\bigr)+\bigo(d^4),
 \\
 \left.\delta_\sigma(d^3\nabla_tW_{tntn})\right|_{\Upsilon W}={}&
 d^3\bigl(2\Upsilon^\mu n^\nu t^\rho n^\lambda W_{\mu\nu\rho\lambda} + 2ct^\mu\Upsilon^\nu t^\rho n^\lambda W_{\mu\nu\rho\lambda}
 -4\Upsilon_tW_{tntn}\bigr)+\bigo(d^4),
 \\
 \left.
\delta_\sigma
\left(
d^3c\,\nabla_nW_{tntn}
\right)
\right|_{\Upsilon W}={}&
 d^3\bigl(2c^2\Upsilon^\mu n^\nu t^\rho n^\lambda W_{\mu\nu\rho\lambda} + 2ct^\mu\Upsilon^\nu t^\rho n^\lambda W_{\mu\nu\rho\lambda}
 -4c\Upsilon_nW_{tntn}\bigr)+\bigo(d^4).
 \end{split}
 \label{sec4_delta_Wstr}
\end{equation}
For the remaining structure, one first finds
\begin{equation}
\delta_\sigma\left(\frac{dc}{r}\right)
=
\left(
\frac{d^2c}{r}-\frac d2
\right)\Upsilon_t
-\frac{d^2}{2r}\Upsilon_n
+\cdots .
\label{sec4_delta_dcr}
\end{equation}
and hence
\begin{equation}
\begin{split}
\left.
\delta_\sigma
\left(
\frac{d^3c}{r}W_{tntn}
\right)
\right|_{\Upsilon W}
={}&
-\frac12d^3\Upsilon_tW_{tntn}
+d^3c\,\Upsilon_nW_{tntn}
\\
&-
d^3c\,t^\mu\Upsilon^\nu t^\rho n^\lambda W_{\mu\nu\rho\lambda}
+\bigo(d^4).
\end{split}
\label{sec4_delta_dcrW}
\end{equation}

Substituting \eqref{sec4_delta_Wstr} and
\eqref{sec4_delta_dcrW} into the Weyl-tensor part of
\eqref{sec4_J}, all terms proportional to
$\Upsilon_tW_{tntn}$ and $\Upsilon_nW_{tntn}$ cancel.  The remaining
variation is
\begin{equation}
\begin{aligned}
\delta_\sigma\!\left(\left.J^{(1)}\right|_W\right)
={}&
\frac{\alpha_1\alpha_2}{6(\Delta_3+1)}d^2r
\Big(
t^\mu\Upsilon^\nu t^\rho n^\lambda W_{\mu\nu\rho\lambda}
-c\Upsilon^\mu n^\nu t^\rho n^\lambda W_{\mu\nu\rho\lambda}
\Big)
\\
&+
\frac{\alpha_1\alpha_2d^3}
     {6(\Delta_3+1)(\Delta_3+2)}
\Big[
(\alpha_2+1)(1-2c^2)
\Upsilon^\mu n^\nu t^\rho n^\lambda W_{\mu\nu\rho\lambda}
\\
&\hspace{4cm}
-(2\alpha_1-\alpha_2+1)c\,
t^\mu\Upsilon^\nu t^\rho n^\lambda W_{\mu\nu\rho\lambda}
\Big].
\end{aligned}
\label{sec4_delta_JW}
\end{equation}
On the other hand, using only the inhomogeneous transformation
\eqref{sec4_deltaCproj}, the Cotton part varies as
\begin{equation}
\begin{split}
\delta_\sigma\!\left(\left.J^{(1)}\right|_C\right)={}&
 \frac{\alpha_1\alpha_2}{6(\Delta_3+1)}d^2r
 \left(c\Upsilon^\mu n^\nu t^\rho n^\lambda W_{\mu\nu\rho\lambda}-t^\mu\Upsilon^\nu t^\rho n^\lambda W_{\mu\nu\rho\lambda}\right)
 \\
 &+\frac{\alpha_1\alpha_2d^3}
 {6(\Delta_3+1)(\Delta_3+2)}
 \bigg[
 (\alpha_2+1)(2c^2-1)\Upsilon^\mu n^\nu t^\rho n^\lambda W_{\mu\nu\rho\lambda}
 \\
 &\hspace{4.0cm}
 +(2\alpha_1-\alpha_2+1)ct^\mu\Upsilon^\nu t^\rho n^\lambda W_{\mu\nu\rho\lambda}
 \bigg].
 \end{split}
 \label{sec4_delta_JC}
\end{equation}
Equations \eqref{sec4_delta_JW} and \eqref{sec4_delta_JC} cancel term
by term.  At the retained curvature-jet order, we have therefore shown
\begin{equation}
 \left.\delta_\sigma J^{(1)}\right|_{\Upsilon W}=\bigo(d^4).
 \label{sec4_Weyl_check}
\end{equation}
This is a check of the inhomogeneous $\nabla W$--Cotton mixing at the retained curvature-jet order. Terms proportional to $\Upsilon C$ are part of the omitted covariant $\nabla C$ completion. Thus, \eqref{sec4_Weyl_check} should not be interpreted as a construction of the complete all-jet trilocal invariant.  In even dimensions, anomalous contact terms and anomalous operator mixing are likewise outside this
separated-point kinematic check.

The cancellation provides two independent checks on the preceding
calculation. The $d^2r$ terms fix the coefficient of $(Y_{tt})_C$ in appendix~\ref{app:cosine-law}, while the constant $d^3C_{ntn}$ term tests the trace response in the last line of \eqref{sec4_descC}. For $D=3$, the Weyl tensor vanishes identically, so the Cotton tensor has no inhomogeneous Weyl-tensor mixing. The displayed Cotton contribution is separately covariant at this conformal-curvature jet order, up to the omitted covariant $\nabla C$
completion.  For $D\geq4$, the displayed $\nabla W$ and Cotton mixing terms cancel only in their sum.

\bibliographystyle{SciPost_bib}
\bibliography{cov}

\end{document}